\documentclass[12pt,a4paper]{article}
\pdfoutput=1

\newif{\ifarxiv}
\newif{\ifdraft}
\newif{\ifremarks}

\arxivtrue % for stuff that should only go into the arxiv preprint
\remarkstrue % turn remarks on/off

\ifdraft\else\fi

\ifremarks
\newcommand{\remarktb}[1]{{\renewcommand{\bfdefault}{b}\color[RGB]{0,150,0}{\textbf{#1}}}}
\newcommand{\remarktf}[1]{{\renewcommand{\bfdefault}{b}\color[RGB]{150,0,0}{\textbf{#1}}}}
\newcommand{\remarkvg}[1]{{\renewcommand{\bfdefault}{b}\color[RGB]{0,0,150}{\textbf{#1}}}}
\fi
\providecommand{\remarktb}[1]{\ignorespaces}
\providecommand{\remarktf}[1]{\ignorespaces}
\providecommand{\remarkvg}[1]{\ignorespaces}

\usepackage[utf8]{inputenc}
\usepackage[bookmarks=true]{hyperref}
\usepackage{xcolor}
\usepackage{graphicx}
\definecolor{darkblue}{cmyk}{0.9,0.9,0,0}
\usepackage{cite}

\usepackage{listings}
\definecolor{mathgreen}{RGB}{0,90,39}
\makeatletter
\let\orig@item\item
\def\item{%
    \@ifnextchar{[}%
        {\lstinline@item}%
        {\orig@item}%
}
\begingroup
\catcode`\]=\active
\gdef\lstinline@item[{%
    \setbox0\hbox\bgroup
        \catcode`\]=\active
        \let]\lstinline@item@end
}
\endgroup
\def\lstinline@item@end{%
    \egroup
    \orig@item[\usebox0]%
}
\makeatother

\providecommand{\hypersetup}[1]{}
\providecommand{\hyperref}[2][]{#2}
\providecommand{\texorpdfstring}[2]{#1}
\providecommand{\pdfbookmark}[3][]{}
\newcommand{\email}[1]{\href{mailto:#1}{\nolinkurl{#1}}}

\hypersetup{plainpages=false}
\hypersetup{pdfpagemode=UseOutlines}
\hypersetup{bookmarksnumbered=true}
\hypersetup{bookmarksopen=true}
\hypersetup{pdfstartview=FitH}
\hypersetup{colorlinks=true}
\hypersetup{citecolor=[rgb]{0 .4 0}}
\hypersetup{urlcolor=[rgb]{.4 0 0}}
\hypersetup{linkcolor=[rgb]{0 0 .5}}
\usepackage[T1]{fontenc}
\usepackage{lmodern}
\usepackage{microtype}

\usepackage[left=2.3cm,right=2.3cm,top=2cm,bottom=2.5cm]{geometry}
\usepackage{booktabs}
\usepackage{amsmath,amssymb,mathtools}
\usepackage[mathbold,autobold,greekcaps,greeklower]{mathfixs}
\usepackage{accents} % for under-character accents
\usepackage{siunitx}
\providecommand{\mathbold}{\mathbf}
\providecommand{\mathbbm}{\mathbf}
\newcommand{\nn}{\nonumber}
\allowdisplaybreaks
\numberwithin{equation}{section}

\usepackage{xspace}
\makeatletter
\DeclareRobustCommand\onedot{\futurelet\@let@token\@onedot}
\def\@onedot{\ifx\@let@token.\else.\null\fi}
\makeatother
\def\etal.{et\penalty50\ al.}
\newcommand*{\eg}{\textit{e.g.}\@\xspace}
\newcommand*{\ie}{\textit{i.e.}\@\xspace}
\newcommand*{\etc}{\textit{etc}\onedot}

\usepackage{graphbox}

\usepackage[font=small,labelfont=bf,width=0.88\textwidth]{caption}

\newcommand{\namedref}[2]{\hyperref[#2]{#1~\ref*{#2}}}
\newcommand{\secref}[1]{\namedref{Section}{#1}}
\newcommand{\appref}[1]{\namedref{Appendix}{#1}}
\newcommand{\tabref}[1]{\namedref{Table}{#1}}
\newcommand{\figref}[1]{\namedref{Figure}{#1}}

\makeatletter
\def\mr@ignsp#1 {\ifx\:#1\@empty\else #1\expandafter\mr@ignsp\fi}%
\newcommand{\multiref}[1]{\begingroup%\let\protect\string%
\xdef\mr@no@sparg{\expandafter\mr@ignsp#1 \: }%
\def\mr@comma{}%
\@for\mr@refs:=\mr@no@sparg\do{\mr@comma\def\mr@comma{,\,}\ref{\mr@refs}}%
\endgroup}
\makeatother
\renewcommand{\eqref}[1]{(\multiref{#1})}

\makeatletter
\let\@myabstract\@empty
\let\@keywords\@empty
\let\@subject\@empty
\providecommand{\affiliation}[1]{\gdef\@affiliation{#1}}
\providecommand{\myabstract}[1]{\gdef\@myabstract{#1}}
\providecommand{\keywords}[1]{\gdef\@keywords{#1}}
\providecommand{\subject}[1]{\gdef\@subject{#1}}
\def\thetitle{\@title}
\def\theauthor{\@author}
\def\theaffiliation{\@affiliation}
\def\theabstract{\@myabstract}
\def\thesubject{\@subject}
\def\thedate{\@date}
\def\thekeywords{\@keywords}
\makeatother
\AtBeginDocument{
\hypersetup{pdftitle={\thetitle}}%
\hypersetup{pdfauthor={\theauthor}}%
\hypersetup{pdfsubject={\thesubject}}%
\hypersetup{pdfkeywords={\thekeywords}}%
}

\newcommand{\suprm}[1]{^{\text{#1}}}
\newcommand{\subrm}[1]{_{\text{#1}}}

\newcommand{\grp}[1]{\mathrm{#1}}
\newcommand{\rep}[1]{\mathbf{#1}}

\newcommand{\mathematica}{\textsc{Mathematica}\@\xspace}
\newcommand{\maple}{\textsc{Maple}\@\xspace}
\newcommand{\sagemath}{\textsc{SageMath}\@\xspace}
\newcommand{\GAP}{\textsc{GAP}\@\xspace}
\newcommand{\hyperint}{\textsc{HyperInt}\@\xspace}
\newcommand{\filename}[1]{\texttt{#1}}

\RequirePackage[extdef]{delimset}
\makeatletter
\providecommand{\brkleft}[1][r]{\begingroup\def\dlm@use{\delim(.}%
\if r#1 \def\dlm@use{\delim(.}\fi%
\if s#1 \def\dlm@use{\delim[.}\fi%
\if c#1 \def\dlm@use{\delim\{.}\fi%
\if a#1 \def\dlm@use{\delim<.}\fi%
\expandafter\endgroup\dlm@use}
\providecommand{\brkright}[1][r]{\begingroup\def\dlm@use{\delim.)}%
\if r#1 \def\dlm@use{\delim.)}\fi%
\if s#1 \def\dlm@use{\delim.]}\fi%
\if c#1 \def\dlm@use{\delim.\}}\fi%
\if a#1 \def\dlm@use{\delim.>}\fi%
\expandafter\endgroup\dlm@use}
\makeatother

\renewcommand{\setcond}{\delimpair\{{[m]|}\}}

\DeclareMathOperator{\tr}{tr}
\DeclareMathOperator{\aut}{Aut}

\newcommand{\threedoublebox}{B}
\newcommand{\fivedoublebox}{B}
\newcommand{\doublebox}{B}
\newcommand{\fivepentabox}{\Pi}
\newcommand{\pentabox}{\Pi}
\newcommand{\fivedoublepenta}{\Delta}
\newcommand{\doublepenta}{\Delta}
\newcommand{\stepfcn}{\Theta}
\newcommand{\op}[1]{\mathcal{#1}}
\newcommand{\order}[1]{\mathcal{O}(#1)}

\newcommand{\sfrac}[2]{{\textstyle\frac{#1}{#2}}}
\newcommand{\half}{\sfrac{1}{2}}
\newcommand{\quarter}{\sfrac{1}{4}}

\newcommand{\ii}{\mathrm{i}}
\newcommand{\eps}{\varepsilon}

\newcommand{\superN}{\mathcal{N}}
\newcommand{\Nc}{N\subrm{c}}
\newcommand{\gym}{g_{\scriptscriptstyle\mathrm{YM}}}

\newcommand{\Reals}{\mathbb{R}}

\newcommand{\dd}[2][]{\mathinner{d\ifx#1\empty\else{^#1}\fi#2}}

\newcommand{\lagr}{\mathcal{L}}

\newcommand{\tp}{\rep{20'}}

\makeatletter
\def\bibalias#1#2{%
    \global\@namedef{b@#1}{%
        {\@ifundefined{b@#2}{\textsf{?}}{}\csname b@#2\endcsname\nocite{#2}}}}
\makeatother
\title{Higher-Point Integrands \texorpdfstring{\\}{}in \texorpdfstring{$\superN=4$}{N=4} super Yang--Mills Theory}

\author{%
  Till Bargheer\texorpdfstring{$^\text{\tiny 1}$}{},
  Thiago Fleury\texorpdfstring{$^\text{\tiny 2}$}{},
  Vasco Gon\c{c}alves\texorpdfstring{$^\text{\tiny 3}$}{}%
}

\myabstract{We compute the integrands of
five-, six-, and seven-point correlation functions of twenty-prime operators
with general polarizations
at the two-loop order in $\superN=4$ super Yang--Mills theory.
In addition, we compute the integrand of the five-point function at
three-loop order.
Using the operator product expansion, we extract
the two-loop
four-point function of one Konishi operator and three twenty-prime operators.
Two methods were used for computing the integrands.
The first method is based on constructing an ansatz,
and then numerically fitting for the coefficients using the twistor-space reformulation of
$\superN=4$ super Yang--Mills theory. The second method is based
on the OPE decomposition.
Only very few correlator integrands for more than four points were known before.
Our results can be used to test conjectures, and to make progresses
on the integrability-based hexagonalization approach for correlation functions.}

\subject{Mathematical Physics}


\begin{document}

\pdfbookmark[1]{Title Page}{title}

\renewcommand{\thefootnote}{\fnsymbol{footnote}}
\setcounter{footnote}{0}

\hfill
DESY-22-195

\mbox{}
\vfill

\begin{center}

{\Large\textbf{\mathversion{bold}\thetitle}\par}

\vspace{1cm}

\textsc{\theauthor}

\bigskip

\begingroup
\footnotesize\itshape

$^{\text{\tiny 1}}$\,Deutsches Elektronen-Synchrotron DESY, Notkestr.
85, 22607 Hamburg, Germany

\smallskip

$^\text{\tiny 2}$ International Institute of Physics, Federal University of Rio Grande do Norte, \\
Campus Universitário, Lagoa Nova, Natal, RN 59078-970, Brazil

\smallskip

$^\text{\tiny 3}$\,Centro de Fisica do Porto e Departamento de
Fisica e Astronomia,\\
Faculdade de Ciencias da Universidade do Porto, Porto 4169-007, Portugal

\endgroup

\bigskip

{\small\ttfamily
\email{till.bargheer@desy.de},
\email{tsi.fleury@gmail.com},
\email{vasco.dfg@gmail.com}
}
\par

% \bigskip

% \today

\vspace{1cm}

\textbf{Abstract}\vspace{5mm}

\begin{minipage}{12cm}\small
\theabstract
\end{minipage}

\end{center}

\vfill
\vfill

\thispagestyle{empty}
\setcounter{page}{0}

% \setcounter{page}{-1}
% \newpage
% \thispagestyle{empty}
% \mbox{}
% \vfill
% %
% \begin{center}
% \textit{in memory of Silke Bargheer}\\
% *\,1.8.1951 \quad \textdagger\,5.11.2022
% \end{center}
% %
% \vfill

\newpage
\renewcommand{\thefootnote}{\arabic{footnote}}

\providecommand{\microtypesetup}[1]{}
\microtypesetup{protrusion=false}
\setcounter{tocdepth}{2}
\pdfbookmark[1]{\contentsname}{contents}
\tableofcontents
\microtypesetup{protrusion=true}

%%%%%%%%%%%%%%%%%%%%%%%%%%%%%%%%%%%%%%%%%%%%%%%%%%%%%%%%%%%%
%%%%%%%%%%%%%%%%%%%%%%%%%%%%%%%%%%%%%%%%%%%%%%%%%%%%%%%%%%%%
\section{Introduction}

Correlation functions of locals operators are among the most
interesting physical observables in a conformal gauge theory. In
four-dimensional $\superN=4$ supersymmetric Yang--Mills theory (SYM),
they are even
more relevant, due to a rich structure of dualities between correlation
functions, scattering amplitudes, and null polygon Wilson
loops~\cite{Alday:2010zy,Eden:2010zz,Eden:2011yp,Alday:2013cwa,Bercini:2021jti,Bercini:2020msp}.

A particularly important class of operators in $\superN=4$ SYM are scalar
single-trace half-BPS operators. Their scaling dimensions and
three-point functions are protected from quantum corrections due
to supersymmetry~\cite{Intriligator:1998ig}. They appear to be the simplest
operators in the theory, yet their higher-point correlation functions
encode a wealth of information about more complex operators and
observables via the operator product expansion, null limits \etc. Correlation functions of
these operators have served as an essential laboratory for holography
and the development of computational methods since the early days of the AdS/CFT duality
(see~\cite{Heslop:2022qgf} for a recent review).
Their all-loop integrands are conjectured to take part in a
hidden ten-dimensional symmetry that extends the correlator/amplitude
duality to Coulomb-branch amplitudes~\cite{Caron-Huot:2018kta,Caron-Huot:2021usw}.
Correlation functions of single-trace half-BPS operators are also most
amenable to the integrability-based hexagonalization
approach~\cite{Basso:2015zoa,Eden:2016xvg,Fleury:2016ykk,Coronado:2018ypq,Coronado:2018cxj},
especially at higher points and higher
genus~\cite{Eden:2017ozn,Bargheer:2017nne,Fleury:2017eph,Bargheer:2018jvq,Fleury:2020ykw}.

Despite their essential role, correlators of half-BPS operators beyond
one-loop order have only been computed for four points, with few
exceptions~\cite{Drukker:2008pi,Bargheer:2018jvq,Goncalves:2019znr,Fleury:2020ykw}.
For the further exploration of new techniques, symmetries, and
dualities, more perturbative data is highly needed. The goal of this
paper is to produce exactly such higher-point and higher-loop data.
For half-BPS operators of lowest charge (called $\tp$
operators), it is possible to write a compact formula for correlation
functions for any number of operator insertions at tree and one-loop
level. These results are reviewed in \appref{app:treeone}. The main
result of this paper is the computation of the integrand for the
five-, six-, and seven-point function of $\tp$ operators at two
loops, as well as the five-point function at three loops. In
all cases, we are assuming general polarizations for the external
operators.

Perturbative computations are, in general, notoriously hard in quantum
field theory. The number of diagrams that have to be dealt with using
standard methods grows factorially with the loop order. Another issue
is that symmetries of a given theory might not be manifest within
conventional approaches. Fortunately, there is a more efficient
alternative in $\superN=4$ SYM that exploits the underlying symmetries
of the theory. The paradigmatic example is the computation of
astoundingly high loop order integrands for four-point functions of
half-BPS operators~\cite{Eden:2012tu,Bourjaily:2016evz}. This method
takes advantage of the so-called Lagrangian insertion procedure to
relate the integrand of an $\ell$-loop correlation function of
$n$~operators with the tree-level correlator of $n+\ell$ operators,
where the additional $\ell$ operators are Lagrangian insertions. The
structure of these tree-level correlators is highly constrained;
they are given just by rational functions of the
positions, which in turn makes it possible to construct an ansatz for
this object.%
\footnote{In \cite{Eden:2010zz,Eden:2010ce}, the Lagrangian insertion
procedure together with super-Feynman rules was used to obtain the
integrand for five- and six-point functions of $\tp$ operators
at two loops, in the limit where consecutive points are light-like
separated and in a special polarization.}
Then the problem boils down to fixing the coefficients in the ansatz.

For four-point correlation functions of half-BPS operators, it was
possible to use consistency conditions (\ie a bootstrap approach) to
completely determine the coefficients in the ansatz up to high loop
orders~\cite{Chicherin:2018avq}. For five or more operators, this
strategy is in general more complicated (however it is simple at
one-loop order, as we show below). We
attempt such a bootstrap approach, with
success for five points at two-loop order, see \secref{sec:OPElimit}.
One reason for the increased
difficulty at higher points lies in our lack of understanding of the
consequences of superconformal Ward identities for more than four
points. Notice that these identities can be
written in a very compact notation using superconformal invariants as
in \cite{Chicherin:2015bza, Eden:2017fow}.%
\footnote{A complete
analysis of $G_{6,1}$, \ie six-points at Grassmann degree 4,
was
performed in \cite{Chicherin:2015bza}.}
Solving these superconformal Ward identities
would also be very helpful in the strong-coupling limit, as emphasized
in~\cite{Goncalves:2019znr}.

Given this state of affairs, we take a different approach to compute the
undetermined coefficients in the ansatz, taking advantage of the
reformulation of $\superN=4$ SYM in twistor space.
The chiral part of correlation functions of stress-tensor multiplets
can be computed using a perturbative prescription based on twistor
methods~\cite{Chicherin:2014uca}.%
\footnote{It is not known in general how to reconstruct the full
super-correlator from the knowledge of its chiral part alone. For
four-point functions, the procedure is described in
\cite{Korchemsky:2015ssa}. For half-BPS operators of charge~$k$, one can
use the composite operators of
\cite{Koster:2016ebi,Chicherin:2016soh,Koster:2016fna}.}
This prescription allows a numerical treatment firstly applied
in~\cite{Fleury:2019ydf} to compute the four-loop non-planar integrand
of four $\tp$ operators. Notice that the integrand was known
before (by using symmetries and dynamical constraints) only up to four
unknown coefficients multiplying four polynomials~\cite{Eden:2012tu}.
The same numerical method is used in this work
for computing several new correlators. In fact, it can be used in many
more cases, and the main difficulty at the moment is the construction of
an ansatz (basis of integrals) with not too many undetermined
coefficients.

For computing correlation functions using the twistor
methods of~\cite{Chicherin:2014uca}, it is necessary to introduce an
auxiliary twistor that breaks some of the symmetries at
intermediate stages of the computation. Nevertheless, the resulting
correlators are independent of the auxiliary twistor, and this happens
only after summing all Feynman diagrams. A different method of
computation that maintains
all the symmetries manifests at all
steps was proposed in~\cite{Chicherin:2015bza} (again, see also the
review~\cite{Heslop:2022qgf}). Namely, the chiral correlators
transform covariantly under the $\superN=4$ superconformal
transformations, see \secref{sec:twistors}. Thus, it is possible
to construct superconformal invariants and multiply them by all
possible polynomials, imposing $\grp{S}_n$ permutation symmetry. The
result is a basis for the integrand, and many of the coefficients can be fixed by
imposing physical constraints.
For example, the six-point tree-level correlator could be
computed by this procedure~\cite{Chicherin:2015bza}.
In this particular case, the correlator could be completely
fixed by Ward identities and the light-like limit.
In this work, we only write explicitly results which would correspond to particular fermionic components in such a basis
of invariants. Nevertheless, it would be great to understand the
interplay between our results and the invariants
further and to look for simplifications, for example from Ward identities.%
\footnote{A conjecture about the
role of the 10d symmetry of \cite{Caron-Huot:2021usw}
for higher-point functions written in terms of invariants was put
forward in~\cite{Heslop:2022qgf}.}

In our computation, we restrict ourselves to the parity-even part of
the integrand. The action of $\superN=4$ SYM theory will also
generate parity-odd terms. However, these parity-odd terms will
integrate to zero, since the only parity-odd term in the Lagrangian is
the topological term $\ii F\mspace{-2mu}\tilde{F}$, which is a total derivative.%
\footnote{For a discussion of parity and the fate of parity-odd terms
in the $\superN=2$ superspace approach, see Appendix~A of~\cite{Eden:2010zz}.}
For this reason, we exclude parity-odd terms from the very beginning,
by not including them in our ansatz. This is easily done, since all
parity-odd terms will involve the four-dimensional totally symmetric
epsilon tensor, which we simply disregard. The twistor computation
produces the parity-even as well as the parity-odd terms. Since we
work in Lorentzian signature, the parity-odd terms numerically appear
as imaginary parts and are easily dropped. That being said, it would
in principle be possible to reconstruct also the parity-odd part by
matching the imaginary terms against a suitable ansatz. We refrain
from doing so in this paper.

The paper is organized as follows. \secref{sec:ansatz}
contains the ansatz for the basis of integrals used for the numerical
fitting. In \secref{sec:twistors}, the twistor reformulation of
$\superN=4$ SYM is reviewed, in particular the perturbative
prescription and the twistor-space Feynman rules. Our results for the
five-, six- and seven-points two-loop correlators as well as the five-point
three-loop correlator are given in \secref{sec:results}. In
\secref{sec:OPElimit}, we describe a bootstrap
approach, and we fix the two-loop five-point function using it. We also
analyze the OPE of the integrated correlator, and extract the two-loop
four-point function of one Konishi operator and three $\tp$
operators from our data. We end the paper with a discussion in
\secref{sec:discussion}. Additional details, some
definitions and a review of some results in the literature
appear in several Appendices.

%%%%%%%%%%%%%%%%%%%%%%%%%%%%%%%%%%%%%%%%%%%%%%%%%%%%%%%%%%%%
%%%%%%%%%%%%%%%%%%%%%%%%%%%%%%%%%%%%%%%%%%%%%%%%%%%%%%%%%%%%
\section{Ansatz}
\label{sec:ansatz}

%%%%%%%%%%%%%%%%%%%%%%%%%%%%%%
\paragraph{Correlation Functions.}

We are interested in correlation functions
\begin{equation}
\mathcal{G}_n=
\avg{\op{O}_1 \op{O}_2 \dots \op{O}_n}=
\sum_{\ell=0}^\infty g^{2\ell} \mathcal{G}_n^{(\ell)}
\,, \quad \quad {\rm{with}} \quad \quad
g^2 = \frac{\gym^2\Nc}{4 \pi^2} \, ,
\label{eq:Gnloop}
\end{equation}
of local single-trace scalar half-BPS operators
\begin{equation}
\op{O}_i=
\tr\brk[s]{\brk{{Y_i}\cdot\Phi(x_i)}^{k_i}}
\,,\qquad
\Phi=\brk{\phi_1,\dots,\phi_6}
\,,\qquad
{Y_i}\cdot{Y_i}=0
\,,
\label{BPS}
\end{equation}
where $\Nc$ is the rank of the gauge group $\grp{SU}(\Nc)$,%
\footnote{Whether the gauge group is $\grp{U}(\Nc)$ or $\grp{SU}(\Nc)$
does not make a difference for the correlators we consider in this
work. This is generally true when all external operators are half-BPS
operators of charge two ($\tp$
operators)~\cite{Fleury:2019ydf}.}
$\gym$ is
the Yang--Mills coupling constant, $\phi_i$~are the six scalar fields of $\superN=4$
SYM, and $Y_i$~are six-dimensional null polarization vectors.
Let us comment on the dependence on the number of colors $\Nc$:
By absorbing a factor of $\Nc$ in the coupling $g$, we are
anticipating the 't~Hooft planar large-$\Nc$ limit, in which all
dependence on $\Nc$ is contained in $g$.
In the
full finite-$\Nc$ theory, the coefficients
$\mathcal{G}_n^{(\ell)}$ still have a
non-homogeneous dependence on $\Nc$, with subleading terms in $1/\Nc$
signifying non-planar corrections.
In all our computations, we make no assumption on planarity.
However, we still find that all terms $\mathcal{G}_n^{(\ell)}$ computed in
this paper are homogeneous in $\Nc$, \ie are free of non-planar
corrections and only consist of the planar contribution.
At one- and two-loop order, this is understood, since we focus on
correlators of lowest-charge operators ($\tp$ operators) whose dependence
on $\Nc$ is particularly simple, and because there are no non-planar
Feynman integrals at one- and two-loop order. At three loops,
non-planar terms may start contributing, but we find that they are
absent in our results (see also \secref{sec:planarity} below).

One way to compute loop corrections to correlation functions in
perturbation theory is via the
Lagrangian insertion method. Here, the $\ell$-loop contribution
is given by an integral over the spacetime
positions of $\ell$ insertions of the Lagrangian
operator~\cite{Eden:2011we}
\begin{gather}
g^{2\ell}
\mathcal{G}_n^{(\ell)}=
\int\brk*{\prod\nolimits_{i=1}^{\ell}\dd[4]x_{n+i}} G_n^{(\ell)}
\,,\nn\\
G_n^{(\ell)}=
\avg{\op{O}_1 \dots \op{O}_n \lagr_{n+1} \dots \lagr_{n+\ell}}\subrm{tree}
\,,\qquad
\lagr_i=\lagr(x_i)
\,,
\label{eq:lagrangianinsertion}
\end{gather}
where $G_n^{(\ell)}$ is the $(n+\ell)$-point correlator with $\ell$
insertions of the chiral on-shell Lagrangian operator~$\lagr$, evaluated at leading
order in perturbation theory.%
\footnote{This formula naively follows from differentiating the path
integral expression for $\mathcal{G}_n$ with respect to the coupling
constant. The effect of the differentiation on the coupling-dependent
operators is canceled by contact terms. The chiral on-shell Lagrangian is
obtained from the full Lagrangian by applying equations of motion. See
\eg~\cite{CaronHuot:2010ek} for a careful treatment.}
For supersymmetric theories, the Lagrangian insertion method was first
introduced in~\cite{Intriligator:1998ig}.
It has played a key role in the study of $\superN=4$ SYM correlation
functions, especially in
constructing the four-point integrand to high loop
orders~\cite{Petkou:1999fv,Fleury:2019ydf,Eden:2000mv,Eden:2000bk,Eden:2011yp,Eden:2011ku,Eden:2011we,Eden:2012tu,Chicherin:2014uca,Chicherin:2015edu,Bourjaily:2016evz,Chicherin:2018avq}.

The correlators~$\mathcal{G}_n^{(\ell)}$
are functions of both the operator
positions~$x_i$ and the polarization vectors $Y_i$. Due to the
internal $\grp{SO}(6)$ invariance, $\mathcal{G}_n^{(\ell)}$ (as well
as $G_n^{(\ell)}$) are polynomials in the basic invariants
$Y_i\cdot{Y_j}$.
From the operator definition~\eqref{BPS}, it follows
that the correlators are homogeneous in each $Y_i$ with weight $k_i$.
Writing the invariants $Y_i\cdot{Y_j}$ in terms of propagator factors
\begin{equation}
d_{ij}=\frac{{Y_i}\cdot Y_j}{x_{ij}^2}
\,,\qquad
x_{ij}^2=(x_i-x_j)^2
\,,
\label{escalarpropagators}
\end{equation}
the tree-level $(n+\ell)$-point function can be unambiguously
decomposed into a finite sum%
\footnote{A similar decomposition for the loop correlators
$\mathcal{G}_n^{(\ell)}$ directly follows.}
\begin{equation}
G_n^{(\ell)}=
C_{k_1\dots k_n}g^{2\ell}
\sum_{\mathbold{a}}
\brk*{\prod\nolimits_{1\leq i<j\leq n}d_{ij}^{a_{ij}}}
f_{\mathbold{a}}^{(\ell)}(x_{ij}^2)
\,, \qquad
\mathbold{a}=\brk[c]{a_{ij}\,|\,1\leq{i}<{j}\leq{n}}
\,.
\label{eq:Gnlstructure}
\end{equation}
Here, we have pulled out an overall constant prefactor $C_{k_1\dots k_n}$ that
depends on $\Nc$ and the charges $k_i$. The explicit factor
$g^{2\ell}$ arises from the Lagrangian insertions and is required for
consistency.
The sum over $\mathbold{a}$ is a finite sum
over polarization structures $\prod_{ij}d_{ij}^{a_{ij}}$ that absorbs
all dependence on the polarization vectors $Y_i$, such that the
coefficient functions $f_{\mathbold{a}}^{(\ell)}$ only depend on the
coordinates $x_i$.%
\footnote{The functions $f_{\mathbold{a}}^{(\ell)}$ in general may
also carry a non-trivial dependence on $\Nc$. However, all examples
computed in this paper are free of subleading terms in $1/\Nc$, and
therefore all dependence on $\Nc$ can be absorbed in the overall
factor $C_{k_1\dots{k_n}}$.}
The polarization structures that can occur in the sum over
$\mathbold{a}$ are constrained by the charges of the operators:
\begin{equation}
\text{For all }
i=1,\dots,n:
\quad
k_i=\sum_{i\neq j=1}^na_{ij}
\qquad
(a_{ij}\equiv{a_{ji}})
\,.
\end{equation}
%

%%%%%%%%%%%%%%%%%%%%%%%%%%%%%%
\paragraph{Coefficient Functions.}

Each polarization structure is multiplied by a rational function
$f_{\mathbold{a}}^{(\ell)}$ of the $n+\ell$ positions~$x_i$. Due to
Lorentz invariance, the functions $f_{\mathbold{a}}^{(\ell)}$ only depend on
squared distances $x_{ij}^2$.%
\footnote{The positions could also appear in invariant contractions
$\eps_{\mu\nu\sigma\tau}x_i^{\mu}x_j^{\nu}x_k^{\sigma}x_{\ell}^{\tau}$
with the totally antisymmetric tensor $\eps_{\mu\nu\sigma\tau}$.
However,
because of parity symmetry,
our results do not contain such terms.
Nevertheless, the twistor action is chiral, and therefore generates
such terms at the integrand level,
but they are always total derivatives because
they follow from a topological term in the action. See
\secref{sec:twistors} for more details.}
We can further constrain their form by
considering their singularity structure, which is constrained by the
operator product expansions (OPEs) among the external half-BPS operators $\op{O}_i$
as well as the internal Lagrangian operators $\lagr$. Since the
functions $f_{\mathbold{a}}^{(\ell)}$ constitute components of a
tree-level correlation function~\eqref{eq:Gnlstructure}, we only need
to consider the tree-level OPEs. Let us first consider the case of two
external operators $\op{O}_i$ and $\op{O}_j$. In their tree-level OPE, all
inverse powers of $x_{ij}^2$ originate in Wick contractions of the
constituent fields $Y_i\cdot\Phi(x_i)$, and hence only occur in the
combination $d_{ij}=Y_i\cdot{Y_j}/x_{ij}^2$. The OPE therefore takes
the form%
\footnote{See~\cite{Chicherin:2015edu} for a more
detailed discussion of this OPE.}
\begin{equation}
\op{O}_i\times\op{O}_j\sim
\sum_{0\leq{k}}d_{ij}^k\times(\text{regular})
\,,
\label{eq:OPEOO}
\end{equation}
where (regular) stands for terms of order $\order{x_{ij}^0}$.
This shows that all inverse powers of $x_{ij}$ in the OPE
are accompanied by numerators $Y_i\cdot Y_j$.
But all dependence of the correlation function $G_n$ on the polarizations $Y_i$ is
absorbed in the propagator products
$\prod{d_{ij}^{a_{ij}}}$, hence the coefficient functions
$f_{\mathbold{a}}^{(\ell)}$ must be regular:
\begin{equation}
f_{\mathbold{a}}^{(\ell)}
\xrightarrow{\;x_i\to x_j\;}
\order{x_{ij}^{0}}
\,,\qquad
1\leq i,j\leq n
\,.
\label{eq:constraint1}
\end{equation}
Next, consider the OPE between one external operator $\op{O}_i$ and
one internal Lagrangian operator $\lagr_j$. A basic analysis shows
that it has the form%
\footnote{If there was a lower-dimension operator in the OPE, it would
involve three or four contractions between~$\op{O}_i$ and~$\lagr$. The
relevant term in the Lagrangian is
$\tr\brk{\comm{\phi_n}{\phi_m}\comm{\phi^n}{\phi^m}}$, hence such
contractions evaluate to zero.}
\begin{equation}
\op{O}_i\times\lagr_j\sim
\frac{C_{\op{O}\lagr\op{O}}}{x_{ij}^4}\op{O}_i+\order{x_{ij}^{-2}}
\,.
\label{eq:OPEOLag}
\end{equation}
However, the coefficient of the leading singularity is proportional to
the three-point function
$C_{\op{O}\lagr\op{O}}\sim\avg{\op{O}_i\lagr\op{\bar{O}}_i}$, which
vanishes since the two-point function $\avg{\op{O}_i\op{\bar{O}}_i}$ is
protected. Hence only the subleading term contributes:
$\op{O}_i\times\lagr_j\sim\order{x_{ij}^{-2}}$. Finally the OPE of two
Lagrangian operators reads%
\footnote{A simple explanation for the form of this OPE is as follows: The chiral
Lagrangian $\lagr\simeq{\mathfrak{Q}^4}\op{O}^{(2)}$ is a
superdescendant of the chiral primary $\op{O}^{(2)}$.
The fact that $\mathfrak{Q} \mathcal{L}=0$ without total derivatives implies that the contact term
proportional to $\delta^4(x_{ij})$ is also proportional to the chiral Lagrangian \cite{CaronHuot:2010ek}.
Moreover the
three-point functions $\avg{\lagr\lagr\op{\bar{O}}}$ are invariant
under a $\grp{U}(1)\subrm{Y}$ ``bonus
symmetry''~\cite{Intriligator:1998ig,Heslop:2001gp}, under which
$\mathfrak{Q}$ has charge $+1$. Thus $\op{\bar{O}}$ must be of the form
$\bar{\mathfrak{{Q}}}^8\op{P}$ for a chiral primary $\op{P}$, that is it
must be part of a long multiplet. The lowest-dimension long multiplet
is the Konishi multiplet, in which case $\op{\bar{O}}$ has dimension
$2+8/2=6$. Therefore all operators in the regular part of the
$\lagr\times\lagr$ OPE have dimension at least six, which means the
divergence is at most $1/x_{ij}^2$. We thank Paul Heslop for
clarifying this point.}
\begin{equation}
\lagr_i\times\lagr_j\sim
\delta^4(x_{ij})(...)+\order{x_{ij}^{-2}} \, .
\label{eq:OPELagLag}
\end{equation}
The dominant contributions are contact terms proportional to
$\delta^4(x_{ij})$.
We only consider the tree-level
correlation function~\eqref{eq:Gnlstructure} for Lagrangians at
distinct points, hence we can ignore these terms and are only left
with the subleading $\order{x_{ij}^{-2}}$ term.%
\footnote{These contact terms however play an important role for the consistency of the
Lagrangian insertion method~\cite{Petkou:1999fv,Eden:2000bk,CaronHuot:2010ek}.}
The two relations~\eqref{eq:OPEOLag} and~\eqref{eq:OPELagLag} imply that
\begin{align}
f_{\mathbold{a}}^{(\ell)}
&\xrightarrow{\;x_i\to x_j\;}
\order{x_{ij}^{-2}}
\,,\qquad
1\leq i\leq n+\ell
\,,\quad
n<j\leq n+\ell
\,.
\label{eq:constraint2}
\end{align}
Combining
the two constraints~\eqref{eq:constraint1} and~\eqref{eq:constraint2},
the rational functions $f_{\mathbold{a}}$, when written as single
fractions, take the form
\begin{equation}
f_{\mathbold{a}}^{(\ell)}=
\frac{P_{\mathbold{a}}^{(\ell)}}{\prod_{(i,j)\in I}x_{ij}^2}
\,,\qquad
I=\setcond{(i,j)}{1\leq i\leq n+\ell,\,n<j\leq n+\ell,\,i<j}
\,,
\label{eq:fform}
\end{equation}
where the numerators $P_{\mathbold{a}}^{(\ell)}$ are \emph{polynomials} in
the squared distances $x_{ij}^2$.
The degree of these polynomials in the
various $x_{i}$ is fixed by the conformal weights of the respective
operators: The external BPS operators $\op{O}_i$ have conformal
weights $k_i$, and the Lagrangian $\lagr$ has conformal weight $4$.
Hence also the correlator $G_n$ has these weights in the
respective points $x_i$. The squared distances $x_{ij}^2$ have conformal weights $-1$
at points $x_i$ and $x_j$, the products $\prod{d_{ij}^{a_{ij}}}$ have
weights $k_i$ at all external points, and the denominator factor
in~\eqref{eq:fform} has weights $\ell$ at external points, and weights
$n+\ell-1$ at internal points. It follows that
$P_{\mathbold{a}}^{(\ell)}$ must have conformal weights $-\ell$ at
all external points, and conformal weights $5-n-\ell$ at all internal
points. Hence for all external points $i$, the total degree in all
$x_{ij}^2$, $j\neq i$ must be $\ell$, whereas for all internal points $i$,
the total degree in all $x_{ij}^2$, $j\neq i$ must be $(n+\ell-5)$.

%%%%%%%%%%%%%%%%%%%%%%%%%%%%%%
\paragraph{Correlators of $\tp$ Operators.}

For $\tp$ operators, the charge is $k=2$. In this case, there
are only few possible terms ($Y$-structures) in the sum over
$\mathbold{a}$ in equation~\eqref{eq:Gnlstructure}. Up to permutations of the external points, the only
possibilities for the five-point function are
\begin{equation}
a_{12}=2
\,,\quad
a_{34}=a_{45}=a_{15}=1
\,,\quad
\text{all other }a_{ij}=0
\,,
\label{eq:a2x3}
\end{equation}
which we call the \textbf{$2\times3$} component, and
\begin{equation}
a_{12}=a_{23}=a_{34}=a_{45}=a_{15}=1
\,,\quad
\text{all other }a_{ij}=0
\,,
\label{eq:acyclic}
\end{equation}
which we call the \textbf{$5$} component. We will label the
functions $f_{\mathbold{a}}$ multiplying the respective propagator
products as $f_{2\times3}\equiv{f_{23}}$ and $f_5$. All other possible
configurations $\mathbold{a}$ are obtained
from~\eqref{eq:a2x3,eq:acyclic} by permutations of the five external
points. By invariance of $G_n^{(\ell)}$ under such permutations, the
component functions $f_{23}$ and $f_5$ uniquely
determine all other component functions $f_{\mathbold{a}}$.
In other words, the five-point correlator of $\tp$ operators can be written as
\begin{equation}
G_5^{(\ell)}=
C_{22222}\,g^{2\ell}
\avg*{
d_{12}^2 \, d_{34}d_{45}d_{53} \, f_{23}^{(\ell)}(x_{ij}^2)
+d_{12}d_{23}d_{34}d_{45}d_{51} \, f_{5}^{(\ell)}(x_{ij}^2)
}_{\grp{S}_5}
\,,
\label{eq:G5}
\end{equation}
where $\avg{\cdot}_{\grp{S}_5}$ denotes averaging over permutations of the
five external points. Similarly, the six-point and seven-point
functions can be written as
\begin{align}
\label{eq:G6G7}
G_6^{(\ell)}&=
C_{2\dots2}\,g^{2\ell}
\brkleft[a]*{
d_{12}^2 \, d_{34}^2 \, d_{56}^2 \, f_{222}^{(\ell)}(x_{ij}^2)
+d_{12}^2 \, d_{34}d_{45}d_{56}d_{63} \, f_{24}^{(\ell)}(x_{ij}^2)
} \\ & \mspace{90mu} \brkright[a]*{ \mathord{}
+d_{12}d_{23}d_{31} \, d_{45}d_{56}d_{64} \, f_{33}^{(\ell)}(x_{ij}^2)
+d_{12}d_{23}d_{34}d_{45}d_{56}d_{61} \, f_{6}^{(\ell)}(x_{ij}^2)
}_{\grp{S}_6}
\,, \nn \\
G_7^{(\ell)}&=
C_{2\dots2}\,g^{2\ell}
\brkleft[a]*{
d_{12}^2 \, d_{34}^2 \, d_{56}d_{67}d_{75} \, f_{223}^{(\ell)}(x_{ij}^2)
+d_{12}^2 \, d_{34}d_{45}d_{56}d_{67}d_{73} \, f_{25}^{(\ell)}(x_{ij}^2)
} \nn \\ & \mspace{90mu} \brkright[a]*{ \mathord{}
+d_{12}d_{23}d_{31} \, d_{45}d_{56}d_{67}d_{74} \, f_{34}^{(\ell)}(x_{ij}^2)
+d_{12}d_{23}d_{34}d_{45}d_{56}d_{67}d_{71} \, f_{7}^{(\ell)}(x_{ij}^2)
}_{\grp{S}_7}
\,.\nn
\end{align}
For the correlators computed in this paper, we find
\begin{equation}
C_{2\dots{2}}=\frac{\Nc^2-1}{(2\pi)^{2n+2\ell}}
\,.
\label{eq:C222}
\end{equation}
Some words on the notation:
For a correlator of a general number $n$ of $\tp$ operators, all
possible polarization structures $\prod{d_{ij}^{a_{ij}}}$ take the
form of a set of polygons, where each vertex represents one
$\tp$ operator, and each edge represents a propagator $d_{ij}$.
A polarization is therefore uniquely specified by a monotonically
increasing set of integers
\begin{equation}
\brk{r_1,r_2,\dots,r_s}
\,,\qquad
r_i\leq{r_{i+1}}
\,,\qquad
\sum_{i=1}^sr_i=n
\,,
\end{equation}
where $s$ is the number of polygons, and $r_i$ is the size of the
$i$'th polygon. As can be seen in~\eqref{eq:G6G7}, we label the
corresponding coefficient functions
$f_{\mathbold{a}}\equiv{f_{r_1\dots{r_s}}}$ by these sequences. By
convention, the first $r_1$ operators populate the first polygon, the
next $r_2$ operators populate the second polygon, and so on.

%%%%%%%%%%%%%%%%%%%%%%%%%%%%%%
\paragraph{Graph Counting.}

\begin{table}
\centering
\begin{tabular}[t]{lrr}
\toprule
& $\ell=2$ & $\ell=3$ \\
\midrule
$P_{23}^{(\ell)}$ & $64$ & $3286$ \\
$P_{5}^{(\ell)}$ & $66$ & $3576$ \\
\midrule
$P_{222}^{(\ell)}$ & $235$ & $46873$ \\
$P_{24}^{(\ell)}$ & $572$ & $137596$ \\
$P_{33}^{(\ell)}$ & $173$ & $32701$ \\
$P_{6}^{(\ell)}$ & $657$ & $174074$ \\
\bottomrule
\end{tabular}
\qquad
\begin{tabular}[t]{lr}
\toprule
& $\ell=2$ \\
\midrule
$P_{223}^{(\ell)}$ & $2435$ \\
$P_{25}^{(\ell)}$ & $4637$ \\
$P_{34}^{(\ell)}$ & $2435$ \\
$P_{7}^{(\ell)}$ & $6143$ \\
\bottomrule
\end{tabular}
\qquad
\begin{tabular}[t]{lr}
\toprule
& $\ell=2$ \\
\midrule
$P_{2222}^{(\ell)}$ & $4170$  \\
$P_{224}^{(\ell)}$ &  $21709$ \\
$P_{233}^{(\ell)}$ &  $10808$ \\
$P_{26}^{(\ell)}$ &   $48419$ \\
$P_{35}^{(\ell)}$ &   $21153$ \\
$P_{44}^{(\ell)}$ &   $11264$ \\
$P_{8}^{(\ell)}$ &    $68453$ \\
\bottomrule
\end{tabular}
\qquad
\begin{tabular}[t]{lr}
\toprule
& $\ell=2$ \\
\midrule
$P_{2223}^{(\ell)}$ &  $68255$ \\
$P_{225}^{(\ell)}$  & $205851$ \\
$P_{234}^{(\ell)}$  & $197580$ \\
$P_{27}^{(\ell)}$   & $547435$ \\
$P_{333}^{(\ell)}$  &  $17474$ \\
$P_{36}^{(\ell)}$   & $232772$ \\
$P_{45}^{(\ell)}$   & $205851$ \\
$P_{9}^{(\ell)}$    & $818180$ \\
\bottomrule
\end{tabular}
\caption{Numbers of independent coefficients in various ansatz
polynomials $P_{\mathbold{a}}^{(\ell)}$ that enter the
correlator~\protect\eqref{eq:Gnlstructure} via~\protect\eqref{eq:fform}.
(The double occurrence of $2435$ is not a typo.) See also
\protect\tabref{tab:reducedansatzsizes} for further reduced ansatz sizes.}
\label{tab:ansatzsizes}
\end{table}

To complete the construction of the ansatz, it remains to find the
most general polynomials $P_{\mathbold{a}}^{(\ell)}$
in~\eqref{eq:fform} for the various Y-structures $\mathbold{a}$. By
mapping each factor $x_{ij}^2$ to an edge between vertices $i$ and
$j$, we can identify each monomial in $x_{ij}^2$ with a multi-graph
(\ie a graph that admits ``parallel'' edges between the same
vertices~$i$ and~$j$). Finding the most general polynomials hence
amounts to listing all multi-graphs with $n$ external vertices with
valency $\ell$, and $\ell$ internal valencies with valency $n+\ell-5$,
and taking a general linear combination of the corresponding
monomials.

Here, we can make use of permutation symmetry: Each propagator
structure $\prod_{ij}d_{ij}^{a_{ij}}$ typically is invariant under a
residual group $K_{\mathbold{a}}\subset\grp{S}_n$ of permutations of
the external points $\set{1,\dots,n}$. By the total $\grp{S}_n$
permutation symmetry of the full correlator $G_n^{(\ell)}$, the
respective component polynomial $P_{\mathbold{a}}$ must also respect
that residual permutation symmetry. Moreover, all polynomials
$P_{\mathbold{a}}$ must be fully symmetric under $\grp{S}_\ell$
permutations of the Lagrangian insertion points
$\set{n+1,\dots,{n+\ell}}$. We can thus impose these symmetries on the
ansatz polynomials from the beginning, which significantly reduces the
numbers of undetermined coefficients.

For the correlators of $\tp$ operators, we list the numbers of
independent terms for the various polynomials
in \tabref{tab:ansatzsizes}. For more details on the construction of
these ansatz polynomials, see \appref{app:graphs}.

%%%%%%%%%%%%%%%%%%%%%%%%%%%%%%
\paragraph{Gram Identities.}

An $n$-point correlation function in a four-dimensional
Lorentz-invariant theory has $4n-10$ kinematic degrees of freedom.
On the other hand, there are $n(n-1)/2$ different squared distances $x_{ij}^2$.
Hence the $x_{ij}^2$ must satisfy some non-trivial relations. One way
to construct such relations is as follows.

Points $x^\mu$ in Minkowski space $\Reals^{1,3}$ can be identified
with null rays $X\in\Reals^{2,4}$, $X^2=0$,
$tX\cong{X}$~\cite{Dirac:1963ta},
for example via
\begin{equation}
X=\brk*{\half(1+x^2),x^\mu,\half(1-x^2)}
\,.
\end{equation}
The fundamental two-point Lorentz invariants can then be written as
${x_{ij}^2}\propto{X_i}\cdot{X_j}$. Now consider the matrix
\begin{equation}
\mathbold{X}=\brk[s]*{X_{ij}}_{i,j=1}^{n+\ell}
\,,\qquad
X_{ij}=\begin{cases}
0 & i=j\,,\\
x_{ij}^2 & i\neq{j}\,.
\end{cases}
\label{eq:Xmatrix}
\end{equation}
Then $X_{ij}\propto{X_i\cdot{X_j}}$. In six dimensions, at most six
vectors $X_i$ can be linearly independent, hence all $7\times7$ minors
of $\mathbold{X}$ must be zero. This introduces non-linear relations
among the fundamental invariants $x_{ij}^2$ called \emph{Gram
determinant relations}. These relations are obviously polynomial.
However, by multiplying each Gram determinant relation with suitable
monomials,%
\footnote{Finding all suitable monomials for a given polynomial
relation is another exercise in graph enumeration.}
we obtain non-trivial \emph{linear} relations among the
various terms in the ansätze for the
polynomials~$P_{\mathbold{a}}^{(\ell)}$. These relations can be used
to reduce the number of undetermined coefficients in the ansätze for
$P_{\mathbold{a}}^{(\ell)}$.

Concretely, we construct the independent Gram relations as follows:
First, we list all $7\times7$ minors of the matrix $\mathbold{X}$.
There are $p(p+1)/2$ such minors, where
$p=\mathrm{Binomial}(n+\ell,7)$. Next, we canonicalize each of these
minors over $\grp{S}_\ell$ permutations of $x_{n+1},\dots,x_{n+\ell}$,
thereby identifying expressions that only differ by permutations of the
$\ell$ integration labels. The numbers of minors that remain for
$(n,\ell)=(5,2),(6,2),(7,2),(5,3)$ are $(1,29,463,22)$, as shown in
the first row of \tabref{tab:gramstatistics}.
\begin{table}
\centering
\begin{tabular}{rrrrrrrrrr@{\qquad}rr}
\toprule
$P_{{23}}^{(2)}$ & $P_{{5}}^{(2)}$ & $P_{{222}}^{(2)}$ & $P_{{24}}^{(2)}$ & $P_{{33}}^{(2)}$ & $P_{{6}}^{(2)}$ & $P_{{223}}^{(2)}$ & $P_{{25}}^{(2)}$ & $P_{{34}}^{(2)}$ & $P_{{7}}^{(2)}$ & $P_{{23}}^{(3)}$ & $P_{{5}}^{(3)}$ \\
\midrule
$1$              & $1$             & $29$              & $29$             & $29$             & $29$            & $463$             & $463$            & $463$            & $463$           & $22$             & $22$ \\
$1$              & $1$             & $6$               & $10$             & $6$              & $7$             & $55$              & $59$             & $55$             & $49$            & $9$              & $6$ \\
$1$              & $1$             & $7$               & $13$             & $7$              & $9$             & $199$             & $241$            & $199$            & $259$           & $177$            & $154$ \\
$1$              & $1$             & $5$               & $9$              & $5$              & $7$             & $81$              & $102$            & $81$             & $115$           & $103$            & $91$ \\
\bottomrule
\end{tabular}
\caption{Statistics of Gram determinant relations at various stages.
\emph{First row:} Numbers of $7\times7$ minors of $\mathbold{X}$, up
to permutations of integration labels. \emph{Second row:} Numbers of
relations that remain after canonicalizing each relation over
$K_{\mathbold{a}}$ permutations. \emph{Third row:} Numbers of
relations after saturating the weights by multiplying with all
possible monomials and canonicalizing \emph{each term} over
$K_{\mathbold{a}}$ permutations. \emph{Last row:} Final numbers of
linearly independent relations that can be used to reduce the
manifestly $K_{\mathbold{a}}$-symmetric ansätze for the polynomials
$P_{\mathbold{a}}^{(\ell)}$.}
\label{tab:gramstatistics}
\end{table}
Each of these expressions is a non-trivial polynomial in $x_{ij}^2$
that evaluates to zero, by construction. To compare to our ansatz
polynomials for $P_{\mathbold{a}}^{(\ell)}$, we still need to
symmetrize over the respective permutations $K_{\mathbold{a}}$ of
external points, and we need to saturate the weights of the relations.
We do this in three steps. First, we canonicalize each relation over
the permutation group $K_{\mathbold{a}}$. This reduces the numbers of
relations to the second row in \tabref{tab:gramstatistics}. Next, we
saturate the weights to $\ell$ for the external points and to
$(n+\ell-5)$ for the integration points by multiplying each relation
with all possible monomials that yield the desired weights.%
\footnote{For some relations, this is not possible. For example, one
Gram determinant relation at $(n,\ell)=(6,2)$ has weight deficits $-1$
in $x_7$ and $-3$ in $x_8$. Obviously, there is no monomial in
$x_{ij}^2$ with such weights. We drop relations whose weights cannot
be saturated by monomials.}
Finally, we expand the weight-completed relations and canonicalize
\emph{each term} over the permutations $K_{\mathbold{a}}$ of external
points. Some relations become manifestly zero, others become identical
to each other. The resulting numbers of relations are listed in the
third row of \tabref{tab:gramstatistics}. Not all of these relations
are linearly independent. To find the ambiguity in our ansatz
polynomials, we need to pick a linearly independent set. The sizes of
the maximal linearly independent sets are listed in the last row
of \tabref{tab:gramstatistics}.

The sizes of these ambiguities might not seem big compared to the
numbers of terms in the ansätze listed in \tabref{tab:ansatzsizes}.
However, when matching these ansätze to data by comparison at many
numerical points $x_i$, any ambiguity easily leads to very
``unnatural'' solutions, with arbitrary numerical coefficients. Having
a good handle on the ambiguities is essential to resolve, or even
better avoid, such ``arbitrary'' solutions.

%%%%%%%%%%%%%%%%%%%%%%%%%%%%%%
\paragraph{Ancillary File.}

We provide the complete ansatz expressions for the polynomials
$P_{\mathbold{a}}^{(\ell)}$ as well as lists with all
Gram relations for $(n,\ell)\in\set{(5,2),(6,2),(7,2),(5,3)}$ in the
attached \mathematica file \filename{ansatzAndGram.m}.

%%%%%%%%%%%%%%%%%%%%%%%%%%%%%%%%%%%%%%%%%%%%%%%%%%%%%%%%%%%
%%%%%%%%%%%%%%%%%%%%%%%%%%%%%%%%%%%%%%%%%%%%%%%%%%%%%%%%%%%%
\section{Twistors}
\label{sec:twistors}

In this section, we briefly review the reformulation of $\superN=4$ SYM in supertwistor space and
the procedure for computing the correlation functions of the chiral stress-tensor supermultiplet using this formalism.
In addition, we explain the numerical methods used for the computations.
The Feynman rules in twistor space depend on the choice of an auxiliary supertwistor,
however the final results for the correlators are independent of the choice of this auxiliary
supertwistor and  they can be
expressed in terms of $\superN=4$ superconformal invariants. These invariants are discussed
in~\cite{Chicherin:2015bza,Eden:2017fow} and we also briefly review them here.

%%%%%%%%%%%%%%%%%%%%%%%%%%%%%%%%%%%%%%%%%%%%%%%%%%%%%%%%%%%%
\subsection{Computing Correlation Functions using Twistors}

The relevant superspace for studying $\superN=4$ SYM has sixteen odd variables $\theta^{a \alpha}$, $\bar{\theta}^{\dot{\alpha}}_a$, with $a=1,\ldots,4$ and $\alpha, \dot\alpha=1,2$. The stress-tensor supermultiplet is a short half-BPS supermultiplet
and it depends on only eight of the odd variables, four chiral and four anti-chiral variables~\cite{Belitsky:2014zha}. In what follows,
we are mostly interested in its chiral part, \ie all the anti-chiral variables $\bar{\theta}^{\dot{\alpha}}_a$ are going to be set to zero.
In order to describe this supermultiplet, it is convenient to introduce the so called harmonic variables
\begin{equation}
u^b_a \equiv (u^{+ \mathfrak{b}}_a, u_a^{- \mathfrak{b}^{\prime}}) \, , \quad {\rm{parametrizing}}  \quad  \frac{\grp{SU}(4)}{\grp{SU}(2) \times \grp{SU}(2)^{\prime}\times \grp{U}(1)} \, ,
\end{equation}
with $\mathfrak{b},\mathfrak{b}^{\prime}=1,2$ being fundamental representation indices of the $\grp{SU}(2), \, \grp{SU}(2)^{\prime}$ respectively and
the $\pm$ indicates the charge under the $\grp{U}(1)$ factor. The harmonic variables obey several constraints due
to the fact that they belong to $\grp{SU}(4)$, see~\cite{Chicherin:2014uca} for example. Using these variables and
their complex conjugates $\bar{u}$,
we can decompose
the odd variables as follows
\begin{equation}
\theta^a_{\alpha} = \theta^{+ \mathfrak{b}}_{\alpha} \, \bar{u}^a_{+ \mathfrak{b}} +
\theta^{- \mathfrak{b}^{\prime}}_{\alpha} \, \bar{u}^a_{- \mathfrak{b}^{\prime}} \, ,
\end{equation}
where
\begin{equation}
\theta^{+ \mathfrak{b}}_{\alpha} = \theta^a_{\alpha} \, u_{a}^{+ \mathfrak{b}} \, , \quad \quad
\theta^{- \mathfrak{b}^{\prime}}_{\alpha} = \theta^a_{\alpha} \, u_{a}^{- \mathfrak{b}^{\prime}} \, ,
\label{thetaplusthetaminus}
\end{equation}
and similarly for $\bar{\theta}^{\dot{\alpha}}_a$. The odd variables defined above are useful for writing down an expansion
of the
stress-tensor supermultiplet $\widehat{\mathcal{T}}(x, \theta, \bar{\theta}, u)$
keeping the $\grp{SU}(4)$ symmetry manifest. The bottom state of the supermultiplet
is the operator $\mathcal{O}_{20^{\prime}}(x,u)$, which is annihilated by the following
eight supersymmetries (and also by all the superconformal generators)
\begin{equation}
Q^{\alpha}_{- \mathfrak{a}^{\prime}}  \cdot \mathcal{O}_{20^{\prime}}(x,u) = (\bar{Q}^{+})^{\dot\alpha \, \mathfrak{a}}  \cdot \mathcal{O}_{20^{\prime}}(x,u) =0 \, ,
\quad {\rm{with}} \quad
Q^{\alpha}_{- \mathfrak{a}^{\prime}} =  \bar{u}^a_{- \mathfrak{a}^{\prime}} Q^{\alpha}_a \, , \quad
(\bar{Q}^{+})^{\dot\alpha \, \mathfrak{a}} = u^{+ \mathfrak{a}}_{a} \bar{Q}^{\dot\alpha \, a} \, .
\end{equation}

Then the supermultiplet is given by
\begin{equation}
\widehat{\mathcal{T}}(x, \theta, \bar{\theta}, u) = {\rm{exp}} \left( \theta^{+ \mathfrak{b}}_{\alpha} Q^{\alpha}_{+ \mathfrak{b}} + \bar{\theta}^{\dot\alpha}_{- \mathfrak{a}^{\prime}} \bar{Q}^{- \mathfrak{a}^{\prime}}_{\dot\alpha} \right) \cdot \mathcal{O}_{20^{\prime}}(x,u) \, .
\end{equation}

In particular, its chiral part $\mathcal{T}(x, \theta^+, u)$ only depends on the four $\theta^{+ \mathfrak{b}}_{\alpha}$ variables of  \eqref{thetaplusthetaminus}. We have schematically
\begin{equation}
\mathcal{T}(x, \theta^+, u)  \equiv \widehat{\mathcal{T}}(x, \theta, 0, u) = \mathcal{O}^{++++}(x) + \ldots+ (\theta^+)^4 \mathcal{L}(x) \, ,
\label{stresstensor}
\end{equation}
and in the expansion above only the relevant terms for this work were written down.
The top operator $\mathcal{L}(x)$ is the chiral on-shell Lagrangian and the bottom one is the $\tp$ operator given by
\begin{equation}
\mathcal{O}^{++++}(x) = {\rm{Tr}} (\phi^{++} \phi^{++}) \, , \quad \quad {\rm{with}} \quad \quad \phi^{++} = \phi^{ab} u_{a}^{+ \mathfrak{b}}
u_{b}^{+ \mathfrak{c}} \epsilon_{\mathfrak{b} \mathfrak{c}} \, ,
\label{phiplusnotation}
\end{equation}
with $\epsilon_{\mathfrak{b} \mathfrak{c}}$ the usual antisymmetric tensor with $\epsilon_{1 2} =1$
and $\phi^{ab}= - \phi^{ba}$ are combinations of the six real scalar
fields $\Phi^I$ of $\superN=4$ SYM, \ie $\phi^{ab}= (\sigma_I)^{ab}
\Phi^I$ with $(\sigma_I)^{ab}$ the $\grp{SO}(6)$ Pauli matrices.
Usually, the length-two half-BPS operator is written in terms of 6d null polarization vectors $Y_I$ in the form~\eqref{BPS},
\begin{equation}
\op{O}_i=\tr\brk[s]{\brk{{Y_i}\cdot\Phi(x_i)}^{2}}
\,,\qquad
{Y_i}\cdot{Y_i}=0
\,,
\label{eq:null}
\end{equation}
and the scalar propagator was given in \eqref{escalarpropagators}.
The connection between the two
descriptions is made easily by choosing a
convenient parametrization for the harmonic variables and their complex conjugates as the one in~\cite{Chicherin:2014uca}
\begin{equation}
u^{+ \mathfrak{a}}_{b} = (\delta^{\mathfrak{a}}_{\mathfrak{b}}, y^{\mathfrak{a}}_{\mathfrak{b}^{\prime}} ) \, , \quad
u^{- \mathfrak{a}^{\prime}}_{b} = (0, \delta^{\mathfrak{a}^{\prime}}_{\mathfrak{b}^{\prime}} ) \, , \quad
\bar{u}^b_{+ \mathfrak{a}} =(\delta^b_{\mathfrak{a}},0) \, ,
 \quad \bar{u}^b_{- \mathfrak{a}^{\prime}} =(- y^{\mathfrak{b}}_{\mathfrak{a}^{\prime}},0) \, .
 \label{harmonicdefinitions}
\end{equation}
We have for the propagator in the representation \eqref{phiplusnotation}
\begin{equation}
 \langle  \phi_1^{++} \phi_2^{++} \rangle \propto \frac{1}{x^2_{12}} \epsilon^{abcd}
 (u_1)_{a}^{+ \mathfrak{a}}
(u_1)_{b}^{+ \mathfrak{b}} \epsilon_{\mathfrak{a} \mathfrak{b}}
(u_2)_{c}^{+ \mathfrak{c}}
(u_2)_{d}^{+ \mathfrak{d}} \epsilon_{\mathfrak{c} \mathfrak{d}}
 \propto \frac{1}{x^2_{12}} (y_{12})^{\mathfrak{b}}_{\mathfrak{a}^{\prime}} (y_{12})^{\mathfrak{a}^{\prime}}_{\mathfrak{b}}
 \propto \frac{y^2_{12}}{x^2_{12}} \, ,
\end{equation}
where we have used
\begin{equation}
y_{12}^2 = - (y_{12})^{\mathfrak{b}}_{\mathfrak{a}^{\prime}} (y_{12})^{\mathfrak{a}^{\prime}}_{\mathfrak{b}}/2 \, , \quad
(y_{12})^{\mathfrak{b}}_{\mathfrak{a}^{\prime}} = (y_{1})^{\mathfrak{b}}_{\mathfrak{a}^{\prime}} -
(y_{2})^{\mathfrak{b}}_{\mathfrak{a}^{\prime}} \, , \quad
y^{\mathfrak{a}^{\prime}}_{\mathfrak{b}}=
y^{\mathfrak{a}}_{\mathfrak{b}^{\prime}} \epsilon^{\mathfrak{b}^{\prime} \mathfrak{a}^{\prime}}
\epsilon_{\mathfrak{a} \mathfrak{b}} \, .
\label{eq:defymatrix}
\end{equation}
Notice that in this notation the null condition \eqref{eq:null} is automatic.
In this work, we are interested in computing both the components $\theta_i^+=0$ and $(\theta^+_j)^4$
of the correlation functions
\begin{equation}
\mathfrak{G}_n = \langle \mathcal{T}(1) \ldots \mathcal{T}(n) \rangle \, , \quad {\rm{with}} \quad
\mathcal{T}(i) = \mathcal{T}(x_i, \theta_i^+, u_i) \, ,
\label{Gncorrelator}
\end{equation}
for several values of $n$. Notice that for $\theta_i^+=0$ for all $i$, we have
\begin{equation}
\mathfrak{G}_n \Big|_{\theta_i^+=0} = \mathcal{G}_n  =
\langle \mathcal{O}_1 \mathcal{O}_2 \ldots  \mathcal{O}_n  \rangle \, .
\end{equation}

Despite the fact that we are interested in correlation functions
of twenty prime operators $\mathcal{O}_i$, the components with  $(\theta^+_j)^4$ of $\mathfrak{G}_n $
are useful for computing loop corrections via the Lagrangian insertion method (see the previous section for a complete discussion; in this section we have an additional $\theta^{+}$ integration when compared with similar formulas of \eqref{eq:lagrangianinsertion})
\begin{equation}
\frac{1}{m!} \frac{\partial^m \mathfrak{G}_n}{\partial \gym^{2 m}} = \int \prod_{i=1}^m \dd[4]{x_{n+i}} \dd[4]{\theta^+_{n+i}} \mathfrak{G}_{n+m} \, .
\label{eq:twistorlag}
\end{equation}

The correlation function $\mathfrak{G}_n$ in
\eqref{Gncorrelator} admits an expansion in $\theta^{\alpha \, \mathfrak{b} +}_i$ of the form
\begin{equation}
\mathfrak{G}_n = \mathcal{G}_{n} + \mathcal{G}_{n;1} + \ldots + \mathcal{G}_{n; n} \, ,
\label{eq:expansionG}
\end{equation}
where by definition $\mathcal{G}_{n;k}$ has Grassmann degree $4k$ and the other orders in $\theta$ are zero by symmetry.
Moreover, the correlators transform covariantly under the $\superN=4$ superconformal transformations.
Consider the following combination of sixteen generators
\begin{equation}
( \epsilon \cdot Q) = \epsilon^a_{\alpha} \, Q^{\alpha}_a \, , \quad \quad (\bar{\xi} \cdot \bar{S}) = \bar{\xi}^{\dot\alpha a} \, \bar{S}_{\dot\alpha a} \, ,
\end{equation}
Note that the generators above form an anti-commuting subalgebra
\begin{equation}
\{ Q , \bar{S} \} = \{ Q, Q \} = \{ \bar{S}, \bar{S} \} =0 \, .
\end{equation}
It is possible to show that the variables transform as \cite{Belitsky:2014zha}
\begin{equation}
\delta x
 \, \propto \, \bar{\theta} \quad \quad
\delta u^+ \, \propto \, \bar{\theta} \quad \quad
\delta \bar{\theta}  \, \propto \, \bar{\theta}^2 \, ,
\label{thetabar}
\end{equation}
and
\begin{equation}
\theta^{\alpha \, \mathfrak{b} +}_i \rightarrow \hat\theta^{\alpha \, \mathfrak{b} +}_i
= \theta^{\alpha \, \mathfrak{b} +}_i
+ (\epsilon^{\alpha a} + x_i^{\alpha \dot{\alpha}} \xi_{\dot{\alpha}}^a) u_{i a}^{+ \mathfrak{b}} = e^{( \epsilon \cdot Q) +  (\bar{\xi} \cdot \bar{S})} \theta^{\alpha \, \mathfrak{b} +}_i \, .
\label{thetaplus}
\end{equation}

Inspecting the transformations above, we see that the value $\bar\theta=0$ is left unchanged and only the
$\theta^{\alpha \, \mathfrak{b} +}_i$ transforms.
Therefore the sixteen transformation parameters
$\{ \epsilon^a_{\alpha} , \bar{\xi}^{\dot\alpha a} \}$ can be used to set for example (for any $\alpha$ and $\mathfrak{b}$)
\begin{equation}
\theta^{\alpha \, \mathfrak{b} +}_{n-3}=\theta^{\alpha \, \mathfrak{b} +}_{n-2}=\theta^{\alpha \, \mathfrak{b} +}_{n-1}
=\theta^{\alpha \, \mathfrak{b} +}_{n}=0 \, .
\label{gaugetheta}
\end{equation}
Thus, the expansion \eqref{eq:expansionG} in fact truncates for this choice:
\begin{equation}
\mathfrak{G}_n = \mathcal{G}_{n} + \mathcal{G}_{n;1} + \ldots + \mathcal{G}_{n; n-4} \, .
\end{equation}
A special case is $n=4$, which implies that all the  $\theta_i^+$'s can be set to zero in this case.

Because the set of fermionic generators considered above
commute among themselves and are nilpotent, it is possible
to define the following set of invariants $\mathcal{I}_{n;k}$, see \cite{Chicherin:2015bza,Eden:2017fow,Heslop:2022qgf}
\begin{equation}
\mathcal{I}_{n;k} (x,y,\theta^+)= Q^8 \bar{S}^8 \mathcal{J}_{n,k+4}(x,y,\theta^+) =
\int d \epsilon \, d \xi \, \mathcal{J}_{n,k+4}(x,y,\hat\theta^+)   \, ,
\label{invariantsJ}
\end{equation}
with $\mathcal{J}_{n,k}(x,y,\theta^+)$ completely unconstrained, $\hat\theta^+$ was defined
in \eqref{thetaplus} and $\epsilon^{\alpha a}, \xi_{\dot{\alpha}}^a$ are the fermionic parameters of the transformations appearing in that formula.
As before the second index $k$ indicates that the object has Grassmann degree $4k$ and the fermionic generators
remove sixteen $\theta_i^+$'s in all possible ways (not necessary the ones in \eqref{gaugetheta}, but that particular case is among the resulting terms).
Notice that the fermionic generators annihilates $x$ and $y$ because of \eqref{thetabar}.

The components of the correlation functions
can then be expanded as follows
\begin{equation}
\mathcal{G}_{n,k} (x,y,\theta^+)= \sum_i \left( \mathcal{I}_{n;k} \right)_i (x, y, \theta^{+}) f_{n;k;i} (x) \, ,
\label{thefunctionsf}
\end{equation}

It is non-trivial that the functions $f_{n;k;i}(x)$ do not depend on the polarizations $y$'s.
However, this follows from considering the behavior of the correlators under inversion.
All the dependency on the $y$'s will come from the invariants which also depend on the positions~$x$.
The sum over $i$ goes through all the independent invariants. It is not easy to count this number precisely apart from extremal cases.
In addition under an arbitrary permutation $\sigma$ of the points $(x_i, y_i, \theta^+_i) \rightarrow (x_{\sigma \cdot i}, y_{\sigma \cdot i}, \theta^+_{\sigma \cdot i})$, the functions satisfy%
\footnote{We are using a very compact notation here. The index $i$ labels the invariants, but can also depend on the
operator labels. This is the reason why the index transforms under an arbitrary permutation $\sigma$.}
\begin{equation}
f_{n;k;i}(x_{1}, \ldots, x_{n}) = f_{n;k; \sigma \cdot i}(x_{\sigma \cdot1}, \ldots,  x_{\sigma \cdot n}) \, .
\end{equation}

As mentioned in the Introduction, it will be nice to write our results in the form \eqref{thefunctionsf}. In~\cite{Chicherin:2014uca}, a procedure for computing $\mathfrak{G}_n$
using supertwistor techniques was described, and we are going to perform the calculations
using the numerical version of it firstly used in~\cite{Fleury:2019ydf}.
The procedure uses the twistor $\superN=4$ SYM action given in~\cite{Witten:2003nn,Boels:2006ir}.
Supertwistors $\mathcal{Z}^A$ live in the complex projective superspace $\mathbb{CP}^{3|4}$, and
they are parametrized as ($\chi^a$ are fermionic coordinates)
\begin{equation}
\mathcal{Z}^A =  (Z^I, \chi^{a}) \, , \quad {\rm{with}} \quad Z^I = (\lambda_{\alpha}, \mu^{\dot{\alpha}})\, .
\label{bostwistor}
\end{equation}
and $Z^I$ are bosonic twistors. A spacetime point $x^{\dot{\alpha} \beta}$ corresponds to a line in bosonic twistor space,
more precisely, the relation of these twistor variables with the $\superN=4$ superspace variables when
all the $\bar{\theta}^{\dot\alpha}_a$'s are zero is given by the following incidence relations%
\footnote{The relations are more complicated when $\bar{\theta}^{\dot\alpha}_a \neq 0$.
They can be found for example in~\cite{Berkovits:2009by,Bullimore:2013jma}.}
\begin{equation}
\mu^{\dot\alpha} = i x^{\dot{\alpha} \beta} \lambda_{\beta} \, , \quad \quad \chi^a = \theta^{a \alpha} \lambda_{\alpha} \, .
\end{equation}

For computing correlation functions perturbatively, we need both an expression for the superfield $\mathcal{T}(x, \theta^+, u)$ and the Feynman rules in twistor space.
We only summarize the formulas here and we refer the reader to~\cite{Chicherin:2014uca}
for details and derivations. The chiral stress-tensor supermultiplet $\mathcal{T}(x, \theta^+, u)$ in this language is given by
\begin{equation}
\mathcal{T}(x, \theta^+, u)= \int \dd[4]{\theta^-} L_{\rm{int}} (x, \theta) \, ,
\end{equation}
where $L_{\rm{int}} (x, \theta)$ is the interaction Lagrangian in twistor space.
This follows as a consequence of the Lagrangian insertion procedure.
The interaction term $L_{\rm{int}} (x, \theta)$ in a particular gauge is a sum of infinitely many terms
containing the one-form superfield $\mathcal{A}(\mathcal{Z}_{1,2})$
\begin{equation}
\begin{aligned}
\mathcal{A}(\mathcal{Z}_{1,2}) & = a(Z_{1,2}) + \chi^a \psi_a(Z_{1,2}) + \frac{1}{2} \chi^a \chi^b \phi_{ab}(Z_{1,2}) \\
& + \frac{1}{3!} \epsilon_{a b c d}
\chi^a \chi^b \chi^c \psi^{\prime d}(Z_{1,2}) + \frac{1}{4!} \epsilon_{a b c d} \chi^a \chi^b \chi^c \chi^d a^{\prime}(Z_{1,2}) \, .
\label{superA}
\end{aligned}
\end{equation}
In the formula above, the fields $a(Z_{1,2})$ and $a^{\prime}(Z_{1,2})$ are the two helicity gluons,
$\psi_a(Z_{1,2})$ and $\psi^{\prime d}(Z_{1,2})$ are the gluinos and
$\phi_{ab}(Z_{1,2})$ are the six scalars. The bosonic twistors
$Z_{1,2} = \{ Z_1 , Z_2 \}$ are two independent twistors parametrizing a line
given by the spacetime point where the fields are defined. Since $L_{\rm{int}}(x, \theta)$ has terms with arbitrary
many superfields $\mathcal{A}(\mathcal{Z}_{1,2})$, a general order perturbative calculation of the correlation functions
can have in principle vertices with arbitrary valences. Given a set of operators $\mathcal{T}(i)$ at space-time
positions $x^{\mu}_i$ determining lines in twistor space parametrized by two independent twistors $Z_{i,1}, Z_{i,2}$, the Feynman rules in the so called
axial gauge are summarized below.
This gauge choice is defined by the vanishing
of the superfield $\mathcal{A}(\mathcal{Z}_{1,2})$ of \eqref{superA} in the direction of an auxiliary twistor $\mathcal{Z}_{\diamond}$.
It is possible to take the fermionic part of $\mathcal{Z}_{\diamond}$ to zero without losing generality.
The bosonic part of it will be denoted by $Z_{\diamond}$. Of course, any correlation function result is expected
to be independent of the $Z_{\diamond}$ choice.
Notice that the individual diagrams depend on the value of $Z_{\diamond}$, and this breaks
the manifest $\superN=4$ superconformal invariance at intermediate
steps, however there are cancellations among the graphs, see~\cite{Chicherin:2014uca}.

The Feynman rules are the following.
If the lines $i$ and $j$ are connected by a propagator, the graph is multiplied
by the factor $(y^2_{i j}/x_{ij}^2) \delta^{a_i a_j}$ where the delta function is a color delta function.
The $m$-valence vertex
connecting the line $i$ to the lines $j_1, \ldots, j_m$ is given by
\begin{equation}
V^i_{j_1, \ldots, j_m}=R^i_{j_1, \ldots, j_m} {\rm{Tr}}(T^{a_1} \ldots T^{a_m}) \, ,
\end{equation}
where $T^a$ are $\grp{U}(\Nc)$ generators%
\footnote{In this paper we are going to consider only planar correlation functions and
for this case the $\grp{U}(\Nc)$ and $\grp{SU}(\Nc)$ groups give the same results. This is also true in general when all the half-BPS external operators
have length two~\cite{Fleury:2019ydf}.}
and
\begin{equation}
R^i_{j_1 j_2 \ldots j_k} = -\int \frac{\dd[4]{\theta^-_i}}{(2\pi)^2} \frac{\delta^2(\langle \sigma_{i j_1} \theta^-_i \rangle+A_{i j_1})
\delta^2(\langle \sigma_{i j_2} \theta^-_i \rangle+A_{i j_2}) \ldots
\delta^2(\langle \sigma_{i j_k} \theta^-_i \rangle+A_{i j_k})}{\langle \sigma_{i j_1} \sigma_{i j_2} \rangle
\langle \sigma_{i j_2} \sigma_{i j_3} \rangle \ldots \langle \sigma_{i j_k} \sigma_{i j_1} \rangle} \, ,
\label{eq:TheR}
\end{equation}
with
\begin{equation}
\sigma^{\alpha}_{ij} = \epsilon^{\alpha \beta} \frac{\langle Z_{i, \beta} Z_{\diamond} Z_{j,1} Z_{j,2} \rangle}{\langle
Z_{i, 1} Z_{i,2} Z_{j,1} Z_{j,2}\rangle} \, , \quad \quad  {\rm{and}} \quad \quad \langle \sigma_i \sigma_j \rangle = \epsilon_{\alpha \beta} \sigma_i^{\alpha} \sigma_j^{\beta} \, .
\end{equation}
Finally, each diagram is multiplied by an explicit factor
\begin{equation}
\brk2{\frac{\gym^2}{4\pi^2}}^p
\,,
\label{eq:twistorprefactor}
\end{equation}
where $p=P-V$, with $P$ the total number of propagators and $V$ the
total number of vertices of the diagram.%
\footnote{From the definition of $R$~\eqref{eq:TheR}, one can see
that, with this definition of $p$, the diagram is of order~$\theta^{4p}$.}

In the formulas above,
\begin{equation}
A^{\mathfrak{a}^{\prime}}_{ij} = [ \langle \sigma_{ji} \theta^{+ \mathfrak{b}}_j \rangle
+ \langle \sigma_{ij} \theta^{+ \mathfrak{b}}_i \rangle] (y^{-1}_{ij})^{\mathfrak{a}^{\prime}}_{\mathfrak{b}} \, .
\label{theA}
\end{equation}
The matrix $(y_{ij})^{\mathfrak{b}}_{\mathfrak{a}^{\prime}}$ was defined in \eqref{eq:defymatrix}.
Its inverse appearing above satisfies $ (y^{-1}_{ij})^{\mathfrak{c}^{\prime}}_{\mathfrak{b}}  (y_{ij})^{\mathfrak{b}}_{\mathfrak{a}^{\prime}} = \delta^{\mathfrak{c}^{\prime}}_{\mathfrak{a}^{\prime}}$
and $(y_{ij})^{\mathfrak{c}}_{\mathfrak{a}^{\prime}}  (y^{-1}_{ij})^{\mathfrak{a}^{\prime}}_{\mathfrak{b}} = \delta^{\mathfrak{c}}_{\mathfrak{a}}$.
The four bracket is defined as
\begin{equation}
\langle Z_{1} Z_{2} Z_{3} Z_{4} \rangle =  \epsilon_{I J K L}  Z_{1}^I  Z_{2}^J
 Z_{3}^K  Z_{4}^L \, .
\end{equation}
In particular, using the parametrization of a bosonic twistor given in \eqref{bostwistor}, one has
\begin{equation}
\langle Z_{i,1} Z_{i,2} Z_{j,1} Z_{j,2} \rangle =  (\epsilon^{\alpha \beta} \lambda_{i, \alpha}  \lambda_{i, \beta})
(\epsilon^{\gamma \delta} \lambda_{j, \gamma}  \lambda_{j,\delta}) \, x_{ij}^2 \, .
\end{equation}

In this work, we are going to perform the twistor calculations using a Lorentzian signature metric.
As long as the external points span the full four-dimensional space,
terms of the form
$\epsilon_{\mu \nu \rho \sigma} x_i^\mu x_j^\nu x_k^\rho x_l^\sigma$ can be nonzero.
In the twistor reformulation of $\superN=4$ SYM, the action has the
total derivative term $\ii F\mspace{-2mu}\tilde{F}$ where $F_{\mu \nu}$ is the field strength and $\tilde{F}_{\mu \nu}$ its dual.
The term $\ii F\mspace{-2mu}\tilde{F}$ can potentially generate such $\epsilon$ terms to the integrands, however
the integrated correlator is insensitive to these contributions because they are produced by a total derivative term.
Using a Lorentzian metric, the $\epsilon$ terms appear as imaginary contributions
to the correlation functions, and therefore can easily be isolated numerically.%
\footnote{These terms also change sign when some of the $(x_i)_{\mu}  \, \rightarrow \,  - (x_i)_{\mu}$. So even using Euclidean signature
it is possible to isolate these kind of contributions by taking special combinations of points.}
It is also possible to numerically bootstrap the $\epsilon$ terms by writing a
basis of integrals and performing a numerical fitting, see~\cite{Fleury:2019ydf}
for an example. In this work,
we are going to consider only the real part of the numerical results for the correlation functions obtained by
the twistor method.

In order to compute loop corrections for the $n$-point functions $\mathcal{G}_n$ of $\tp$ operators, we are going to use the formula \eqref{eq:twistorlag}.
At $\ell$ loops, we need to compute $G_{n+\ell}$ at order $4\ell$ on the $\theta^+_i$'s for $i > n$.
The diagrams contributing for this case contain $n+\ell$ vertices (lines in twistor space) and $n+ 2\ell$ propagators.
This follows from the Feynman rules described above, as the $n$-valence vertex
with $n>2$ contributes with $2(n-2)$ $\theta$'s, see \eqref{eq:TheR}.
It then follows that the diagram is of order $\theta^{4\ell}$. The
correlator $G_{n+\ell}$ of $n$ $\tp$ operators and $\ell$
Lagrangian insertions is extracted by setting $\theta_i^+=0$ for
$i \leq n$, which projects to the $\prod_{i=n+1}^{n+\ell}(\theta_i^+)^4$
component, whose coefficient is the desired correlator. Collecting
powers of $2\pi$ and $\gym$, and doing the color algebra, one finds
that every diagram contains an overall factor
\begin{equation}
\frac{\Nc^2-1}{(2\pi)^{2n+2\ell}}\brk2{\frac{\gym^2\Nc}{4\pi^2}}^\ell
=\frac{\Nc^2-1}{(2\pi)^{2n+2\ell}}g^{2\ell}
=C_{2\dots2}\,g^{2\ell}
\,.
\label{eq:C222fromtwistors}
\end{equation}
This is exactly the overall prefactor~\eqref{eq:C222}.

For the connected part of the correlator, only connected diagrams are important, because it is possible to show
that disconnected twistor diagrams contribute only to lower-point correlators.
Notice that \eqref{eq:TheR} implies that two operators can at most be
connected by a single propagator,
otherwise the factor $R^i_{j_1 j_2 \ldots j_k}$ vanishes as it is
antisymmetric. Moreover, all the vertices must have valence at least two. It is possible to
generate all the necessary skeleton graphs very efficiently using \sagemath~\cite{sagemath}.
For example, all the six-point two-loop skeleton graphs can be generated by using the code
of \tabref{tablesix}.
\begin{table}
\begin{center}
\begin{tabular}{cc}
\toprule
\texttt{degree\_sequence}  &  $\#$\,Graphs \\
\midrule
\texttt{[2,2,2,2,2,2,2,6]} & $1$ \\
\texttt{[2,2,2,2,2,2,3,5]} & $7$ \\
\texttt{[2,2,2,2,2,2,4,4]} & $9$ \\
\texttt{[2,2,2,2,2,3,3,4]} & $31$ \\
\texttt{[2,2,2,2,3,3,3,3]} & $28$ \\
\bottomrule
\end{tabular}
\end{center}
\caption{There are in total 76 six-point two-loop skeleton graphs (some of them are disconnected).
At this loop order, there must be 8 vertices (length of the sequence of numbers) and 10 propagators
(half the sum of the numbers).
The command in Sage is \texttt{graphs(8, degree\_sequence = (...))}, with
\texttt{(...)} replaced by an entry of the table. The graphs can be transformed
to a list of adjacency matrices and saved in a file.}
\label{tablesix}
\end{table}

Using the skeleton graphs and the Feynman rules, it is possible
to generate all the graphs. The total number increases fast with the number of points $n$ and loops $\ell$,
and it is hard even for the simplest cases to do any analytical simplification.
Thus we have evaluated all the graphs numerically. A great simplification
is that even numerically it is possible to select a particular polarization
from the beginning, which projects out many permutations and graphs.
The vertices $R^i_{j_1 j_2 \ldots j_k}$ contain the factors of $(y^{-1}_{ij})^{\mathfrak{a}^{\prime}}_{\mathfrak{b}}$
inside the $A^{\mathfrak{a}^{\prime}}_{ij}$, see \eqref{theA}, and in principle
these factors can change the factors of $y^2_{ij}$ coming from the propagators.
However, the factors $(y^{-1}_{ij})^{\mathfrak{a}^{\prime}}_{\mathfrak{b}}$ multiply a $\theta^+_i$ or a $\theta^+_j$ and
because we are only interested in the contributions with $\theta^+_i =0$ for $i \leq n$
the $R$-charge coming from the external propagators are never canceled.

For example, in the case of five- and six-point functions, it is possible to select the operator polarizations in such way that
only the cyclic contribution ($y^2_{12} y^2_{23} \ldots y^2_{i1}$), or any other disconnected contribution, is non vanishing.
This is not true for $n>6$ points.
The operators for the cyclic case are given in~\cite{Bargheer:2019kxb} and they read for five points
\begin{equation}
\mathcal{O}_1 = {\rm{Tr}} (X X) \, , \quad
\mathcal{O}_2 = {\rm{Tr}} (\bar{X} \bar{Y}) \, , \quad
\mathcal{O}_3 = {\rm{Tr}} (\bar{Z} Y) \, , \quad
\mathcal{O}_4 = {\rm{Tr}} (Z Z) \, , \quad
\mathcal{O}_5 = {\rm{Tr}} (\bar{Z} \bar{X}) \, , \quad
\end{equation}
where the bar means complex conjugation, and the fields $\set{X,Y,Z}$ are defined in terms of the real scalars $\Phi^I$ as
\begin{equation}
X = \frac{1}{\sqrt{2}} ( \Phi^1 + i \Phi^2) \, , \quad  Y = \frac{1}{\sqrt{2}} ( \Phi^3 + i \Phi^4)  \, , \quad  Z = \frac{1}{\sqrt{2}} ( \Phi^5 + i \Phi^6) \, .
\end{equation}
For the six-point case, one has,
\begin{equation}
\begin{aligned}
& \mathcal{O}^{\prime}_1 = {\rm{Tr}} (X X) \, , \quad
\mathcal{O}^{\prime}_2 = {\rm{Tr}} (\bar{X} \bar{Y}) \, , \quad
\mathcal{O}^{\prime}_3 = {\rm{Tr}} (Y Y) \, , \quad \\
& \quad \quad \mathcal{O}^{\prime}_4 = {\rm{Tr}} (\bar{Z} \bar{Y}) \, , \quad
\mathcal{O}^{\prime}_5 = {\rm{Tr}} (Z Z) \, , \quad
\mathcal{O}^{\prime}_6 = {\rm{Tr}} (\bar{Z} \bar{X}) \, . \quad
\end{aligned}
\end{equation}
These polarizations can be reproduced by selecting particular values for the matrices $(y_i)^{\mathfrak{a}}_{\mathfrak{b}^{\prime}}$ appearing in the definitions
of the harmonic variables in \eqref{harmonicdefinitions}. We define the function
\begin{equation}
y^{\prime}(a,b,c,d) = a  \, \mathbbm{e}_{11} + b  \, \mathbbm{e}_{12} + c  \, \mathbbm{e}_{21} + d  \, \mathbbm{e}_{22} \, ,
\end{equation}
where $ \mathbbm{e}_{ij}$ is a $2 \times 2$ matrix with a single nonzero component at position $\{i,j\}$ with value~$1$.
The cyclic six-point polarization, for example, can be taken to be
\begin{align}
(y_1)^{\mathfrak{a}}_{\mathfrak{b}^{\prime}} &= y^{\prime}(1,0,0,0) \,, &
(y_2)^{\mathfrak{a}}_{\mathfrak{b}^{\prime}} &= y^{\prime}(0,0,1,1) \,, &
(y_3)^{\mathfrak{a}}_{\mathfrak{b}^{\prime}} &= y^{\prime}(0,1,0,0) \,,  \\
(y_4)^{\mathfrak{a}}_{\mathfrak{b}^{\prime}} &= y^{\prime}(0,0,1,0) \,, &
(y_5)^{\mathfrak{a}}_{\mathfrak{b}^{\prime}} &=y^{\prime}(1,0,-1,1) \,, &
(y_6)^{\mathfrak{a}}_{\mathfrak{b}^{\prime}} &=y^{\prime}(0,1,1,1/2) \,.
\end{align}
The explicit polarizations that we used in our computation are given
in \secref{sec:method} below.

Finally, the numerical results for the correlators obtained with the twistors were fitted against the ansatz of integrals described in \secref{sec:ansatz}.
The positions of the operators were generated randomly, and the number of equations were always greater
than the number of unknowns coefficients in the basis.

%%%%%%%%%%%%%%%%%%%%%%%%%%%%%%%%%%%%%%%%%%%%%%%%%%%%%%%%%%%%
%%%%%%%%%%%%%%%%%%%%%%%%%%%%%%%%%%%%%%%%%%%%%%%%%%%%%%%%%%%%
\section{Results}
\label{sec:results}

%%%%%%%%%%%%%%%%%%%%%%%%%%%%%%%%%%%%%%%%%%%%%%%%%%%%%%%%%%%%
\subsection{Method}
\label{sec:method}

%%%%%%%%%%%%%%%%%%%%%%%%%%%%%%
\paragraph{Strategy.}

We fix the free coefficients in the ansatz polynomials
$P_{\mathbold{a}}^{(\ell)}$ constructed in \secref{sec:ansatz} by
matching the ansatz correlators~\eqref{eq:G5,eq:G6G7} against the
$(n+\ell)$-point tree correlator computed from twistors
(\secref{sec:twistors}) on many numerical points $(x_i,y_i)$. The
numerical data points provide a linear system for the free
coefficients in the ansatz that we solve numerically. The solution is
not unique due to non-trivial Gram determinant relations among the
various terms in the ansatz. However, at least at two loops, we notice
that once we restrict the ansätze for the component functions
$f_{\mathbold{a}}^{(2)}$ to a certain set of conformal integrals, all
ambiguity is removed, and the solution becomes unique.

%%%%%%%%%%%%%%%%%%%%%%%%%%%%%%
\paragraph{Computational Aspects.}

In the computation of the data points and finding the solution for the
ansatz parameters, there are two main bottlenecks:
\begin{itemize}
%%%%%%%%%%%%%%%
\item The symbolic algebra of Grassmann-odd variables $\theta_i$ in
the twistor computation, especially when the number of contributing
twistor diagrams becomes large,
%%%%%%%%%%%%%%%
\item Solving large and dense numerical linear systems.
%%%%%%%%%%%%%%%
\end{itemize}
For the first point, we could boost the performance by representing
homogeneous polynomials in a finite number of Grassmann-odd variables
as component vectors. Multiplication of two or more homogeneous
polynomials can then be implemented by precomputed numerical tensors,
such that the computation becomes completely numerical.
With this and some other optimizations as well as parallelization, we
could compute a few thousand data points per day on a $48$-core machine.

The second point is somewhat more essential: Our method inevitably
produces large and completely dense linear systems, whose coefficients
are either high-precision floats, or large rationals. Solving such
systems is a hard computational problem. Using \mathematica's
\lstinline"Nsolve", we could solve such systems up to size
${\sim}10\,000$, but would run out of memory beyond that, even on a
machine with $256$\,GB memory.

%%%%%%%%%%%%%%%%%%%%%%%%%%%%%%
\paragraph{Polarizations.}

In order to contain the sizes of the linear systems, as well as the
numbers of graphs that contribute to the twistor computation, we
identify a few numerical choices for the polarizations $Y_i$,
$1\leq{i}\leq{n+\ell}$ that set all but a few of the polarization
structures $\prod{d_{ij}^{a_{ij}}}$ to zero. For each polarization
choice, we then evaluate the ansatz and compute the twistor correlator
for many numerical values of the coordinates $x_i$, $1\leq{i}\leq{n+\ell}$.
At five points, we use the following two polarizations:
\begin{equation}
\mathbold{Y}_{23}=\begin{pmatrix}
-1 & 0    & 1 & \ii  & 0      & -\ii \\
 1 & 0    & 1 & \ii  & 0      & -\ii \\
 0 & -1   & 1 & 0    & -\ii   & -\ii \\
 0 & -1   & 1 & 0    &  \ii   & -\ii \\
 0 & -1/3 & 2 & 2\ii & -\ii/3 & 0
\end{pmatrix}
\,,\qquad
\mathbold{Y}_{5}=\begin{pmatrix}
-1 & 0  & 1 & \ii  & 0    & -\ii \\
 1 & 0  & 1 & \ii  & 0    & -\ii \\
 1 & -1 & 3 & 3\ii &  \ii &  \ii \\
 0 & -1 & 2 & 2\ii & -\ii & 0 \\
 2 & -3 & 2 & 4\ii & -\ii & 0
\end{pmatrix}
\,,
\end{equation}
where we collected the external $Y_i$, $1\leq{i}\leq5$ into a vector
$\mathbold{Y}$. In all cases, the polarizations of the integration points
$Y_{i}$, $i>n$ is set to arbitrary fixed values,
such that all $d_{ij}$, ${i}\leq{n}$, $j>n$,
are non-zero.
For the choice $\mathbold{Y}_{23}$, the only non-zero
polarization structure is
\begin{equation}
d_{12}^2d_{34}d_{45}d_{53}
=\frac{128}{3x_{12}^4x_{34}^2x_{45}^2x_{35}^2}
\,,
\end{equation}
hence we can use this polarization to determine the $f_{23}$ component function.
For the choice $\mathbold{Y}_5$, the only non-zero polarization
structure is
\begin{equation}
d_{12}d_{23}d_{34}d_{45}d_{51}
=\frac{-64}{x_{12}^2x_{23}^2x_{34}^2x_{45}^2x_{15}^2}
\,,
\end{equation}
hence we can use this polarization to determine the $f_5$ component function.
We list the numbers of twistor graphs that contribute to the two- and
three-loop correlators for the various choices of polarizations in
\tabref{tab:twistorgraphs}.
\begin{table}
\centering
\begin{tabular}{cccccccc}
\toprule
         & $\mathbold{Y}_{23}$ & $\mathbold{Y}_{5}$ & $\mathbold{Y}_6$ & $\mathbold{Y}_{7,1.1}$ & $\mathbold{Y}_{7,1.2}$ & $\mathbold{Y}_{7,2}$ & $\mathbold{Y}_{7,3}$ \\
\midrule
$\ell=2$ & $\num{557}$         & $\num{1790}$       & $\num{221910}$   & $\num{21154}$          & $\num{21154}$          & $\num{24502}$        & $\num{18688}$ \\
$\ell=3$ & $\num{73380}$       & $\num{167430}$     &                  &                        &                        &                      & \\
\bottomrule
\end{tabular}
\caption{The numbers of twistor graphs that contribute to the two- and
three-loop correlators for the various choices of polarizations.}
\label{tab:twistorgraphs}
\end{table}
At six points, we use the relatively random polarization
\begin{equation}
\mathbold{Y}_6=\begin{pmatrix}
\frac{622}{7} & -\frac{274}{13}  & \frac{1101}{91} & \frac{624}{7}\ii  & -\frac{272}{13}\ii  & \frac{919}{91}\ii \\
0             & -2               & 1               & 2\ii              & 0                   & -\ii \\
2             & -4               & 6               & 6\ii              & 2\ii                & 4\ii \\
-4            & -\frac{1171}{78} & \frac{333}{26}  & 8\ii              & -\frac{1169}{78}\ii & \frac{281}{26}\ii \\
-15           & -12              & -30             & 7\ii              & -14\ii              & -32\ii \\
\frac{111}{2} & 80               & -127            & -\frac{113}{2}\ii & -76\ii              & -129\ii
\end{pmatrix}
\,.
\end{equation}
With this polarization, all $d_{ij}$ are non-zero, so all component
functions contribute. At two loops, the total number of unknowns in
the ansatz then is (see \tabref{tab:ansatzsizes})
$235+572+173+657=1637$. A linear system of this size is still easily
solvable. At seven points, we use four different polarizations:
\begin{gather}
\mathbold{Y}_{7,1.1}=\begin{pmatrix}
-1 & 0  & 1 & \ii  & 0    & -\ii \\
 1 & -1 & 1 & \ii  &  \ii & -\ii \\
 0 & -1 & 1 & 0    & -\ii & -\ii \\
 0 & -1 & 1 & 0    &  \ii & -\ii \\
 0 & -1 & 2 & 2\ii &  \ii & 0 \\
 1 & -2 & 0 & \ii  & 0    & -2\ii \\
 1 & 0  & 1 & \ii  & 0    & -\ii
\end{pmatrix}
\,,\qquad
\mathbold{Y}_{7,1.2}=\begin{pmatrix}
-1 & 0  & 1 & \ii   & 0      & -\ii \\
1  & -1 & 1 & \ii   & \ii    & -\ii \\
0  & -1 & 1 & 0     & -\ii   & -\ii \\
0  & -1 & 1 & 0     & \ii    & -\ii \\
0  & 2  & 2 & 2 \ii & -2 \ii & 0 \\
-1 & -2 & 0 & -\ii  & 0      & -2 \ii \\
1  & 0  & 1 & \ii   & 0      & -\ii
\end{pmatrix}
\,,\\
\mathbold{Y}_{7,2}=\begin{pmatrix}
-1   & 0    & 1   & \ii   & 0      & -\ii \\
1    & -1   & 1   & \ii   & \ii    & -\ii \\
0    & -1   & 1   & 0     & -\ii   & -\ii \\
0    & -1   & 1   & 0     & \ii    & -\ii \\
-\sfrac{1}{2} & -\sfrac{3}{2} & \sfrac{1}{2} & \sfrac{1}{2}\ii & -\sfrac{1}{2}\ii & -\sfrac{3}{2}\ii \\
-1   & -2   & 0   & -\ii  & 0      & -2 \ii \\
1    & 1    & 3   & 3 \ii & \ii    & \ii
\end{pmatrix}
\,,\qquad
\mathbold{Y}_{7,3}=\begin{pmatrix}
-1    & 0        & 1    & \ii        & 0           & -\ii \\
1     & -1       & 1    & \ii        & \ii         & -\ii \\
0     & -1       & 1    & 0          & -\ii        & -\ii \\
1     & -1       & 1    & \ii        & \ii         & -\ii \\
3     & 10       & 8    & 11 \ii     & -4 \ii      & 6 \ii \\
\sfrac{16}{99} & -\sfrac{265}{198} & \sfrac{17}{9} & \sfrac{160}{99}\ii & \sfrac{329}{198}\ii & -\sfrac{1}{9}\ii \\
12    & 11       & 13   & 12 \ii     & 13 \ii      & 11 \ii
\end{pmatrix}
\,.\nn
\end{gather}
For the first polarization $\mathbold{Y}_{7,1.1}$, the only non-zero
polarization structures are
\begin{equation}
d_{23}^2d_{45}^2d_{16}d_{67}d_{71}
=\frac{-256}{x_{23}^4x_{45}^4x_{16}^2x_{67}^2x_{71}^2}
\,,\qquad
d_{12}^2d_{67}^2d_{34}d_{45}d_{53}
=\frac{256}{x_{12}^4x_{67}^4x_{34}^2x_{45}^2x_{53}^2}
\,,
\end{equation}
hence we can use this polarization to determine the $2435$
coefficients of the $f_{223}$ component function.
For the next polarization $\mathbold{Y}_{7,1.2}$, the only non-zero
polarization structures are
\begin{equation}
d_{12}^2d_{67}^2d_{34}d_{45}d_{53}
=\frac{-128}{x_{12}^4x_{67}^4x_{34}^2x_{45}^2x_{53}^2}
\,,\quad
d_{12}d_{23}d_{34}d_{45}d_{56}d_{67}d_{71}
=\frac{128}{x_{12}^2x_{23}^2x_{34}^2x_{45}^2x_{56}^2x_{67}^2x_{71}^2}
\,.
\end{equation}
Plugging in the known answer for the $f_{223}$ component function, we
can therefore use this polarization to determine the $6143$
coefficients of the $f_{7}$ component function. For the
polarization $\mathbold{Y}_{7,2}$, the only contributing components
are one $f_{223}$ function, one $f_{7}$ function, and two $f_{25}$
functions. With the final polarization $\mathbold{Y}_{7,3}$, the
contributing components are two $f_{223}$ functions and one $f_{34}$
function. We can thus use these polarizations to independently
determine the remaining components $f_{25}$ and $f_{34}$.
Recall that the polarizations are expressed using two-dimensional indices in \secref{sec:twistors}. It is possible
to solve for all the $(y_{i})^{\mathfrak{b}}_{\mathfrak{a}^{\prime}}$ in \eqref{eq:defymatrix} allowing complex solutions and imposing that all the
$y_{ij}^2\sim{Y_i}\cdot{Y_j}$ obtained here are reproduced.

%%%%%%%%%%%%%%%%%%%%%%%%%%%%%%%%%%%%%%%%%%%%%%%%%%%%%%%%%%%%
\subsection{Two-Loop Integrals}
\label{sec:integrals}

\begin{figure}
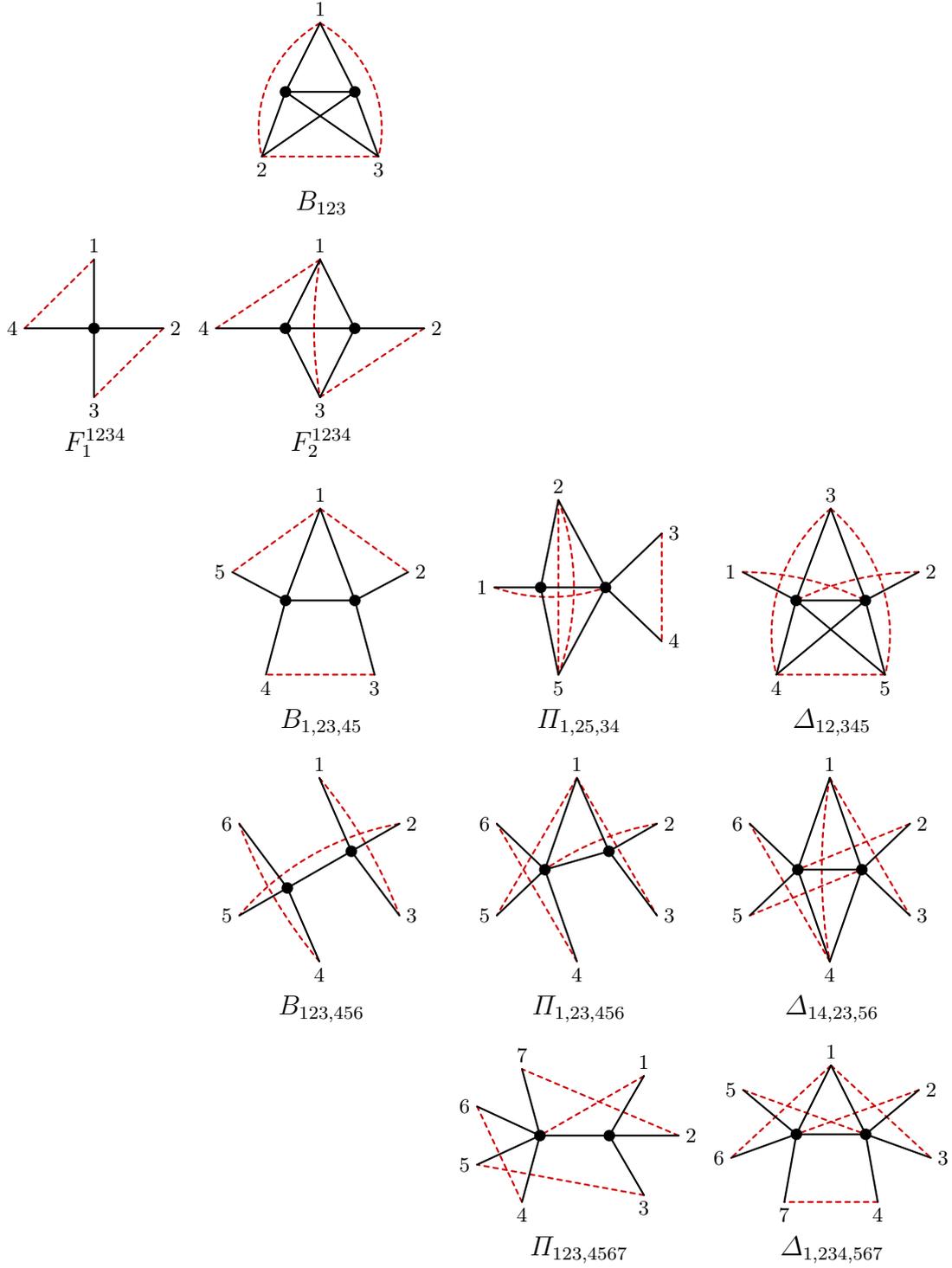

\centering
\begin{tabular}{c@{\;\;}c@{\;\;}c@{\;\;}c}
                             & \includegraphics{FigIntYY}   &                             &                             \\
                             & $\threedoublebox_{123}$      &                             &                             \\[1.5ex]
\includegraphics{FigLadder1} & \includegraphics{FigLadder2} &                             &                             \\
$F_1^{1234}$                 & $F_2^{1234}$                 &                             &                             \\[1.5ex]
                             & \includegraphics{FigIntLL}   & \includegraphics{FigIntPP}  & \includegraphics{FigIntQQ}  \\
                             & $\fivedoublebox_{1,23,45}$   & $\fivepentabox_{1,25,34}$   & $\fivedoublepenta_{12,345}$ \\[1.5ex]
                             & \includegraphics{FigIntBB}   & \includegraphics{FigIntPB}  & \includegraphics{FigIntDP}  \\
                             & $\doublebox_{123,456}$       & $\pentabox_{1,23,456}$      & $\doublepenta_{14,23,56}$   \\[1.5ex]
                             &                              & \includegraphics{FigIntPB7} & \includegraphics{FigIntDP7} \\
                             &                              & $\pentabox_{123,4567}$      & $\doublepenta_{1,234,567}$
\end{tabular}
\caption{The complete set of conformal integrals that appear in the
correlation functions of $\tp$ operators for up to seven points and two
loops, see~\protect\eqref{eq:integrals}. \emph{Black:} Propagators
$1/x_{ij}^2$. \emph{Red dashed:} Numerator factors $x_{ij}^2$. Further
two-loop conformal integrals are shown
in \protect\figref{fig:extra-integrals}.}
\label{fig:conformal-integrals}
\end{figure}

\begin{figure}[tb]
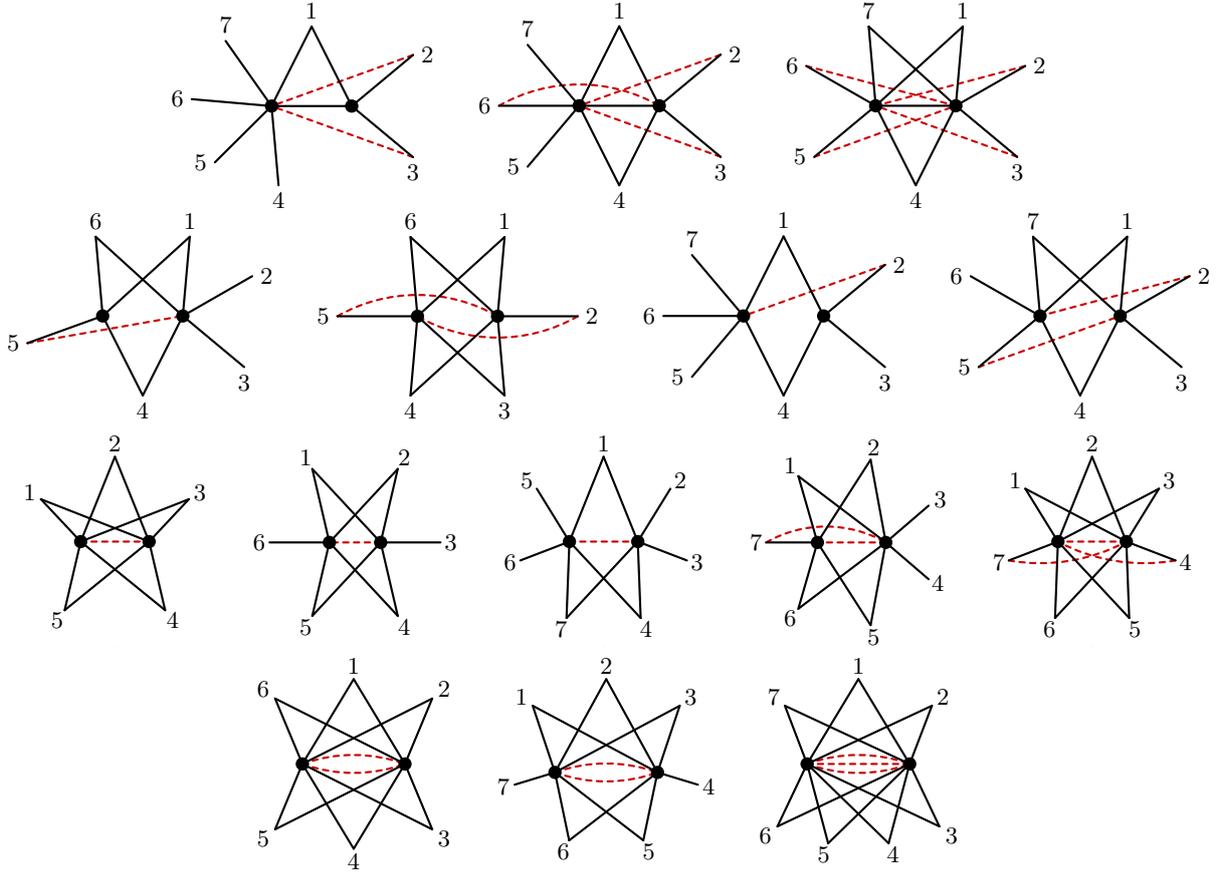

\centering
\includegraphics[align]{FigIntSevenHexaBox.mps}
\quad
\includegraphics[align]{FigIntSevenPentaHexa.mps}
\quad
\includegraphics[align]{FigIntSevenHexaHexa.mps}
\\
\includegraphics[align]{FigIntSixPentaBoxNoProp.mps}
\quad
\includegraphics[align]{FigIntSixPentaPentaNoProp.mps}
\quad
\includegraphics[align]{FigIntSevenPentaBoxNoProp.mps}
\quad
\includegraphics[align]{FigIntSevenPentaPentaNoProp.mps}
\\[1ex]
\includegraphics[align]{FigIntFiveDoublePenta2}
\quad
\includegraphics[align]{FigIntSixDoublePenta2}
\quad
\includegraphics[align]{FigIntSevenDoublePenta2}
\quad
\includegraphics[align]{FigIntSevenPentaHexaNum}
\quad
\includegraphics[align]{FigIntSevenDoubleHexa}
\\[0.5ex]
\includegraphics[align]{FigIntSixDoubleHexaNum.mps}
\quad
\includegraphics[align]{FigIntSevenDoubleHexa2}
\quad
\includegraphics[align]{FigIntSevenDoubleHeptaNum.mps}
\caption{Two-loop conformal integrals with up to seven external points that
appear in the ansätze as constructed in \protect\secref{sec:ansatz},
but do not occur in the final two-loop correlation functions of $\tp$
operators. Unlike in \protect\figref{fig:conformal-integrals}, numerator
factors between external points are not drawn.}
\label{fig:extra-integrals}
\end{figure}

Up to two loops and seven points, we find that the correlators of
$\tp$ operators can be expressed in terms of the following conformally
invariant integrals (see~\figref{fig:conformal-integrals}):
\begin{align}
F_1^{1243}&\equiv
x_{13}^2x_{24}^2
\int\frac{\dd[4]x_5}{x_{15}^2x_{25}^2x_{35}^2x_{45}^2}
\,,\nn\\
\threedoublebox_{123}&\equiv
x_{12}^2x_{13}^2x_{23}^2
\int\frac{\dd[4]{x_4}\dd[4]{x_5}}
{\brk{x_{14}^2x_{24}^2x_{34}^2}x_{45}^2\brk{x_{15}^2x_{25}^2x_{35}^2}}
=\doublebox_{123,231}
\,,\nn\\
F_2^{1243}&\equiv
x_{13}^2x_{24}^2x_{14}^2
\int\frac{\dd[4]{x_5}\dd[4]{x_6}}{\brk{x_{15}^2x_{25}^2x_{45}^2}x_{56}^2\brk{x_{46}^2x_{36}^2x_{16}^2}}
=\doublebox_{124,341}
\,,\nn\\
\fivedoublebox_{1,23,45}&\equiv
x_{12}^2x_{15}^2x_{34}^2
\int\frac{\dd[4]{x_6}\dd[4]{x_7}}{\brk{x_{46}^2x_{56}^2x_{16}^2}x_{67}^2\brk{x_{17}^2x_{27}^2x_{37}^2}}
=\doublebox_{132,541}
\,,\nn\\
\fivepentabox_{1,25,34}&\equiv
x_{25}^4x_{34}^2
\int\frac{x_{16}^2\dd[4]{x_6}\dd[4]{x_7}}
{\brk{x_{26}^2x_{36}^2x_{46}^2x_{56}^2}x_{67}^2\brk{x_{57}^2x_{17}^2x_{27}^2}}
=\pentabox_{125,3245}
\,,\nn\\
\fivedoublepenta_{12,345}&\equiv
x_{34}^2x_{35}^2x_{45}^2
\int\frac{x_{17}^2x_{26}^2\dd[4]{x_6}\dd[4]{x_7}}
{\brk{x_{16}^2x_{36}^2x_{46}^2x_{56}^2}x_{67}^2\brk{x_{37}^2x_{47}^2x_{57}^2x_{27}^2}}
=\doublepenta_{3,254,145}
\,,\nn\\
\doublebox_{123,456}&\equiv
x_{13}^2x_{46}^2x_{25}^2
\int\frac{\dd[4]{x_7}\dd[4]{x_8}}{\brk{x_{17}^2x_{27}^2x_{37}^2}x_{78}^2\brk{x_{48}^2x_{58}^2x_{68}^2}}
\,,\nn\\
\pentabox_{1,23,456}&\equiv
x_{46}^2x_{13}^2x_{15}^2
\int\frac{x_{28}^2\dd[4]{x_7}\dd[4]{x_8}}{\brk{x_{27}^2x_{37}^2x_{17}^2}x_{78}^2\brk{x_{18}^2x_{48}^2x_{58}^2x_{68}^2}}
=\pentabox_{231,4561}
\,,\nn\\
\doublepenta_{14,23,56}&\equiv
x_{13}^2x_{14}^2x_{46}^2
\int\frac{x_{28}^2x_{57}^2\dd[4]{x_7}\dd[4]{x_8}}{\brk{x_{17}^2x_{27}^2x_{37}^2x_{47}^2}x_{78}^2\brk{x_{48}^2x_{58}^2x_{68}^2x_{18}^2}}
=x_{14}^2x_{46}^2\brk[s]*{\frac{\doublepenta_{1,234,567}}{x_{16}^2x_{47}^2}}_{7\to4}
\,,\nn\\
\pentabox_{123,4567}&\equiv
x_{27}^2x_{35}^2x_{46}^2
\int\frac{x_{19}^2\dd[4]{x_8}\dd[4]{x_9}}{\brk{x_{18}^2x_{28}^2x_{38}^2}x_{89}^2\brk{x_{49}^2x_{59}^2x_{69}^2x_{79}^2}}
\,,\nn\\
\doublepenta_{1,234,567}&\equiv
x_{13}^2x_{16}^2x_{47}^2
\int\frac{x_{29}^2x_{58}^2\dd[4]{x_8}\dd[4]{x_9}}{\brk{x_{18}^2x_{28}^2x_{38}^2x_{48}^2}x_{89}^2\brk{x_{19}^2x_{59}^2x_{69}^2x_{79}^2}}
\,.
\label{eq:integrals}
\end{align}
Here, $F_1$ and $F_2$ are the one-loop and two-loop ladder integrals,
$\doublebox$ are double-box integrals, $\pentabox$ are penta-box (pentaladder)
integrals, and $\doublepenta$ are double-penta integrals.
Additional conformal integrals that do appear in the ansätze,
but whose coefficients are set to zero in the actual functions
$f_{\mathbold{a}}^{(2)}$ are shown in \figref{fig:extra-integrals}.
The integrals that \emph{do} contribute to the functions
$f_{\mathbold{a}}^{(2)}$ can be characterized as follows: Either they
are products of one-loop box integrals $F_1$, or they have one
loop-loop propagator factor $1/x_{n+1,n+2}^2$ and at most one numerator factor
$x_{ij}^2$ per integration point $j\in\set{n+1,n+2}$. Conversely, the
integrals in \figref{fig:extra-integrals} that \emph{do not} contribute fall into three classes:
\begin{itemize}
%%%%%%%%%%%%%%%
\item Two-loop integrals with two or more numerator factors connecting
to the same integration point (first row),
%%%%%%%%%%%%%%%
\item Products of one-loop integrals that include numerator factors
(second row),
%%%%%%%%%%%%%%%
\item Two-loop integrals that include loop-loop numerator factors
$x_{n+1,n+2}^2$ (last two rows).
%%%%%%%%%%%%%%%
\end{itemize}
The fact that such integrals do not contribute is an observation for
which we do not have a direct derivation at this point. What we can
say is that excluding all these integrals completely removes all
Gram-relation ambiguity. In other words, there is no linear
combination of Gram relations that is free of these excluded
integrals. This means that once these integrals are excluded, the
ansatz becomes free of redundant parameters,
\ie all coefficients can be uniquely fixed by matching to data or any
other type of constraints. Moreover, and perhaps even more importantly
for future bootstraps, the sizes of the ansatz polynomials greatly
reduce by dropping these integrals, especially at higher points,
see \tabref{tab:reducedansatzsizes}.
\begin{table}
\centering
\begin{tabular}[t]{lr}
\toprule
$P_{23}^{(2)}$ & $59$ \\
$P_{5}^{(2)}$  & $60$ \\
\midrule
$P_{222}^{(2)}$ & $160$ \\
$P_{24}^{(2)}$  & $400$ \\
$P_{33}^{(2)}$  & $117$ \\
$P_{6}^{(2)}$   & $465$ \\
\bottomrule
\end{tabular}
\qquad
\begin{tabular}[t]{lr}
\toprule
$P_{223}^{(2)}$ &  $974$ \\
$P_{25}^{(2)}$  & $1850$ \\
$P_{34}^{(2)}$  &  $974$ \\
$P_{7}^{(2)}$   & $2457$ \\
\bottomrule
\end{tabular}
\qquad
\begin{tabular}[t]{lr}
\toprule
$P_{2222}^{(2)}$ & $  724$  \\
$P_{224}^{(2)}$ &  $ 3666$ \\
$P_{233}^{(2)}$ &  $ 1886$ \\
$P_{26}^{(2)}$ &   $ 7781$ \\
$P_{35}^{(2)}$ &   $ 3500$ \\
$P_{44}^{(2)}$ &   $ 1919$ \\
$P_{8}^{(2)}$ &    $10793$ \\
\bottomrule
\end{tabular}
\qquad
\begin{tabular}[t]{lr}
\toprule
$P_{2223}^{(2)}$ &  $ 3948$ \\
$P_{225}^{(2)}$  & $ 10370$ \\
$P_{234}^{(2)}$  & $ 11138$ \\
$P_{27}^{(2)}$   & $ 25812$ \\
$P_{333}^{(2)}$  &  $ 1116$ \\
$P_{36}^{(2)}$   & $ 11996$ \\
$P_{45}^{(2)}$   & $ 10370$ \\
$P_{9}^{(2)}$    & $ 37083$ \\
\bottomrule
\end{tabular}
\caption{Numbers of independent coefficients in the two-loop ansatz
polynomials $P_{\mathbold{a}}^{(2)}$ after excluding classes of
integrals that do not appear for $n=5,6,7$. As can be seen by
comparing to~\protect\tabref{tab:ansatzsizes}, the numbers are greatly
reduced, especially for $n=8,9$.}
\label{tab:reducedansatzsizes}
\end{table}
%

%%%%%%%%%%%%%%%%%%%%%%%%%%%%%%%%%%%%%%%%%%%%%%%%%%%%%%%%%%%%
\subsection{Two-Loop Results}
\label{sec:resultstwoloops}

%%%%%%%%%%%%%%%%%%%%%%%%%%%%%%
\paragraph{Five Points.}

We can express the final answers for the two-loop component functions
$f_\mathbold{a}^{(2)}$ in terms of the conformal
integrals~\eqref{eq:integrals}. Pulling out the overall
prefactor~\eqref{eq:C222fromtwistors} from the twistor result,
we find
\begin{align}
f_{23}^{(2)}&=
12\,\Big\langle
2 \brk*{1-\frac{u_4}{u_3u_5}} \fivedoublebox_{1,23,45}
+ 4 u_1 u_3 \brk{\fivedoublebox_{3,14,25} - \fivedoublebox_{3,12,45}}
- \threedoublebox_{123}
\nn\\ & \mspace{60mu}
+ 2 \fivepentabox_{3,45,12}
- 2 u_1 u_4 \fivepentabox_{3,14,25}
+ \brk{1 - 2 \frac{1}{u_1 u_4}} \fivepentabox_{3,12,45}
+ \fivedoublepenta_{34,125}
\nn\\ & \mspace{60mu}
+ \frac{u_1\brk{2+u_1u_3-u_2u_4}}{u_2^2u_5} F_1^{1234} F_1^{1235}
+ 2 F_2^{1324}
+ \frac{u_1 u_3}{u_2} \brk{6 F_2^{1234} - F_2^{3142}}
\Big\rangle_{\mspace{-5mu}23}
\,,
\label{eq:f23}
\end{align}
where $\avg{\cdot}_{23}$ means the average over permutations in the
symmetry group $K_{23}\equiv\grp{S}_2\times\grp{S}_3$ of the respective propagator
structure. The second (``cyclic'') component function is given by
\begin{align}
f_{5}^{(2)}&=
10\,\Big\langle
\brk2{4 + \frac{2 u_2 - 4 - 1/u_5}{u_1 u_3}} \fivedoublebox_{1,23,45}
+ \brk2{\frac{4 - 2 u_4 - 1/u_1}{u_3 u_5} - 2} \fivedoublebox_{1,24,35}
\nn\\ & \mspace{60mu}
+ 2\brk*{\frac{1}{u_1} - u_3 u_5} \fivedoublebox_{1,25,34}
+ 2\brk*{1 + \frac{1}{u_2 u_4}} \fivepentabox_{1,23,45}
+ 2\brk*{u_3 -1 - u_2 u_4} \fivepentabox_{1,24,35}
\nn\\ & \mspace{60mu}
+ \frac{2\brk{1-u_2 u_4-u_3}}{u_3}  \fivepentabox_{1,25,34}
+ \frac{1+u_2 u_4-u_3}{u_3} \fivepentabox_{1,34,25}
- \fivedoublepenta_{12,345}
\nn\\ & \mspace{60mu}
+ \frac{1}{u_2}\brk2{\frac{u_1 u_4 - 2 + 1/u_2}{u_5}-2u_2 - 2 u_1 u_3 + 2} F_1^{1234} F_1^{1235}
+ \frac{4\brk{1-u_2}}{u_2} F_2^{1234}
\nn\\ & \mspace{60mu}
+ \brk*{u_2 + u_1 u_3 -1} F_2^{1243}
- 2\brk*{u_1 + 2 u_2 u_5} F_2^{1253}
- 2 u_1 u_4 F_2^{1254}
+ \threedoublebox_{123}
\Big\rangle_5
\,,
\label{eq:f5}
\end{align}
where $\avg{\cdot}_5$ means averaging over the symmetry group $K_5\equiv{\grp{D}_5}$ of
dihedral permutations. In both expressions, we have used the five-point cross ratios
defined in~\eqref{eq:ui5}.

The above expressions are not unique due to the existence of Gram
relations among the terms in the ansätze for the component functions
$f_{\mathbold{a}}$. At five points and two loops, there is one Gram
determinant relation (see \tabref{tab:gramstatistics}). After reducing
over permutations of the integration points $\set{x_6,x_7}$, the
relation has $442$ terms. Further reducing each term over the
permutation symmetry group $K_{23}$ ($K_5$), the number of terms in
the relation reduces to $64$ ($66$). However, the relation unavoidably
includes a conformal integral that is not in the
list~\eqref{eq:integrals}, namely
\begin{equation}
\int\frac{\dd{x_6}\dd{x_7}x_{67}^2}{x_{16}^2x_{26}^2x_{36}^2x_{46}^2x_{56}^2\,x_{17}^2x_{27}^2x_{37}^2x_{47}^2x_{57}^2}
\,,
\end{equation}
which is the first integral in the third row of \figref{fig:extra-integrals}.
Excluding this integral makes the expressions above unique.

We find that the two component functions $f_{23}^{(2)}$ and $f_{5}^{(2)}$
can alternatively be written in terms of the following four monomials:
\begin{alignat}{2}
q_{1234567}^{1} &=x_{16}^2 x_{17}^2 x_{26}^2 x_{27}^2 x_{34}^2 x_{35}^2 x_{45}^2
\,, & \qquad
q_{1234567}^{2} &=x_{16}^2 x_{17}^2 x_{25}^2 x_{27}^2 x_{34}^2 x_{36}^2 x_{45}^2
\,, \nn \\
q_{1234567}^{3} &= x_{15}^2 x_{16}^2 x_{27}^4 x_{34}^2 x_{36}^2 x_{45}^2
\,, & \qquad
q_{1234567}^{4} &=x_{17}^4 x_{26}^4 x_{34}^2 x_{35}^2 x_{45}^2
\,.
\end{alignat}
This compares with four different polynomials that appear in the
three-loop four point function of $\tp$ operators~\cite{Eden:2012tu}.
Multiplying with the common denominator as in~\eqref{eq:fform},
\begin{equation}
P_{\mathbold{a}}^{(2)}=
x_{16}^2x_{17}^2
x_{26}^2x_{27}^2
x_{36}^2x_{37}^2
x_{46}^2x_{47}^2
x_{56}^2x_{57}^2
x_{67}^2
\,f_{\mathbold{a}}^{(2)}
\,,
\end{equation}
we find that the coefficient of the $2\times 3$ polarization
$d_{12}^2d_{34}d_{35}d_{45}$ is given by
\begin{align}
P_{23}&= \big[
 2q_{1345672}^{4}
- q_{1345267}^{3}
-2q_{1345627}^{1}
+ q_{1346725}^{2}
+3q_{1354627}^{2}
+ q_{1356427}^{2}
-2q_{1364257}^{3}
\nn \\ & \mspace{50mu}
+4q_{1364527}^{2}
+2q_{1647325}^{2}
+2q_{3164572}^{3}
-4q_{3412567}^{1}
+2q_{3412567}^{4}
-8q_{3412756}^{1}
\nn \\ & \mspace{50mu}
+4q_{3412756}^{4}
-2q_{3612547}^{1}
-2q_{1436275}^{3}
\big]
+\text{perm} \,,
\end{align}
while the coefficient of the cyclic polarization
$d_{12}d_{23}d_{34}d_{45}d_{15}$ reads
\begin{align}
P_{5} &= \big[
 2q_{1234567}^{1}
- q_{1234567}^{2}
+ q_{1234567}^{3}
- q_{1234567}^{4}
+ q_{1234657}^{1}
-2q_{1234657}^{4}
+ q_{1234675}^{3}
+ q_{1235467}^{3}
\nn \\ & \mspace{50mu}
-2q_{1243567}^{2}
- q_{1243567}^{3}
-2q_{1243657}^{2}
+4q_{1246357}^{2}
- q_{1263547}^{2}
+2q_{1324657}^{1}
- q_{1324657}^{4}
\nn \\ & \mspace{50mu}
+2q_{1324675}^{3}
- q_{1342567}^{2}
+ q_{1345267}^{2}
-2q_{1346257}^{2}
+ q_{1362475}^{3}
- q_{1362547}^{2}
- q_{1362574}^{3}
+ q_{1423657}^{4}
\nn \\ & \mspace{50mu}
-2q_{1423675}^{3}
+ q_{1634257}^{2}
+ q_{1634527}^{1}
- q_{1635247}^{2}
-2q_{1643257}^{2}
+ q_{2135467}^{3}
+2q_{2136457}^{3}
+ q_{2143567}^{3}
\nn \\ & \mspace{50mu}
-2q_{2146357}^{3}
+2q_{2314657}^{1}
- q_{2413657}^{4}
-2q_{1236457}^{2}
- q_{1326457}^{2}
-2q_{1423756}^{1}
\big]
+\text{perm} \,.
\end{align}
The permutations in these equations are given by the elements of
$\grp{S}_{7}$ that leave each polarization structure fixed.

%%%%%%%%%%%%%%%%%%%%%%%%%%%%%%
\paragraph{Six Points.}

At six points, there are four independent component functions, as can
be seen in~\eqref{eq:G6G7}. The simplest of them reads
\begin{align}
f_{222}^{(2)}&=
48\,\Big\langle
\frac{u_1u_3u_5}{U_1} B_{123,456}
+ \frac{u_1^2u_3u_5U_1}{2u_2u_6U_3} F_1^{1234} F_1^{1256}
\Big\rangle_{222}
\,,
\label{eq:f222}
\end{align}
where $\avg{\cdot}_{222}$ means the average over permutations in the
symmetry group $K_{222}=\grp{S}_3\ltimes\brk{\grp{S}_2\times{\grp{S}_2}\times{\grp{S}_2}}$
that stabilizes the propagator structure, \ie the
pairings $(1,2)$, $(3,4)$, $(5,6)$ of external points. The remaining six-point
component functions are presented in \appref{app:results}. We express
the functions in terms of the six-point cross ratios~\eqref{eq:ui6}.

%%%%%%%%%%%%%%%%%%%%%%%%%%%%%%
\paragraph{Seven Points.}

At seven points, there are again four independent component
functions~\eqref{eq:G6G7}, the simplest of them being
\begin{align}
f_{223}^{(2)}&=
48\,\brkleft[a]!{
2 u_{1234} u_{2456} B_{123,456}
+ u_{1256} u_{2643} B_{125,346}
}
\nn\\ & \mspace{80mu}
\brkright[a]!{
+ u_{1243} u_{1256} u_{1562} F_1^{1234} F_1^{1256}
+ u_{1275} u_{3465} F_1^{1257} F_1^{3456}
}_{223}
\,.
\label{eq:f223}
\end{align}
Here, we use the general cross ratios~\eqref{eq:uijkl}. As
before, $\avg{\cdot}_{223}$ means averaging over the permutation group
$K_{223}=\brk{\grp{S}_2\ltimes\brk{\grp{S}_2\times{\grp{S}_2}}}\times{\grp{S}_3}$ that
stabilizes the propagator structure $d_{12}^2 d_{34}^2
d_{56}d_{67}d_{57}$ which multiplies the function $f_{223}$. The
remaining three seven-point component functions are given in \appref{app:results}.

%%%%%%%%%%%%%%%%%%%%%%%%%%%%%%
\paragraph{Any Multiplicity.}

We present some (very preliminary and yet incomplete) guesses for the
component functions $f_{2,n-2}^{(2)}$ and $f_n^{(2)}$ for any
multiplicity $n$ in \appref{sec:npointguesses}.

%%%%%%%%%%%%%%%%%%%%%%%%%%%%%%%%%%%%%%%%%%%%%%%%%%%%%%%%%%%%
\subsection{Three-Loop Results}
\label{sec:three-loop-results}

The ansätze for $f_{23}^{(3)}$ and $f_{5}^{(3)}$
contain $139$ different conformal integrals.%
\footnote{Strictly speaking, we should not call the individual terms
in three-loop ansatz conformal integrals, since we have not checked
their convergence. Of course the full three-loop correlator should be
free of divergences.}
Each integral appears
with various permutations of the external (and internal) points. We
can organize the answers as follows:
\begin{equation}
f_{23}^{(3)}=
9\avg2{
\sum_{i=1}^{139}
\sum_{\sigma\in\pi^{23}_i}
c_{23,i}^{\sigma}
\,I^{(3)}_{i,\sigma}
}_{\mspace{-6mu}23}
\,,\qquad
f_{5}^{(3)}=
30\avg2{
\sum_{i=1}^{139}
\sum_{\sigma\in\pi^{5}_i}
c_{5,i}^{\sigma}
\,I^{(3)}_{i,\sigma}
}_{\mspace{-6mu}5}
\,.
\label{eq:3loopanswer}
\end{equation}
Here, $I_{i,\sigma}^{(3)}$ are three-loop conformal integrals
\begin{equation}
I_{i,\sigma}^{(3)}=
I_i^{(3)}(x_{\sigma_1},x_{\sigma_2},x_{\sigma_3},x_{\sigma_4},x_{\sigma_5})
\,,
\end{equation}
the occurring permutations $\sigma$ of each integral $I_i$ are collected in
the sets $\pi^{23}_i$ and $\pi^{5}_i$, and the coefficients $c_{\mathbold{a},i}^{\sigma}$ are rational
functions of the five conformal cross ratios~\eqref{eq:ui5}. As before, $\avg{\cdot}_{23}$
and $\avg{\cdot}_{5}$ denotes averaging over the permutation
symmetry groups $\grp{S}_2\times\grp{S}_3$ and $\grp{D}_5$
respectively. The sets $\pi^{23}_i$ and $\pi^{5}_i$ contain between zero and eleven
permutations with non-vanishing $c_{\mathbold{a},i}^{\sigma}$. The first~$18$
integrals $\setcond{I_i^{(3)}}{1\leq{i}\leq18}$ are products of
one-loop and two-loop integrals,%
\footnote{Besides the one- and two-loop integrals that appear in the
two-loop ansätze (see \figref{fig:conformal-integrals}
and \figref{fig:extra-integrals}), these products include one
further two-loop integral, which features an external point that only
appears in the numerator, namely:
$\int{\dd{x_6}\dd{x_7}x_{56}^2}/\brk{x_{16}^2x_{26}^2x_{36}^2x_{46}^2x_{67}^2x_{17}^2x_{27}^2x_{37}^2}$.}
the remaining $\setcond{I_i^{(3)}}{19\leq{i}\leq139}$ are genuine three-loop integrals.
For example:
\begin{gather}
I_{65,(12345)}=\int\dd{x_6}\dd{x_7}\dd{x_8}
\frac{x_{12}^2 x_{13}^2 x_{15}^2 x_{23}^2 x_{47}^2}{x_{16}^2 x_{17}^2 x_{18}^2 x_{27}^2 x_{28}^2 x_{37}^2 x_{38}^2 x_{46}^2 x_{57}^2 x_{67}^2 x_{68}^2}
\,,\\[1.5ex]\nn
\begin{aligned}
c_{23,65}^{12345}&=-\frac{37}{2}+\frac{9}{2u_2 u_5}+\frac{7 u_1}{2u_2 u_5}
\,,&\qquad
c_{5,65}^{12345}&=\frac{2}{u_2 u_5}-\frac{2u_1}{u_2 u_5}
\,,\\
c_{23,65}^{13452}&=\frac{13}{2}
\,,&
c_{5,65}^{21345}&=-2 u_2 u_5
\,,\\
c_{23,65}^{31245}&=-\frac{15}{2}
\,,&
c_{5,65}^{31245}&=2+\frac{2}{u_1}-\frac{2u_2 u_5}{u_1}
\,,\\
c_{23,65}^{31452}&=\frac{3}{2}+\frac{1}{2u_2}-\frac{u_1 u_3}{2u_2}
\,,&&
\end{aligned}
\end{gather}
with all other $c_{\mathbold{a},65}^{\sigma}=0$.
Here, we use the five-point conformal cross ratios defined in~\eqref{eq:ui5}.
All integral expressions $I_i^{(3)}$ and
rational coefficients $c_{\mathbold{a},i}^{\sigma}$ are provided in the attached file
\filename{results.m}.
The numbers of various terms in the answers~\eqref{eq:3loopanswer} are
as follows:%
\footnote{Of the $139$ integrals in the ansätze, $28$ do not occur in
either answer.}
\begin{equation}
\begin{tabular}{lrr}
\toprule
number of:                                                                          & $f^{(3)}_{23}$ & $f^{(3)}_{5}$ \\
\midrule
occuring integrals $I^{(3)}_i$ (out of the $139$)                                   & $107$          & $94$ \\
non-zero $c_{i,\sigma}$ (in total)                                                  & $388$          & $445$ \\
terms in expanded expression inside $\avg{\dots}_{\dots}$ in~\eqref{eq:3loopanswer} & $1320$         & $1223$ \\
terms in fully expanded expression in~\eqref{eq:3loopanswer}                        & $13840$        & $11725$ \\
\bottomrule
\end{tabular}
\end{equation}
The representation~\eqref{eq:3loopanswer} is not unique: Some linear
combinations of terms in the ansätze for $f_{23}^{(3)}$ and
$f_{5}^{(3)}$ vanish due to Gram determinant relations
(see \tabref{tab:gramstatistics}).%
\footnote{There might be further linear relations after integration,
which we do not take into account here.}
Hence not all free coefficients in the ansätze are independent.
Indeed, we
find that matching against the twistor data leaves $103$ parameters in
$f_{23}^{(3)}$ unfixed. Similarly, $91$ parameters in $f_{5}^{(3)}$
remain unfixed. We verified that this remaining freedom indeed amounts to
adding Gram relations.
We use this freedom to minimize the number of non-zero
$c_{\mathbold{a},i}^{\sigma}$ in the answer, and to make their rational coefficients
as simple as possible, by setting the remaining coefficients to
particular values (effectively adding terms that sum to zero).

%%%%%%%%%%%%%%%%%%%%%%%%%%%%%%%%%%%%%%%%%%%%%%%%%%%%%%%%%%%%
\subsection{Planarity}
\label{sec:planarity}

Our results show that correlation functions of $\tp$ operators
are free of higher-genus corrections at two loops up to seven points,
and at three loops up to five points. At what loop order will the
first higher-genus corrections show up? For higher-charge operators,
it is clear that higher-genus terms will appear at lower loop orders.
In fact, for sufficiently large charges, already the tree-level
correlator will have higher-genus contributions. Consistently, also
the one-loop corrections contain higher-genus
terms~\cite{Bargheer:2017nne}. These higher-genus terms at low loop
orders do not arise from non-planar Feynman integrals, which do not
exist at one and two loops. Rather, the higher-genus terms at larger
charges originate in non-trivial color factors from the larger number
of propagators that connect to each operator (whereas the color
factors of $\tp$ operators are only delta functions).

We do not know about a rigorous argument for the absence of non-planar
corrections at three loops at higher points.
However, one can argue that for correlators of $\tp$ operators,
the loop order at which non-planar terms start to appear will be
independent of the number of inserted operators: Since their color
structure is so simple, inserting more $\tp$ operators at a
given loop order cannot increase the genus. In other words, any extra
handles in the large $\Nc$ expansion must be formed purely by loop
corrections. Since this does not happen at four points up to three loops,%
\footnote{For the four-point function, it was found that the potential
genus-one term at three loops is proportional to a conformal Gram
relation and thus vanishes~\cite{Eden:2012tu}.}
we find it reasonable to expect that the same will be true at higher
points, \ie that the perturbative order at which higher-genus terms
appear will be
independent of the number of inserted $\tp$ operators. This is
consistent with the fact that we do not observe higher-genus terms in
the three-loop five-point function, even though there exist non-planar
three-point integrals.

It would be interesting to verify any of these arguments
more rigorously.

%%%%%%%%%%%%%%%%%%%%%%%%%%%%%%%%%%%%%%%%%%%%%%%%%%%%%%%%%%%%
\subsection{Guide to the Results File}
\label{sec:guide}

All our results for the correlators of $\tp$ operators are
collected in the attached \mathematica file \filename{results.m}. The
file includes comments alongside every definition.
In the file, the two-loop integrals~\eqref{eq:integrals} are defined
as follows (see the definition of \lstinline!intDef!):
\begin{alignat}{3}
\text{\lstinline!F1[1,2,3,4]!}&=F_1^{1234} \,,&\qquad
\text{\lstinline!YY[1,2,3]!}&=\threedoublebox_{123} \nn\\
\text{\lstinline!F2[1,2,3,4]!}&=F_2^{1234} \,,&\qquad
\text{\lstinline!LL[1,2,3,4,5]!}&=\fivedoublebox_{1,23,45} \nn\\
&&
\text{\lstinline!BB[1,2,3,4,5,6]!}&=\doublebox_{123,456} \nn\\[1ex]
\text{\lstinline!PP[1,2,5,3,4]!}&=\fivepentabox_{1,25,34} \,,&\qquad
\text{\lstinline!QQ[1,2,3,4,5]!}&=\fivedoublepenta_{12,345} \nn\\
\text{\lstinline!PB[1,2,3,4,5,6]!}&=\pentabox_{1,23,456} \,,&\qquad
\text{\lstinline!DP[1,4,2,3,5,6]!}&=\doublepenta_{14,23,56} \nn\\
\text{\lstinline!PB7[1,2,3,4,5,6,7]!}&=\pentabox_{123,4567} \,,&\qquad
\text{\lstinline!DP7[1,2,3,4,5,6,7]!}&=\doublepenta_{1,234,567} \,,
\label{eq:integrals-file}
\end{alignat}
After loading the
file with \lstinline!<<"results.m"!, the results can be accessed
through the following symbols:
\begin{description}
%%%%%%%%%%%%%%%
\item[\lstinline!answer52A!:] The two-loop five-point
component functions $f_{\mathbold{a}}^{(2)}$. Here, the component
$\mathbold{a}$ is specified by choosing
$\mathbold{a}=\text{\lstinline!A!}\in\set{\text{\lstinline!23!},\text{\lstinline!5!}}$.
The answers are written in terms of conformal
integrals~\eqref{eq:integrals-file} and cross ratios~\eqref{eq:uijkl}. Each
term in the expressions is canonicalized over the respective
permutation group $K_{\mathbold{a}}$, that is the answers are identical
to the expressions inside $\avg{\cdot}$ in~\eqref{eq:f23}
and~\eqref{eq:f5} (including the numerical prefactors).
%%%%%%%%%%%%%%%
\item[\lstinline!answer62A!:] The same for six points, that is
$\text{\lstinline!A!}\in\set{\text{\lstinline!222!},\text{\lstinline!24!},\text{\lstinline!33!},\text{\lstinline!6!}}$.
The answers are identical to the expressions inside $\avg{\cdot}$
in~\eqref{eq:f222}, \eqref{eq:f24}, \eqref{eq:f33}, and \eqref{eq:f6}.
%%%%%%%%%%%%%%%
\item[\lstinline!answer72A!:] The same for seven points, that is
$\text{\lstinline!A!}\in\set{\text{\lstinline!223!},\text{\lstinline!25!},\text{\lstinline!34!},\text{\lstinline!7!}}$.
The answers are identical to the expressions inside $\avg{\cdot}$
in~\eqref{eq:f223}, \eqref{eq:f25}, \eqref{eq:f34}, and \eqref{eq:f7}.
%%%%%%%%%%%%%%%
\item[\lstinline!integrandN2A!:] The fully expanded two-loop integrand
of the component function $f_{\mathbold{a}}^{(2)}$, where all
integrands of conformal integrals as well as cross ratios are expanded
in terms of squared distances $x_{ij}^2=\text{\lstinline!x[i,j]!}$,
and the symmetrization (average) over the permutation group
$K_{\mathbold{a}}$ has been carried out. Here,
$\text{\lstinline!N!}\in\set{\text{\lstinline!5!},\text{\lstinline!6!},\text{\lstinline!7!}}$,
and \lstinline!A! as above. The expressions are not explicitly
symmetrized with respect to permutations of the integration points
$\set{x_{n+1},x_{n+2}}$. Calling
\begin{lstlisting}
integrandN2A // Map[symmetrizeInt[N,2]]
\end{lstlisting}
generates a manifestly symmetric expression (\ie the correct
tree-level correlator component of \lstinline!N! $\tp$ operators
and two Lagrangian operators).
%%%%%%%%%%%%%%%
\item[\lstinline!I3def[6,7,8]!:] A list of replacement rules that defines the
$139$ conformal integrals that contribute to the five-point three-loop
function.
%%%%%%%%%%%%%%%
\item[\lstinline!cIAdef!:] A list of replacement rules that defines the non-zero coefficients
$c_{\mathbold{a},i}^{\sigma}$ that enter the three-loop component function
$f_{\mathbold{a}}^{(3)}$ in~\eqref{eq:3loopanswer}, where
$\mathbold{a}=\text{\lstinline!A!}\in\set{\text{\lstinline!23!},\text{\lstinline!5!}}$.
The coefficients are expressed in terms of the five-point cross ratios~\eqref{eq:ui5}.
%%%%%%%%%%%%%%%
\item[\lstinline!answer53A!:] The expressions inside the brackets
$\avg{\cdot}$ in~\eqref{eq:3loopanswer}, including the numerical
prefactors, where
$\text{\lstinline!A!}\in\set{\text{\lstinline!23!},\text{\lstinline!5!}}$.
The coefficients are expressed in terms of cross
ratios~\eqref{eq:ui5}, the integrals $I_{i,\sigma}^{(3)}$ are left as
abstract symbols.
%%%%%%%%%%%%%%%
\item[\lstinline!answer53Ax!:] The same as \lstinline!answer53A!, but
with the integrals (or rather their integrands) as well as the cross
ratios expanded in terms of squared distances
$x_{ij}^2=\text{\lstinline!x[i,j]!}$.
%%%%%%%%%%%%%%%
\item[\lstinline!integrand53A!:] The integrands of the component
functions $f_{\mathbold{a}}^{(3)}$~\eqref{eq:3loopanswer},
$\mathbold{a}=\text{\lstinline!A!}\in\set{\text{\lstinline!23!},\text{\lstinline!5!}}$,
completely expanded in terms of squared distances, and with the
symmetrization (average) over the symmetry group $K_{\mathbold{a}}$
carried out. Not symmetrized over permutations of the integration
points $\set{x_6,x_7,x_8}$. As in the two-loop case,
\begin{lstlisting}
integrand53A // Map[symmetrizeInt[5,3]]
\end{lstlisting}
achieves that symmetrization.
%%%%%%%%%%%%%%%
\end{description}
All expressions \lstinline!answerN2A!,
$\text{\lstinline!N!}\in\set{5,6,7}$ are written in terms of the
general cross ratios~\eqref{eq:uijkl}. To convert all cross ratios in
the five- and six-point expressions to the basis cross
ratios~\eqref{eq:ui5} and~\eqref{eq:ui6}, one can use the following
code:
\begin{lstlisting}
{answer5223, answer525} /. u4Tox /. xToBGV5
{answer62222, answer6224, answer6233, answer626} /. u4Tox /. xToBGV6
\end{lstlisting}
%

%%%%%%%%%%%%%%%%%%%%%%%%%%%%%%%%%%%%%%%%%%%%%%%%%%%%%%%%%%%%
%%%%%%%%%%%%%%%%%%%%%%%%%%%%%%%%%%%%%%%%%%%%%%%%%%%%%%%%%%%%
\section{OPE Limit and Other Constraints}
\label{sec:OPElimit}

In \secref{sec:ansatz} we have described in detail the construction of
an ansatz for higher point ($n\ge 5$) correlation functions of twenty
prime operators at two and three loops. The undetermined coefficients
in the ansatz can be fixed, in principle, in two different ways. One
is based on a twistor reformulation of $\superN=4$ SYM action,
following the same strategy that was applied successfully
in~\cite{Fleury:2019ydf}. This was the approach of
\secref{sec:twistors}. The other is based on imposing OPE and supersymmetry
constraints on the ansatz. This is the direct generalization of the
four point bootstrap~\cite{Eden:2012tu} and it is the approach we
shall pursue in this section. The end result of this analysis is that
we are able to fix the two-loop five-point correlator, and we are able
to considerably reduce the number of undetermined coefficients of the
six-point correlator at two loops. One advantage of this approach is that it can
be applied with small differences to correlators of operators with
higher $R$-charges.

In a second part, we will do an OPE analysis of some of the correlation
functions just obtained, and thus provide new OPE data at two loops.
More concretely, we give the two-loop four-point function involving one
Konishi operator in the $[0,2,0]$ representation and three $\tp$ operators.

%%%%%%%%%%%%%%%%%%%%%%%%%%%%%%%%%%%%%%%%%%%%%%%%%%%%%%%%%%%%
\subsection{Fixing the Correlator at Two Loops }

In the following we will study the OPE limit of correlation functions
involving twenty prime operators and some Lagrangian insertions
\begin{align}
\langle \mathcal{O}(x_1,y_1) \dots \mathcal{O}(x_n,y_n) \mathcal{L}(x_{n+1})\dots \mathcal{L}(x_{n+k}) \rangle = \mathcal{G}_{n,k}(x_i,y_k).\label{eq:partiallyintegrated}
\end{align}
As mentioned in the previous sections, these correlators control the loop
corrections to the correlation functions of twenty prime operators. As
an example, the two-loop correlator $\mathcal{G}_{n,0}$, is given by
integrating the one-loop part of the correlator $\mathcal{G}_{n,1}$
over $x_{n+1}$. The advantage to do this is that it is easier to
analyze the analytic structure of the correlator and interpret it in
terms of OPE data. More concretely, the $\log u_1$ divergence (in the
$x_{12}^2\rightarrow 0$ limit)
of the five-point correlator, where $u_1$ as in~\eqref{eq:ui5},
is controlled by lower-loop information,
and thus can be used to fix the undetermined coefficients.

The ansatz of $\mathcal{G}_{5,1}$, $\mathcal{G}_{5,2}$, and
$\mathcal{G}_{6,1}$ up to one-loop order can be expressed in terms of a
combination of one-loop four-point ladder integrals, as can be seen
from the \appref{app:OneLoopIntegrals}. Moreover, these correlators
should vanish once we impose the chiral algebra
twist~\cite{Beem:2013sza},%
\footnote{The result of~\cite{Beem:2013sza} implies that the loop
corrections of $\mathcal{G}_{n,0}$ should vanish. It is possible to
argue that up to two loops the integrand of $\mathcal{G}_{n,0}$ should
also vanish in the chiral algebra twist limit, which is consistent with the twistor computation. This means that the Born level
$\mathcal{G}_{n,k}$ for $k\leq2$ vanishes in this twist limit.
The three loop integrand for five point twenty prime operators does
not vanish in the chiral algebra twist limit, however it does vanish once one
integration is performed. For this reason,
we expect that the one-loop correlators should vanish. The chiral algebra twist has been
used to bootstrap higher point functions
in~\cite{Goncalves:2019znr,Alday:2022lkk}.}
\ie when the positions $x_1,\dots,x_n$ are placed in a two-dimensional
plane, and the polarizations $y_1,\dots,y_n$ are set to
\begin{align}
y_{ij}^2 =(v_i-v_j)(z_i-z_j)
\,,
\end{align}
with $z_i$ being a two-dimensional complex coordinate for the
point $x_i$ (more explicitly
$x_{ij}^2 = (z_i-z_j)(\bar{z}_i-\bar{z}_j)$), and $v_i$ a generic
parameter. This is not the unique polarization for which the
correlator vanishes, another one is the so-called Drukker-Plefka
twist~\cite{Drukker:2009sf}
\begin{align}
y_{ij}^2=x_{ij}^2.
\end{align}
These last two constraints are powerful enough to completely fix the
tree-level correlators $\mathcal{G}_{5,1}$ and $\mathcal{G}_{6,1}$,
without imposing any other constraint from the OPE (\ie it fixes the
one-loop corrections to $\mathcal{G}_{5,0}$ and $\mathcal{G}_{6,0}$).
This is not the case for the one-loop corrections to these correlators,
as they are not completely fixed by these twists. Nonetheless, the
number of undetermined coefficients in the ansatz spelled out in
\secref{sec:ansatz} is greatly reduced, as can be seen from these
numbers:%
\footnote{We have used, in this section, a slightly improved ansatz
compared to the one described in~\eqref{eq:fform}. This ansatz excludes
polynomials that contain $(x_{ab}^2)^{k}$ for $k\ge \ell$ (where $a$
and $b$ are integration points). It is possible to show, using the
results of \appref{app:OneLoopIntegrals}, that the terms
excluded with this rule give products of one-loop four-point
conformal integrals that are already included through other
polynomials of the ansatz.}
\begin{align}
&\mathcal{G}_{5,1}^{(1)} \, : \, \mathbf{64+66=130} \rightarrow \mathbf{24} \\
&\mathcal{G}_{6,1}^{(1)}\, : \, \mathbf{235+572+173+657=1637} \rightarrow \mathbf{327}
\label{eq:sixpttwoloopTwist}
\end{align}

Another simple constraint to impose comes from studying the leading
term in the Lorentzian OPE between two $\tp$ operators. The
operators have $R$-charge, so it is better to choose wisely the
polarizations of the external operators to avoid degeneracies
in the OPE. One possible choice is setting the polarization to
\begin{align}
Y_1 = \frac{1}{\sqrt{2}}(1,i,\alpha_1,i\alpha_1,0,0)
\,,\qquad
Y_2 = \frac{1}{\sqrt{2}}(1,i,\alpha_2,-i\alpha_2,0,0)
\,,
\label{eq:SL2polarizations}
\end{align}
taking one derivative in each $\alpha_1$ and $\alpha_2$, and
subsequently setting both to zero.
The corresponding operators then become
\begin{equation}
\op{O}_1\to\tr(ZX)
\,,\quad
\op{O}_2\to\tr(Z\bar{X})
\,,\qquad\text{where}\qquad
Z=\frac{\phi_1+\phi_2}{\sqrt{2}}
\,,\quad
X=\frac{\phi_3+\phi_4}{\sqrt{2}}
\,.
\end{equation}
The effect of this is that we are projecting into a
channel where operators appearing at leading order in the OPE will
have the form
\begin{align}
\textrm{Tr}(Z D^{J}Z)+\dots
\,,\qquad
J\geq0
\,,
\label{eq:operatorstwisttwo}
\end{align}
where the dots represent different ways to distribute the
derivatives. For simplicity, we will focus first on $\mathcal{G}_{5,1}$,
and then generalize to higher points. The leading term in the OPE is
just the BPS operator $\textrm{Tr}(Z^2)$, and this reduces the number
of unfixed coefficients to
\begin{align}
&\mathcal{G}_{5,1}^{(1)} \, : \, \mathbf{24} \rightarrow \mathbf{19}
\,,
\end{align}
since the four-point function of twenty prime operators at two loops
is known and should match the leading term in the OPE limit of the
five-point function.

At subleading order in the OPE limit, the operators with $J\geq2$
start to contribute. These operators are unprotected, \ie their
dimension depends non-trivially on the coupling constant, and one
consequence is that they will give rise to $\log x_{12}^2$ terms in the
OPE limit. This new structure will allow to fix more undetermined
coefficients in the ansatz. The $\log$ terms come from the one-loop
conformal integrals reviewed in \appref{app:OneLoopIntegrals}. This is
one of the reasons why we decided to analyze the partially integrated
correlators~\eqref{eq:partiallyintegrated}, since it allows to extract
the $\log x_{12}^2$ divergences while dealing only with one-loop
integrals. More importantly the coefficient of $\log x_{12}^2$ is completely
determined by lower-loop data and the light-cone conformal blocks,%
\footnote{Here we have used the light-cone conformal blocks obtained
in~\cite{Bercini:2020msp} for higher-point functions. The formula
reads
\begin{align}
\langle \mathcal{O}(x_1)\dots \mathcal{O}(x_n) \rangle = \sum_{k}\frac{C_{12k}(x_{12}\cdot \partial_{z})^J}{(x_{12}^2)^{\Delta_{\phi}-(\Delta_k-J)/2}} \int_{0}^{1} [dt]  \langle \mathcal{O}_k(x_2+tx_{12},z) \mathcal{O}(x_3) \dots \mathcal{O}(x_n) \rangle+\dots
\end{align}
where $[dt] = \Gamma(\Delta+J){\dd{t}
(t(1-t))^{\brk{\Delta+J}/{2}-1}}/{\Gamma^2(\brk{\Delta+J}/{2})}$,
$\mathcal{O}_k(x,z)$ is a spin $J$ operator with polarization vector
$z$ satisfying $z^2=0$ and the $\dots$ represent subleading terms in
the light-cone limit $x_{12}^2\rightarrow 0$. The lower-loop data in this
case can be read off directly from the OPE of $\mathcal{G}_{5,1}^{(0)}$.}
reducing the number of unfixed coefficients further to
\begin{align}
&\mathcal{G}_{5,1}^{(1)} \, : \, \mathbf{19} \rightarrow \mathbf{6}
\,.
\end{align}
The remaining six constants can be fixed by looking at the singlet
$R$-charge channel, \ie $[0,0,0]$, in the $(12)$ OPE. Here we have
to be more cautious as there is degeneracy, \ie more than one operator
with the same spin and dimension contribute to the OPE. But luckily there is no degeneracy at
leading order in this limit, and so we can use the lower-loop data
from the Konishi operator and also the stress tensor. We could have
also used the correlator/amplitude duality to fix some of the above
coefficients.

One issue that prevents us to go to higher loops or points is that the
number of terms in the ansatz grows substantially. For this reason, it
is important to look at the ansatz and check if some of the terms
could be dropped based on some physical reasoning. This also has the
advantage that it can be applied to correlators with higher
$R$-charge. In the ansatz there are terms of the form
\begin{align}
\frac{y_{12}^2 y_{15}^2 y_{23}^2 y_{34}^2 y_{45}^2}{x_{34}^2 x_{23}^2 x_{15}^2 {\color{red}{x_{12}^2}} {\color{red}{x_{45}^2}}}
\frac{ {\color{red}{x_{12}^2}} {\color{red}{x_{45}^2}}  x_{12}^2 x_{45}^2 }{ x_{16}^2 x_{17}^2  x_{26}^2 x_{27}^2x_{46}^2 x_{47}^2 x_{56}^2 x_{57}^2}
\label{eq:dropterm1}
\end{align}
with two double propagators in the numerator. We have highlighted in
red terms that need to be generated by the interactions. One possible
reason to eliminate these diagrams is that they are not present in the
four-point function, and the interaction only involves four of
the five points. We have also noticed that the result does not contain
the following diagram:
\begin{align}
&\frac{y_{12}^2 y_{15}^2 y_{23}^2 y_{34}^2 y_{45}^2}{x_{12}^2 x_{15}^2 x_{23}^2 x_{34}^2 \color{red}{x_{45}^2}}
\frac{{\color{red}{x_{45}^2} } x_{16}^2 x_{24}^2 x_{25}^2 x_{37}^2 }{x_{17}^2  x_{26}^2 x_{27}^2 x_{36}^2 x_{46}^2 x_{47}^2 x_{56}^2 x_{57}^2 x_{67}^2}
\label{eq:dropterm2}
\end{align}
This term has five factors of $x_{ij}^2$ in the numerator, and one
possible reason to eliminate such terms is that the interactions would not
be able to generate so many terms in the numerator.%
\footnote{The only terms that can generate numerators in the action
are the ones coming from the field strength and from interactions with
fermions. Such terms however cannot generate five numerators at
two-loop order.}
Notice that both terms~\eqref{eq:dropterm1} and~\eqref{eq:dropterm2}
are not dropped by excluding the integrals in
\figref{fig:extra-integrals}, hence eliminating such terms reduces the
ansatz further even if such integrals
are excluded from the beginning.

The same strategy can be applied to the correlator
$\mathcal{G}_{6,1}^{(1)}$, or in other words to the two-loop six-point
function of twenty prime operators. This time the ansatz
of~\eqref{eq:fform} can produce diagrams with
six propagators in the numerator, which at this loop order should
not be possible from the interactions. Eliminating these diagrams (as well as the two double
propagators mentioned in the $\mathcal{G}_{5,1}^{(1)}$ analysis),
together with OPE constraints (leading $\log$ Lorentzian OPE and
leading Euclidean OPE in the $[0,2,0]$ channel) reduces the number of
undetermined coefficients~\eqref{eq:sixpttwoloopTwist} significantly:
\begin{align}
\mathcal{G}_{6,1}^{(1)}\, : \, \mathbf{327} \rightarrow \mathbf{27}.
\end{align}
Imposing the constraints coming from other OPE channels like the
singlet, did not fix the ansatz completely in this case.%
\footnote{In principle, we could use integrability data
\cite{Bercini:2022gvs,Bercini:2022jxo} for the non-$\log$ part of the
correlator to fix more coefficients of the ansatz.}
One use of this bootstrap exercise is to reduce both efficiently and
substantially the number of undetermined coefficients that enter the
approach of the previous section. This might be useful in the future
at higher loops/points, since generating data with the twistor method
becomes harder. Let us also point out that the bootstrap method can also be
applied directly to correlation functions of operators with other
$R$-charges.

%%%%%%%%%%%%%%%%%%%%%%%%%%%%%%%%%%%%%%%%%%%%%%%%%%%%%%%%%%%%
\subsection{OPE of the Integrated Correlator}

In the previous subsection we have analyzed the OPE of partially
integrated correlation functions (where a subset of the Lagrangian
insertions is not integrated over). While this has proved useful to
study part of the structure of the correlation function, in particular
to fix it, it does not provide all the information contained in this
observable. The goal of this subsection is to decompose the five-point
function at two-loop level in terms of lower-point correlators. This
will give access to new two-loop four-point functions with one non-BPS
operator. For simplicity, we will be working at leading and subleading
orders. We are able to extract the two-loop OPE coefficients with two
spinning operators and compare with a result that has been previously
computed~\cite{Bianchi:2021wre,Bianchi:2019jpy,Bianchi:2018zal}
(see~\cite{Bercini:2022gvs} for an integrability-based computation).

We will take the Lorentzian OPE limit as explained in the last subsection, and
subsequently take the two points to approach each other,
$x_2\to{x_1}$, keeping only the first
three nontrivial orders. This method has been applied for five-point
functions at weak and strong
coupling~\cite{Goncalves:2019znr,Bercini:2020msp,Bercini:2021jti}. One
of the technical difficulties is that the correlation function is
expressed in terms of conformal integrals that have not been computed
before. To overcome this problem (at least partially), we will apply the
method of asymptotic
expansion~\cite{Smirnov:1994tg,Eden:2012rr,Goncalves:2016vir,Georgoudis:2017meq}
to these higher-point integrals. The expansions of the integrals
together with the higher-point light-cone conformal
blocks~\cite{Goncalves:2019znr,Bercini:2020msp,Bercini:2021jti,Antunes:2021kmm}
allows us to read off the following OPE data:%
\footnote{Here we are being schematic. The goal of this equation is to
show the overall structure and what four-point functions can be read off.
We have suppressed the dependence on the OPE coefficients
and some space-time prefactors multiplying the four-point functions.
We also drop the dependence on the polarizations $y_3\dots y_5$ in
\eqref{eq:KonishifourPtFunction} since it gives only an overall
prefactor that does not depend on the coupling.}
\begin{multline}
\avg{\textrm{Tr}(ZX)(x_1) \textrm{Tr}(Z\bar{X})(x_2) \op{O}(x_3,y_3) \dots \op{O}(x_5,y_5)}
=x_{12}^{-2}\avg{\textrm{Tr}(Z^2) \op{O}(x_3,y_3) \dots \op{O}(x_5,y_5)}
\\
+(\text{descendant})
+ \avg{\textrm{Tr}(D^2Z^2) \op{O}(x_3,y_3) \dots \op{O}(x_5,y_5)}
+ \dots
\end{multline}
where the $\dots$ represent subleading operators and the correlator
involving the Konishi is given by
\begin{align}
\langle  \textrm{Tr}(D^2Z^2) \op{O}(x_3,y_3) \op{O}(x_4,y_4) \op{O}(x_5,y_5)  \rangle
=\frac{\left({x_{34}^2}/{x_{41}^2}\right)^{{\gamma_{\mathcal{K}}}/{2}}}{(x_{13}^2)^{2+\gamma_\mathcal{K}/2}(x_{45}^2)^2}\sum_{\ell=0}^{2} V_{1,34}^{2-\ell}V_{1,35}^{\ell} A^{\ell} (u_3,u_4)
\label{eq:KonishifourPtFunction}
\end{align}
where $V_{i,jk}= \brk{z_i\cdot x_{ij} x_{ik}^2-z_i\cdot x_{ik}
x_{ij}^2}/{x_{jk}}$ is a typical tensor structure~\cite{Costa:2011mg} (where
$z_i^\mu$ with $z_i^2=0$ is a spin polarization vector)
that usually appears in spinning
correlators, and
$\gamma_{\mathcal{K}}=12g^2-48g^4+\dots$ is the anomalous dimension of the Konishi
operator.
In the limit $x_2\to{x_1}$, the cross ratios $u_3$ and
$u_5$~\eqref{eq:ui5} become the usual four-point cross ratios.
The coefficients, $A^{\ell} $, in this decomposition are
finite functions of the cross ratios and have a perturbative expansion
in the coupling $g$ (recall that this information is contained in the
two-loop five-point function):
\begin{align}
A^{\ell} (u_3,u_4) &= \sum_{k=0}^{\infty} g^{2k} A^{\ell}_k (u_3,u_4)
\,,\\
A^{0}_2(u_3,u_4) &= \mathcal{I}_1 a_{1,0}+\mathcal{I}_2 a_{2,0}+\mathcal{I}_3 a_{3,0} +b_{1,0} (\Phi^{(1)})^2+b_{2,0} \Phi^{(1)}+\sum_{i=1}^3\sum_{j=0}^1c_{i,j,0}\partial_{u_{3+j}}\mathcal{I}_{i} +d_0
\,,\nn\\
A^{1}_2(u_3,u_4) &= \mathcal{I}_1 a_{1,1}+\mathcal{I}_2 a_{2,1}+\mathcal{I}_3 a_{3,1}+b_{1,1} (\Phi^{(1)})^2+b_{2,1} \Phi^{(1)} +\sum_{i=1}^3\sum_{j=0}^1c_{i,j,1}\partial_{u_{3+j}}\mathcal{I}_{i}+d_1
\,,\nn\\
A^{2}_2(u_3,u_4) &= \mathcal{I}_1 a_{1,2}+\mathcal{I}_2 a_{2,2}+\mathcal{I}_3 a_{3,2}+b_{1,1}(\Phi^{(1)})^2+b_{2,2} \Phi^{(1)} +\sum_{i=1}^3\sum_{j=0}^1c_{i,j,2}\partial_{u_{3+j}}\mathcal{I}_{i}+d_2
\,,\nn
\end{align}
where the integrals
$\mathcal{I}_i$ are just the two loop ladders defined
in~\eqref{eq:I123def}, and $\Phi^{(1)}=F_1^{2354}/(x_{24}^2x_{35}^2)$
is the one-loop box integral~\eqref{eq:integrals}.
Some of the coefficients $a_{i,j}$, $b_{i,k}$, $c_{i,j,k}$, and $d_i$ are
given in \appref{app:CoefficientsKonishi}, and the full four-point function is given in the
auxiliary file \filename{CorrelatorInLimit.m}.
We have omitted the terms with $k<2$ because
they can be read off from tree-level and one-loop five-point functions.

Our two-loop result for the correlator of a Konishi operator with
three $\tp$ operators extends the earlier one-loop
result~\cite{Bianchi:2001cm}. Note that the one-loop result can also
be obtained from integrability and the OPE~\cite{Fleury:2016ykk}.

%%%%%%%%%%%%%%%%%%%%%%%%%%%%%%%%%%%%%%%%%%%%%%%%%%%%%%%%%%%%
%%%%%%%%%%%%%%%%%%%%%%%%%%%%%%%%%%%%%%%%%%%%%%%%%%%%%%%%%%%%
\section{Discussion}
\label{sec:discussion}

We have used numerically the twistor reformulation of $\superN=4$
SYM to completely fix the two-loop five-, six- and seven-point
correlation functions and the three-loop five-point correlator of
twenty prime operators with arbitrary polarizations. It is possible to
generate numerical data for even higher-points and higher loops by
isolating sets of polarizations and keeping a reasonable number of
twistor diagrams. However the ansatz as described in
\secref{sec:ansatz} for those cases has still too many undetermined
coefficients for a fitting.

It is also possible to follow the approach of this paper for
correlators of higher-charge operators. One difficulty is the
increasing size of the ansatz, which however could be mitigated by
bootstrap methods similar to the ones we investigated.

We find that all two-loop as well as the five-point three-loop
correlator of $\tp$ operators are free of non-planar corrections. We
argued that this might remain true for any number of points, which
however remains to be verified.

Our results for the two-loop five-, six-, and seven-point
correlators of $\tp$ operators can be expressed in terms of a
restricted set of conformal integrals, see~\eqref{eq:integrals} and
\figref{fig:conformal-integrals}. All integrals that contribute have a
propagator connecting the two integration points, and at most one
numerator factor per integration point. Beyond seven points, there is
only one further integral of this type:%
\footnote{This integral has five numerators, which is not in
contradiction with the analysis of the previous section. Note that
this is just a conformal integral, and we decided to add prefactors
such that the weight in each point is zero.}
\begin{equation}
\doublepenta_{1234,5678}
\equiv
x_{23}^2x_{67}^2x_{48}^2
\int\frac{
\dd{x\subrm{a}}\dd{x\subrm{b}}x_{1 \mathrm{b}}^2x_{5 \mathrm{a}}^2
}{
x_{1\mathrm{a}}^2x_{2\mathrm{a}}^2x_{3\mathrm{a}}^2x_{4\mathrm{a}}^2
\,x_{\mathrm{a}\mathrm{b}}^2\,
x_{5\mathrm{b}}^2x_{6\mathrm{b}}^2x_{7\mathrm{b}}^2x_{8\mathrm{b}}^2
}
=\includegraphics[align=c]{FigIntEightDoublePenta}
\end{equation}
If the pattern of contributing integrals that we observe for $n\leq7$
continues to higher points, then all two-loop correlation functions of
$\tp$ operators should be expressible in terms of the integrals
in \figref{fig:conformal-integrals} and the above eight-point integral.
If true, this puts an extension of our results to eight points within
reach, since it substantially reduces the numbers of undetermined
coefficients in the ansatz (see \tabref{tab:reducedansatzsizes}).
Of course it is also possible that the absence of other conformal
integrals is a low-$n$ artifact and does not continue to higher
points. It would be nice to understand more systematically which kinds
of integrals can contribute at higher points (and at higher loops,
where the data is much more limited).

With the fresh new data obtained in this paper, we have made some
initial steps towards bootstrapping the integrand of correlation
functions with five or more half-BPS operators. An obvious next step
in this bootstrap game is to explore correlation functions of
operators with different $R$-charge. This is essentially an uncharted
territory and definitely deserves further analysis, specially because
one might wonder if there are hidden structures as there are for four
points~\cite{Coronado:2018cxj,Caron-Huot:2021usw}. The twistor
reformulation of $\superN=4$ SYM can also be very useful for this
generalization.

We have also studied the first non-trivial order in the light cone OPE
of a five point function and have obtained two loop four point
function involving one Konishi operator in the $[0,2,0]$ and three
$\tp$ operators. It would be interesting to further develop this
OPE analysis to subleading terms in the OPE, higher point functions
and higher loops. This would be very important to probe the recently
discovered dualities between three point functions and null polygon
hexagonal Wilson loops~\cite{Bercini:2020msp,Bercini:2021jti}. The
main obstacle to achieve this is the computation of five and six point
conformal integrals (which in this paper we have only computed in a
limit). We hope to make progress on this in the future.

The four point function of $\tp$-operators was important to
analyze many different physical limits, such as the Regge limit or event
shapes~\cite{Costa:2012cb,Costa:2013zra,Belitsky:2013bja}. It would be
very interesting to use our recent data and extensions of it to study
these physical observables with more points.

Another direction is the connection with integrability
\cite{Basso:2015zoa,Eden:2016xvg,Fleury:2016ykk}. We have not managed
to find a closed expression for the $n$-point correlation functions at
two-loop in this work. However, it is possible to organize the results
in terms of integrability contributions. At two-loop and for a
particular set of mirror cuts, the only contributions are strings and
loops of just one mirror particles. There are many relations between
these objects involving several particles with the same objects with
lower number of particles, such as decoupling and flipping. It is
expected that all the necessary integrability contributions can be
fixed. The set of integrals appearing for the correlation functions of
the twenty primes operators forms also a basis of integrals for the
correlators of other length-$k$ half-BPS operators. This follows
because increasing the bridge lengths kill possible diagrams. Thus
fitting the basis against the power series produced by integrability
in a line (the integrals are unknown outside the line) should give the
integrand for different correlators such as the dodecagon. The same
strings and loops can also be used to produce non-planar data.

In addition, notice that the twistor method for computing correlation functions was the
motivation for the proposal of the Correlahedron~\cite{Eden:2017fow},
which is a geometric object computing the
integrands of correlation functions of the stress-tensor multiplet.
One of the properties of this object is that it reduces to the
``squared'' Amplituhedron~\cite{Arkani-Hamed:2013jha} when the light-like
limit is taken.%
\footnote{The proposed ``squared'' Amplituhedron is also a
Grassmannian, but with fewer constraints~\cite{Eden:2017fow} than the
original Amplituhedron.}
The Correlahedron is a Grassmannian, and the external data are
points in chiral Minkowski space. However, there are still some open
questions about the proposal.
An important one is concerning the volume form. The volume form for
correlation functions can be more complicated, having different kinds
of singularities.
All the known results for
correlation function integrands in the literature were shown to have an uplift to the
Correlahedron language. This includes the four-point functions up to
ten loops~\cite{Eden:2012tu,Bourjaily:2016evz}, and the
six-point tree-level correlator mentioned above.
We hope that our new data can help to test the proposal further.

%%%%%%%%%%%%%%%%%%%%%%%%%%%%%%%%%%%%%%%%%%%%%%%%%%%%%%%%%%%%
\section*{Acknowledgments}

We would like to thank Frank Coronado, Paul Heslop, Raul Pereira,
Pedro Vieira for useful discussions. V.G. would like to thank Maria
Nocchi for reading carefully part of \appref{app:Integrals}.
This work was supported by the Serrapilheira Institute (grant number
Serra – R-2012-38185), and funded by the Deutsche
Forschungsgemeinschaft (DFG, German Research Foundation) -- 460391856.
T.F. would like to thank the warm hospitality of the KITP Santa Barbara during the program Integrable22 where
part of this work was done.
This research was supported in part by the National Science Foundation under Grant No. NSF PHY-1748958.
V.G. is supported by Simons Foundation grants
\#488637 (Simons collaboration on the non-perturbative bootstrap).
Centro de F\'{i}sica do Porto is partially funded by Funda\c{c}\~{a}o
para a Ci\^{e}ncia e Tecnologia (FCT) under the grant UID04650-FCUP.

\appendix

%%%%%%%%%%%%%%%%%%%%%%%%%%%%%%%%%%%%%%%%%%%%%%%%%%%%%%%%%%%%
%%%%%%%%%%%%%%%%%%%%%%%%%%%%%%%%%%%%%%%%%%%%%%%%%%%%%%%%%%%%
\section{Tree-Level and One-Loop \texorpdfstring{$n$}{n}-Point Functions}
\label{app:treeone}

This appendix is a review of the known results in the literature about tree-level and one-loop $n$-point functions of length-two half-BPS operators.
At tree level and with general polarizations $Y_i$, one has
\begin{equation}
G_n \big|_0  =  \left( d_{12}d_{23} \ldots d_{n-1,n}d_{n1} + \text{non-cyclic permutations} \right) + \text{disconnected}  \,,
\end{equation}
with propagators $d_{ij}$ as in~\eqref{escalarpropagators}.

At one-loop order, there are two ways of obtaining the results.
The first method is the perturbative calculation of
\cite{Drukker:2008pi}, and the second method uses integrability techniques \cite{Bargheer:2018jvq}.
The starting point of the integrability calculation is to consider
all tree-level diagrams. The perturbative corrections are obtained by adding
the so called mirror particle contributions.
The relevant tree-level diagrams for integrability are the connected ones.%
\footnote{The disconnected diagrams can in principle contribute to the integrability calculation because of the stratification procedure (which is a prescription for treating the boundary graphs of the moduli space)
in hexagonalization. However in \cite{Bargheer:2018jvq},
it was argued that these contributions vanish at one-loop order. At the moment, it is not known if they contribute to higher-loop correlators.}
Considering the sphere, the cyclic graphs divide it into two faces,
where each face forms a polygon with $2n$ edges for $n$ operators
(each operator has a small size). At one-loop order for length-two half-BPS operators, the mirror particles in different faces
do not interact, and the correlator is the product of the value of the
two polygons. At two-loop order, this factorization breaks down
and there are strings and loops of mirror particles connecting the two faces.
From integrability, one has
\begin{equation}
{\rm{polygon}}(1, \ldots, 2n)
=\quad\sum_{\mathclap{\substack{[i,i+1],[j,j+1]:\\\text{non-consecutive edges}}}}\quad
m\brk{z_{ij},\alpha_{ij}}
\,,
\label{polygons}
\end{equation}
where
\begin{equation}
m(z, \alpha) \equiv g^2 \frac{(z+ \bar{z})- (\alpha + \bar{\alpha})}{2} F^{(1)}(z, \bar{z}) \, ,
\label{integrabilitym}
\end{equation}
with the local cross ratios%
\footnote{Notice that the expression for the polygon is valid for any 4d kinematics.
All the local cross ratios $z_{ij}$ and $\bar{z}_{ij}$ depend only on the $n(n-3)/2$ cross ratios of the problem (which is
also the number of terms in the sum in~\eqref{polygons}).}
\begin{alignat}{2}
z_{ij}\bar{z}_{ij} &= \frac{x^2_{i,j+1}x^2_{i+1,j}}{x^2_{i,i+1}x^2_{j+1,j}}
\,,&\qquad
\brk{1-z_{ij}}\brk{1-\bar{z}_{ij}} &= \frac{x^2_{i,j}x^2_{i+1,j+1}}{x^2_{i,i+1}x^2_{j+1,j}}
\,,\nn\\
\alpha_{ij} \bar{\alpha}_{ij} &= \frac{y^2_{i,j+1}y^2_{i+1,j}}{y^2_{i,i+1}y^2_{j+1,j}}
\,,&\qquad
\brk{1-\alpha_{ij}} \brk{1-\bar{\alpha}_{ij}} &= \frac{y^2_{i,j}y^2_{i+1,j+1}}{y^2_{i,i+1}y^2_{j+1,j}}
\,,
\end{alignat}
and
\begin{equation}
F^{(1)}(z, \bar{z}) = \frac{1}{z-\bar{z}} \left( 2 \, {\rm{Li}}_2 (z) - 2 \, {\rm{Li}}_2 (\bar{z}) + {\rm{log}} (z \bar{z}) \, {\rm{log}} \left(\frac{1-z}{1-\bar{z}} \right) \right) \, ,
\label{Ffunction}
\end{equation}
which is the box integral~\eqref{eq:integrals}:
\begin{equation}
F^{(1)}\brk*{\frac{1}{1-z_{13}},\frac{1}{1-\bar{z}_{13}}}=\frac{1}{\pi^2}F_1^{1243}
\,,\qquad
F^{(1)}\brk*{z_{13},\bar{z}_{13}}=\frac{1}{\pi^2}F_1^{1432}
\,.
\end{equation}
The final result for the one-loop correlators is the sum of two of these polygon factors times the
tree-level propagators of each tree-level graph.
The calculation from integrability relies on the two-mirror-particle contribution obtained in \cite{Fleury:2017eph} and the flip
relations explored in \cite{Bargheer:2018jvq} that enable one to
iteratively obtain the contribution of ``strings'' of any number $n$ of interacting mirror particles. The result for the general polygon~\eqref{polygons} also follows from an inductive argument.
In order to compute the correlation functions of operators with length bigger than two, it is also possible to use the formula for the polygons.
Similarly, each tree-level graph divides the surface into faces or polygons. However, the number of tree-level graphs grows substantially with the length $k$ of the operators involved. At the moment, as far as we know there is no closed formula for general $k$ even for the simplest case of five operators.
The function \eqref{integrabilitym} depends on the $R$-charge cross-ratios $\alpha_{ij}$, and therefore can change the original polarization structure of a given tree-level graph. In addition, the function $F^{(1)}(z, \bar{z})$ of \eqref{Ffunction} satisfies several properties, and many simplifications are expected when all graphs are summed. Note that it is possible to generate several correlators with different $k$'s using the expression for the polygons.

%%%%%%%%%%%%%%%%%%%%%%%%%%%%%%%%%%%%%%%%%%%%%%%%%%%%%%%%%%%%
%%%%%%%%%%%%%%%%%%%%%%%%%%%%%%%%%%%%%%%%%%%%%%%%%%%%%%%%%%%%
\section{Ansatz Construction}
\label{app:graphs}

As explained in \secref{sec:ansatz}, by mapping each factor $x_{ij}^2$
to an edge between vertices $i$ and $j$, we can identify each monomial
in $x_{ij}^2$ with a multi-graph (\ie a graph that admits ``parallel''
edges between the same vertices~$i$ and~$j$). Finding the most general
polynomials $P_{\mathbold{a}}^{(\ell)}$ hence amounts to listing all
multi-graphs with $n$ external vertices with valency $\ell$, and
$\ell$ internal valencies with valency $n+\ell-5$, and taking a
general linear combination of the corresponding monomials.

We split the construction of the ansatz for each polynomial
$P_\mathbold{a}$ into three steps. First, we construct all admissible
\emph{unlabeled} graphs with $n+\ell$ vertices.%
\footnote{For $n=5$, all vertices have the same valency $\ell$, hence
we need to explicitly distinguish different partitionings of the
vertex set into external and internal vertices.}
Next, for each graph $g$, we construct a set of inequivalent labelings
of the external vertices. Each such labeling $\sigma$ will correspond
to one independent term in the ansatz that gets multiplied by an
undetermined coefficient $c_{g,\sigma}$. In order to find the minimal
set of inequivalent labelings, we make use of the
permutation symmetry $K_{\mathbold{a}}$ of the respective propagator
factor $\prod_{ij}d_{ij}^{a_{ij}}$. Due to the total $\grp{S}_n$ permutation
symmetry of the correlator, also the polynomial $P_{\mathbold{a}}$
must be invariant under the permutation group $K_{\mathbold{a}}$.
The set of inequivalent labelings
therefore is $K_{\mathbold{a}}\backslash\grp{S}_n/H_g$, where $H_g$ is
the automorphism group of the graph $g$.%
\footnote{To be precise, we do not care about $\grp{S}_\ell$
relabelings of the integration points, since we will symmetrize the ansatz over
$\grp{S}_\ell$ permutations in any case. Hence the relevant group is
$H_g=\aut(g)/\grp{S}_\ell|_n$, where $\aut(g)$ is the automorphism
group of $g$, and $|_n$ means restriction to the points
$\set{1,\dots,n}$. For $n\neq5$, this step is trivial, since internal
and external vertices have different valencies and thus
$\aut(g)\subset{\grp{S}_n}\times{\grp{S}_\ell}$.}
Finally, we symmetrize each labeled graph over the residual symmetry group
$K_{\mathbold{a}}\times{\grp{S}_\ell}$, where $\grp{S}_\ell$ permutes the
integration vertices. Putting everything together, we arrive at
\begin{equation}
P_{a}^{(\ell)}=
\sum_{\mathclap{g\in \Gamma_{n,\ell}}} \mspace{12mu}
\sum_{\sigma\in{K_{\mathbold{a}}\backslash{\grp{S}_n}/H_g}}
\mspace{-12mu} c_{g,\sigma} \mspace{5mu}
\sum_{\mathclap{\pi\in{K_{\mathbold{a}}}\times{\grp{S}_\ell}}}
g_{\pi\circ\sigma}
\,.
\end{equation}
Here, $\Gamma_{n,\ell}$ is the set of all unlabeled multi-graphs with $n$
vertices of valency $\ell$, and $\ell$ vertices of valency $n+\ell-5$.%
\footnote{In practice, the graph vertices are always labeled. However, the
initial labeling of $g\in{\Gamma_{n,\ell}}$ is arbitrary and not relevant.}
For each graph $g\in{\Gamma_{n,\ell}}$, we sum over the labelings
(permutations) $\sigma\in{K_a\backslash{\grp{S}_n}/H_g}$ of the $n$ external
points, where $H_g$ is the automorphism group of $g$, and
$K_{\mathbold{a}}$ is the symmetry group of the respective propagator
factor $\prod_{ij}d_{ij}^{a_{ij}}$. Each such labeling produces one
independent term, \ie comes with one independent coefficient
$c_{g,\sigma}$. At the end, each independent term is symmetrized over
permutations $K_{\mathbold{a}}\times{\grp{S}_\ell}$.
To find all the graphs as well as their inequivalent labelings
$\sigma\in{K_a\backslash{\grp{S}_n}/H_g}$, we use \sagemath~\cite{sagemath}, in
particular its interface to \GAP~\cite{GAP4}.

We find the following numbers of different
multi-graphs for various $n$ and $\ell$, where we distinguish internal
from external vertices, but otherwise treat all vertices
as identical (unlabeled):
\begin{equation}
\begin{tabular}{ccccccc@{\qquad}cc}
\toprule
$\abs{\Gamma_{5,2}}$ & $\abs{\Gamma_{6,2}}$ & $\abs{\Gamma_{7,2}}$ & $\abs{\Gamma_{8,2}}$ & $\abs{\Gamma_{9,2}}$ & $\abs{\Gamma_{10,2}}$ & $\abs{\Gamma_{11,2}}$ & $\abs{\Gamma_{5,3}}$ & $\abs{\Gamma_{6,3}}$ \\
\midrule
$15$                 & $41$                 & $85$                 & $178$                & $327$                & $607$                 & $1051$                 & $429$                & $4105$ \\
\bottomrule
\end{tabular}
\label{eq:graphnumsunlabeled}
\end{equation}
The residual permutation symmetry groups are as follows:
\begin{align}
K_{23} &=\grp{S}_2\times{\grp{S}_3}\,, & K_{222} &=\grp{S}_3\ltimes\brk{\grp{S}_2\times{\grp{S}_2}\times{\grp{S}_2}}\,, & K_{223} &=\brk{\grp{S}_2\ltimes\brk{\grp{S}_2\times{\grp{S}_2}}}\times{\grp{S}_3}\,, \nn\\
K_{5}  &=\grp{D}_5\,,                  & K_{24}  &=\grp{S}_2\times{\grp{D}_4}\,,                                        & K_{25}  &=\grp{S}_2\times{\grp{D}_5}\,, \nn\\
       &                               & K_{33}  &=\grp{S}_2\ltimes\brk{\grp{S}_3\times{\grp{S}_3}}\,,                  & K_{34}  &=\grp{S}_3\times{\grp{D}_4}\,, \nn\\
       &                               & K_{6}   &=\grp{D}_6\,,                                                         & K_{7}   &=\grp{D}_7\,,
\label{eq:Kgroups}
\end{align}
where $\grp{D}_k$ is the dihedral group on $k$ elements. According to the
above procedure, the total number $\aleph_{\mathbold{a}}^{(\ell)}$ of
independent terms (undetermined coefficients) in the ansatz for
$P_{\mathbold{a}}^{(\ell)}$ is
\begin{equation}
\aleph_{\mathbold{a}}^{(\ell)}=
\sum_{\mathclap{g\in \Gamma_{n,\ell}}} \mspace{12mu}
\abs{K_{\mathbold{a}}\backslash{\grp{S}_n}/H_g}
\,.
\end{equation}
These are the numbers shown in \tabref{tab:ansatzsizes}.

%%%%%%%%%%%%%%%%%%%%%%%%%%%%%%%%%%%%%%%%%%%%%%%%%%%%%%%%%%%%
%%%%%%%%%%%%%%%%%%%%%%%%%%%%%%%%%%%%%%%%%%%%%%%%%%%%%%%%%%%%
\section{Kinematics}
\label{sec:kinematics}

We define the general conformally invariant cross ratios:
\begin{equation}
u_{ijkl}=\frac{x_{ij}^2x_{kl}^2}{x_{ik}^2x_{jl}^2}
\,,\qquad
v_{ijkl}=\frac{x_{il}^2x_{jk}^2}{x_{ik}^2x_{jl}^2}
\,.
\label{eq:uijkl}
\end{equation}
A collection of $n$ points in four dimensions has $4n-15$ conformally
invariant degrees of freedom, and therefore as many independent cross
ratios.

%%%%%%%%%%%%%%%%%%%%%%%%%%%%%%
\paragraph{Five Points.}

At $n=5$, there are five independent cross ratios. A convenient choice is
\begin{equation}
u_i=u_{i,i+1,i+2,i+4}
\,,\qquad
1\leq{i}\leq5
\,,
\label{eq:ui5}
\end{equation}
where the point labels are understood modulo $5$.
These are the same cross ratios used in~\cite{Bercini:2020msp}.
One can express any
conformally invariant combination of distances $x_{ij}$ in terms of
the $u_i$ by comparing expressions in a fixed conformal frame. For
example, one can set $x_5=\infty$ and $x_{12}^2=1$. The
relations~\eqref{eq:ui5} then imply
\begin{equation}
x_{13}^2=\frac{1}{u_1}
\,,\quad
x_{14}^2=\frac{1}{u_1u_4}
\,,\quad
x_{23}^2=\frac{u_2u_5}{u_1}
\,,\quad
x_{24}^2=\frac{u_5}{u_1u_4}
\,,\quad
x_{34}^2=\frac{u_3u_5}{u_1u_4}
\,.
\end{equation}
%

%%%%%%%%%%%%%%%%%%%%%%%%%%%%%%
\paragraph{Six Points.}

At six points, there are nine independent cross ratios. As a basis, we
choose
\begin{equation}
u_i=u_{i,i+1,i+2,i+4}
\,,\quad
1\leq{i}\leq6
\qquad \text{and} \qquad
U_i=u_{i,i+2,i+3,i+5}
\,,\quad
1\leq{i}\leq3
\,,
\label{eq:ui6}
\end{equation}
where the point labels are understood modulo $6$.
These are the same cross ratios used in~\cite{Bercini:2021jti} (see
Figure~5 there). Again, one can go to a conformal frame where
$x_6=\infty$ and $x_{12}^2=1$, which fixes all remaining distances
$x_{ij}^2$, $1\leq{i,j}\leq5$ in terms of the $u_i$ and $U_i$.

%%%%%%%%%%%%%%%%%%%%%%%%%%%%%%
\paragraph{Seven Points.}

At seven points, one can pick a basis of $14$ multiplicatively
independent cross ratios. They will not be completely functionally
independent, because seven points in four dimensions have only $13$
conformally invariant degrees of freedom. The $14$ multiplicatively
independent cross ratios reduce to $13$ degrees of freedom via a
conformal Gram relation.%
\footnote{The seven-point conformal Gram relation takes the form
$\det_{i,j}(x_{ij}^2)=0$, see the discussion below~\eqref{eq:Xmatrix}.}
Nonetheless, any ratio of $x_{ij}^2$ can be uniquely written as a
ratio of $14$ multiplicatively independent cross ratios. One
``nice'' basis of $14$ cross ratios appears to be:
\begin{equation}
\setcond{u_{12j7}}{3\leq{j}\leq6}
\cup
\setcond{u_{1i7j}}{2\leq{i}<j\leq6}
\,.
\label{eq:ui7}
\end{equation}
Another potentially useful basis is (this is closer to the six-point
set~\eqref{eq:ui6}):
\begin{equation}
u_i=u_{i,i+1,i+2,i+4}
\,,\quad
1\leq{i}\leq7
\qquad \text{and} \qquad
U_i=u_{i,i+2,i+3,i+6}
\,,\quad
1\leq{i}\leq7
\,,
\label{eq:ui72}
\end{equation}
Again, one can go to a conformal frame where
$x_7=\infty$ and $x_{12}^2=1$, which fixes all remaining distances
$x_{ij}^2$, $1\leq{i,j}\leq6$ in terms of either of the two sets.

%%%%%%%%%%%%%%%%%%%%%%%%%%%%%%%%%%%%%%%%%%%%%%%%%%%%%%%%%%%%
%%%%%%%%%%%%%%%%%%%%%%%%%%%%%%%%%%%%%%%%%%%%%%%%%%%%%%%%%%%%
\section{Explicit Correlator Expressions}
\label{app:results}

%%%%%%%%%%%%%%%%%%%%%%%%%%%%%%%%%%%%%%%%%%%%%%%%%%%%%%%%%%%%
\subsection{Correlator Components}
\label{sec:components}

%%%%%%%%%%%%%%%%%%%%%%%%%%%%%%
\paragraph{Six Points.}

Besides~\eqref{eq:f222}, the remaining six-point two-loop component
functions are quoted in the following. The expressions are also
included in the attached file \filename{results.m}.
\begin{align}
f_{24}^{(2)}&=4\brkleft[a]!{
- 8 B_{123,456} u_{1234} u_{2456}
+   B_{134,256} \brk{2 u_{1246} + 6 u_{1243} u_{2365} - 2 u_{1245} u_{2563}}}
\nn\\ & \mspace{40mu}
+ 8 \pentabox_{3,45,126} u_{1263}
- 8 \pentabox_{3,54,126} u_{1263}
+   \pentabox_{1,34,256} \brk{-8 u_{1246} + 8 u_{1245} u_{1652}}
\nn\\ & \mspace{40mu}
+ \doublepenta_{12,34,65} \brk{-2 u_{1245} + 4 u_{1542}}
+ \doublepenta_{12,34,56} \brk{2 u_{1246} - 4 u_{1642}}
\nn\\ & \mspace{40mu}
+ F_1^{1236} F_1^{1245} \brkleft{
- 4 u_{1243} u_{1462}
+ 4 u_{1253} u_{1462}}
\nn\\ & \mspace{160mu}
\brkright{\mathord{}+ u_{1234} u_{1263} u_{1356} + 3 u_{1236} u_{1254} u_{1362} - u_{1246} u_{1253} u_{1462}}
\nn\\ & \mspace{40mu}
+ F_1^{1234} F_1^{3456} \brk{-4 u_{1243} + 4 u_{1243} u_{3564} - 4 u_{1243} u_{3456} u_{3564}}
\nn\\ & \mspace{40mu}
-  8 \fivedoublebox_{3,12,45} u_{1234}
-  8 \fivedoublebox_{3,12,46} u_{1234}
+ 16 \fivedoublebox_{3,14,26} u_{1234}
+    \fivedoublebox_{1,23,45} \brk{-8 u_{1354} + 8 u_{1453}}
\nn\\ & \mspace{40mu}
+ 8 \fivepentabox_{3,45,12}
- 8 \fivepentabox_{3,14,25} u_{1245}
+   \fivepentabox_{3,12,45} \brk{4 - 8 u_{1425}}
+ 4 \fivedoublepenta_{35124}
\nn\\ & \mspace{40mu}
+ F_1^{1234} F_1^{1236} \brk{4 u_{1243} u_{1263} + 8 u_{1263} u_{1342} - 4 u_{1246} u_{1362}}
+ 24 F_2^{1234} u_{1243}
\nn\\ & \mspace{40mu}
\brkright[a]!{\mathord{}
- 4 F_2^{3142} u_{1243}
+ F_2^{1324} \brk{4 + 2 u_{1243}}
+ F_2^{1325} \brk{4 - 2 u_{1253}}
- 4 \threedoublebox_{123}}_{24}
\,,
\label{eq:f24}
\end{align}
\begin{align}
f_{33}^{(2)}&=36\brkleft[a]!{
B_{124,356} \brk{4 u_{1245} + 2 u_{1643} - 2 u_{1243} u_{2356} - 2 u_{1645} u_{2356} - 2 u_{1642} u_{2653}}
}
\nn\\ & \mspace{40mu}
+ \pentabox_{1,42,356} \brk{4 u_{1356} + 4 u_{1623} - 4 u_{1653} - 4 u_{1325} u_{1653}}
\nn\\ & \mspace{40mu}
+ \pentabox_{1,24,356} \brk{-4 u_{1356} - 4 u_{1643} + 4 u_{1653} + 4 u_{1345} u_{1653}}
\nn\\ & \mspace{40mu}
+ \doublepenta_{14,25,63} \brk{-2 - 2 u_{1354} + 2 u_{1453}}
+ \doublepenta_{14,25,36} \brk{-2 u_{1456} + 4 u_{1654}}
\nn\\ & \mspace{40mu}
+ F_1^{1246} F_1^{1345} \brkleft{2 u_{1264} + 2 u_{1452} - 2 u_{1453} + 2 u_{1254} u_{1364} + 2 u_{1364} u_{1452}}
\nn\\ & \mspace{160mu}
\brkright{\mathord{} - 4 u_{1356} u_{1463} - u_{1364} u_{1436} u_{1452} + u_{1264} u_{1426} u_{1453} + u_{1256} u_{1423} u_{1462}}
\nn\\ & \mspace{40mu}
- 8 \fivedoublebox_{1,23,45} u_{1354}
+   \fivedoublebox_{1,24,35} \brk{4 + 4 u_{1354} - 4 u_{1423} + 4 u_{1453}}
\nn\\ & \mspace{40mu}
+   \fivedoublebox_{1,24,56} \brk{-4 u_{1425} + 4 u_{1524}}
+ 4 \fivepentabox_{1,23,45}
+   \fivepentabox_{1,24,35} \brk{-4 + 4 u_{2543}}
\nn\\ & \mspace{40mu}
- 2 F_2^{1425} u_{1254}
+   F_2^{1243} \brk{-4 + 2 u_{1432}}
+   F_2^{1245} \brk{-2 + 6 u_{1254} + 2 u_{1452}}
\brkright[a]!{\mathord{}}_{33}
\,,
\label{eq:f33}
\end{align}
\begin{align}
f_{6}^{(2)}&=3\big\langle
B_{123,456} \brk{-8 + 4 u_{1436} + 16 u_{2456} - 4 u_{1432} u_{2456} - 4 u_{1536} u_{2456}}
\nn\\ & \mspace{40mu}
+ B_{124,356} \brkleft{-2 - 4 u_{1245} - 2 u_{1346} + 2 u_{1643} - 8 u_{1243} u_{2356}}
\nn\\ & \mspace{140mu}
\brkright{\mathord{}+ 4 u_{1645} u_{2356} + 8 u_{2653} + 8 u_{1246} u_{2653} - 4 u_{1345} u_{2653} - 8 u_{1642} u_{2653}}
\nn\\ & \mspace{40mu}
+ \pentabox_{1,23,456} \brk{16 - 8 u_{1436} - 8 u_{1456} - 16 u_{1654} + 8 u_{1435} u_{1654}}
\nn\\ & \mspace{40mu}
+ \pentabox_{1,25,346} \brk{-4 + 8 u_{1346} - 4 u_{1356} - 4 u_{1643} + 8 u_{1653} - 4 u_{1345} u_{1653}}
\nn\\ & \mspace{40mu}
+ \pentabox_{1,32,456} \brk{-16 + 8 u_{1426} + 8 u_{1456} + 16 u_{1654} - 8 u_{1425} u_{1654}}
\nn\\ & \mspace{40mu}
+ \pentabox_{1,34,256} \brk{-8 - 8 u_{1256} + 4 u_{1642} + 4 u_{1652} - 4 u_{1245} u_{1652}}
\nn\\ & \mspace{40mu}
+ \pentabox_{1,36,245} \brk{4 + 4 u_{1245} - 8 u_{1265} - 8 u_{1542} + 4 u_{1562} + 4 u_{1246} u_{1562}}
\nn\\ & \mspace{40mu}
+ \pentabox_{1,43,256} \brk{8 + 8 u_{1256} - 4 u_{1632} - 4 u_{1652} + 4 u_{1235} u_{1652}}
\nn\\ & \mspace{40mu}
+ \doublepenta_{12,36,45} \brk{-4 + 4 u_{1265} - 4 u_{1562}}
+ \doublepenta_{12,36,54} \brk{4 - 4 u_{1264} + 4 u_{1462}}
\nn\\ & \mspace{40mu}
+ \doublepenta_{14,23,56} \brk{-4 + 4 u_{1436} - 4 u_{1634}}
+ \doublepenta_{14,23,65} \brk{-4 u_{1435} + 8 u_{1534}}
\nn\\ & \mspace{40mu}
+ \doublepenta_{14,25,36} \brk{-2 + 2 u_{1456} - 2 u_{1654}}
+ \doublepenta_{14,25,63} \brk{4 u_{1354} - 2 u_{1453}}
\nn\\ & \mspace{40mu}
+ F_1^{1234} F_1^{1456} \brkleft{4 + 4 u_{1263} + 2 u_{1465} - 4 u_{1564}}
\nn\\ & \mspace{120mu}
\brkright{\mathord{} - 8 u_{1264} u_{2435} + 2 u_{1462} u_{2435} + 4 u_{1364} u_{2534} - 2 u_{1463} u_{2534}}
\nn\\ & \mspace{40mu}
+ F_1^{1236} F_1^{1245} \brkleft{-4 - 6 u_{1254} + 2 u_{1263} - 4 u_{1264} - 4 u_{1362} + 4 u_{1452}}
\nn\\ & \mspace{120mu}
- 4 u_{1264} u_{1352} + 6 u_{1254} u_{1362} + 4 u_{1256} u_{1362} - 2 u_{1263} u_{1452} + 4 u_{1256} u_{1462}
\nn\\ & \mspace{120mu}
\brkright{\mathord{} + 4 u_{1352} u_{1462} - 2 u_{1234} u_{1263} u_{1356} - 6 u_{1236} u_{1254} u_{1362} + 2 u_{1246} u_{1253} u_{1462}}
\nn\\ & \mspace{40mu}
+ F_1^{1245} F_1^{1346} \brkleft{-4 - 4 u_{1254} + 6 u_{1452} - 4 u_{1453} - 2 u_{1463} + 2 u_{1356} u_{1463}}
\nn\\ & \mspace{120mu}
\brkright{\mathord{} + 2 u_{1256} u_{1462} + u_{1364} u_{1436} u_{1452} + u_{1264} u_{1426} u_{1453} - u_{1256} u_{1423} u_{1462}}
\nn\\ & \mspace{40mu}
+ \fivedoublebox_{1,23,45} \brk{20 - 16 u_{1324} - 12 u_{1354} + 8 u_{1423} - 12 u_{1453} + 8 u_{1325} u_{1453}}
\nn\\ & \mspace{40mu}
+ \fivedoublebox_{1,23,46} \brk{16 + 16 u_{1364} + 8 u_{1423} - 8 u_{1463} - 8 u_{1326} u_{1463}}
\nn\\ & \mspace{40mu}
+ \fivedoublebox_{1,23,56} \brk{-8 u_{1523} - 4 u_{1326} u_{1563}}
\nn\\ & \mspace{40mu}
+ \fivedoublebox_{1,24,36} \brk{-12 + 8 u_{1324} - 4 u_{1364} - 4 u_{1423} + 8 u_{1463} - 4 u_{1326} u_{1463}}
\nn\\ & \mspace{40mu}
+ \fivedoublebox_{1,25,34} \brk{-4 + 4 u_{1325} + 16 u_{1345} + 4 u_{1523} - 8 u_{1324} u_{1543}}
\nn\\ & \mspace{40mu}
+ \fivedoublebox_{1,25,36} \brk{-4 + 4 u_{1365} + 4 u_{1563} - 4 u_{1326} u_{1563}}
\nn\\ & \mspace{40mu}
+ \fivedoublebox_{1,26,34} \brk{8 u_{1326} - 8 u_{1346} + 4 u_{1623} - 20 u_{1643} + 4 u_{1324} u_{1643}}
\nn\\ & \mspace{40mu}
+ \fivepentabox_{1,23,45} \brk{8 + 8 u_{2435}}
+ \fivepentabox_{1,23,56} \brk{12 + 4 u_{2536} - 4 u_{2635}}
\nn\\ & \mspace{40mu}
+ \fivepentabox_{1,24,35} \brk{-8 - 8 u_{2345} + 8 u_{2543}}
+ \fivepentabox_{1,25,34} \brk{-8 - 8 u_{2354} + 8 u_{2453}}
\nn\\ & \mspace{40mu}
+ \fivepentabox_{1,25,36} \brk{-4 - 12 u_{2356} + 4 u_{2653}}
+ \fivepentabox_{1,26,35} \brk{-8 - 8 u_{2365} + 8 u_{2563}}
\nn\\ & \mspace{40mu}
+ \fivepentabox_{1,34,25} \brk{-4 + 4 u_{2354} + 4 u_{2453}}
- 4 \fivedoublepenta_{13,456}
\nn\\ & \mspace{40mu}
+ F_1^{1234} F_1^{1236} \brk{-8 - 8 u_{1243} + 8 u_{1342} - 8 u_{1362} + 4 u_{1246} u_{1362} + 4 u_{1342} u_{1362}}
\nn\\ & \mspace{40mu}
+ F_2^{1234} \brk{-16 + 16 u_{1342}}
+ F_2^{1243} \brk{-2 + 6 u_{1234} + 2 u_{1432}}
\nn\\ & \mspace{40mu}
+ F_2^{1245} \brk{-4 + 12 u_{1254} + 4 u_{1452}}
+ F_2^{1246} \brk{-12 + 6 u_{1462}}
\nn\\ & \mspace{40mu}
+ F_2^{1254} \brk{-4 + 4 u_{1245} + 4 u_{1542}}
+ F_2^{1263} \brk{4 - 4 u_{1236} - 20 u_{1632}}
\nn\\ & \mspace{40mu}
+ F_2^{1264} \brk{-4 - 4 u_{1246} + 4 u_{1642}}
- 8 F_2^{1265} u_{1256}
- 16 F_2^{1364} u_{1643}
+ 4 \threedoublebox_{123}
\big\rangle_6
\,.
\label{eq:f6}
\end{align}
As before, $\avg{\cdot}_{24}$, $\avg{\cdot}_{33}$, and $\avg{\cdot}_6$
means averaging over the respective permutation symmetry group
$K_{\mathbold{a}}$, see~\eqref{eq:Kgroups}. The results are expressed
in terms of the conformal integrals~\eqref{eq:integrals} and general
cross ratios~\eqref{eq:uijkl}. They can easily be converted to the independent
basis~\eqref{eq:ui6} of six-point cross ratios by expanding in
$x_{ij}^2$ and setting $x_6=\infty$ and $x_{12}^2=1$ (see \secref{sec:guide}).

%%%%%%%%%%%%%%%%%%%%%%%%%%%%%%
\paragraph{Seven Points.}

At seven points, the two-loop component functions
besides~\eqref{eq:f223} are as follows. Again, all expressions are
included in the attached file \filename{results.m}.
\begin{align}
f_{25}^{(2)}&=10\big\langle
F_1^{1237} F_1^{3456} \brk{-4 u_{1273} - 4 u_{1273} u_{3465} + 4 u_{1273} u_{3564}}
\nn\\ & \mspace{60mu}
- 4 B_{123,456} u_{1234} u_{2456}
+   B_{134,256} \brk{2 u_{1246} + 6 u_{1243} u_{2365} - 2 u_{1245} u_{2563}}
\nn\\ & \mspace{60mu}
- 4 B_{123,457} u_{1237} u_{2754}
+ \pentabox_{1,34,256} \brk{-4 u_{1246} + 4 u_{1245} u_{1652}}
\nn\\ & \mspace{60mu}
+ \pentabox_{1,34,267} \brk{-4 u_{1247} + 4 u_{1246} u_{1762}}
- 4 \pentabox_{3,56,124} u_{1243}
+ 4 \pentabox_{3,45,127} u_{1273}
\nn\\ & \mspace{60mu}
- 4 \pentabox_{3,54,127} u_{1273}
+ 4 \pentabox_{3,56,127} u_{1273}
+ \doublepenta_{12,34,65} \brk{-u_{1245} + 2 u_{1542}}
\nn\\ & \mspace{60mu}
+ \doublepenta_{12,34,56} \brk{2 u_{1246} - 4 u_{1642}}
+ \doublepenta_{12,34,76} \brk{-u_{1246} + 2 u_{1642}}
\nn\\ & \mspace{60mu}
+ F_1^{1237} F_1^{1245} \brkleft{
- 2 u_{1243} u_{1472}
+ 4 u_{1253} u_{1472}
- 2 u_{1257} u_{1472}}
\nn\\ & \mspace{160mu}
\brkright{\mathord{}
+ u_{1234} u_{1273} u_{1357}
+ 3 u_{1237} u_{1254} u_{1372}
- u_{1247} u_{1253} u_{1472}
}
\nn\\ & \mspace{60mu}
+ F_1^{1245} F_1^{3456} \brk{-2 u_{1254} + 2 u_{1253} u_{2346} - 2 u_{1256} u_{2643}}
\nn\\ & \mspace{60mu}
- 4 \fivedoublebox_{3,12,45} u_{1234}
- 4 \fivedoublebox_{3,12,47} u_{1234}
+ 8 \fivedoublebox_{3,14,27} u_{1234}
+ \fivedoublebox_{1,23,45} \brk{-4 u_{1354} + 4 u_{1453}}
\nn\\ & \mspace{60mu}
+ 4 \fivepentabox_{3,45,12}
- 4 \fivepentabox_{3,14,25} u_{1245}
+ \fivepentabox_{3,12,45} \brk{2 - 4 u_{1425}}
+ 2 \fivedoublepenta_{35,124}
\nn\\ & \mspace{60mu}
+ F_1^{1234} F_1^{1237} \brk{2 u_{1243} u_{1273} + 4 u_{1273} u_{1342} - 2 u_{1247} u_{1372}}
+ 12 F_2^{1234} u_{1243}
\nn\\ & \mspace{60mu}
- 2 F_2^{3142} u_{1243}
+ F_2^{1324} \brk{2 + u_{1243}}
+ F_2^{1325} \brk{2 - u_{1253}}
- 2 \threedoublebox_{123}
\big\rangle_{25}
\,,
\label{eq:f25}
\end{align}
\begin{align}
f_{34}^{(2)}&=24\big\langle
  \pentabox_{145,2367} \brk{4 - 2 u_{2365} - 4 u_{2764} + 2 u_{2364} u_{3457}}
\nn\\ & \mspace{50mu}
+ \pentabox_{415,2367} \brk{-4 + 4 u_{1375} + 4 u_{1672} + 4 u_{2365} - 4 u_{1372} u_{2365} - 4 u_{1675} u_{2365}}
\nn\\ & \mspace{50mu}
+ \doublepenta_{1,245,367} \brk{-2 u_{1765} + 2 u_{1546} u_{1765}}
\nn\\ & \mspace{50mu}
+ \doublepenta_{1,245,637} \brk{-4 + 4 u_{1547} + 4 u_{1735} - 4 u_{1534} u_{1745}}
\nn\\ & \mspace{50mu}
+ \doublepenta_{1,425,637} \brk{2 - 4 u_{1527} + 2 u_{1523} u_{1735}}
\nn\\ & \mspace{50mu}
+ \doublepenta_{1,425,736} \brk{-2 + 4 u_{1526} - 2 u_{1523} u_{1635}}
\nn\\ & \mspace{50mu}
+ F_1^{1267} F_1^{1345} \brkleft{
-2
+ 2 u_{1672}
+ 2 u_{2463}
+ 3 u_{1276} u_{1354}}
\nn\\ & \mspace{160mu}
+ u_{1256} u_{1374}
- 4 u_{1374} u_{1652}
+ 2 u_{1674} u_{2364}
- 2 u_{1673} u_{2463}
\nn\\ & \mspace{160mu}
\brkright{\mathord{}
- 2 u_{1453} u_{1675} u_{2365}
+ 4 u_{1354} u_{1675} u_{2465}
- u_{1275} u_{1354} u_{2564}
}
\nn\\ & \mspace{50mu}
+   B_{124,356} \brk{4 u_{1643} + 4 u_{2356} - 4 u_{1243} u_{2356} - 4 u_{1546} u_{2356} + 4 u_{1246} u_{2653} - 4 u_{1642} u_{2653}}
\nn\\ & \mspace{50mu}
+   B_{145,267} \brk{2 - u_{1257} + 2 u_{1752} + 3 u_{1254} u_{2476} - 2 u_{2674} + u_{1256} u_{2674} - 2 u_{1652} u_{2674}}
\nn\\ & \mspace{50mu}
- 4 B_{124,567} u_{1245} u_{2567}
+   \pentabox_{1,24,356} \brk{4 u_{1346} - 4 u_{1356} - 4 u_{1643} + 4 u_{1653}}
\nn\\ & \mspace{50mu}
+   \pentabox_{1,42,356} \brk{-4 u_{1326} + 4 u_{1356} + 4 u_{1623} - 4 u_{1653}}
+   \pentabox_{1,45,267} \brk{-4 + 4 u_{1762}}
\nn\\ & \mspace{50mu}
+   \pentabox_{4,15,237} \brk{4 u_{2374} - 4 u_{2473} + 4 u_{2475} - 4 u_{2453} u_{2574}}
\nn\\ & \mspace{50mu}
+   \pentabox_{4,51,237} \brk{-4 u_{1742} - 4 u_{2374} + 4 u_{1342} u_{2374} + 4 u_{2473}}
+ 4 \pentabox_{4,56,127} u_{1274}
\nn\\ & \mspace{50mu}
- 4 \pentabox_{4,65,127} u_{1274}
+ \doublepenta_{14,25,37} \brk{-2 u_{1457} + 4 u_{1754}}
\nn\\ & \mspace{50mu}
+ \doublepenta_{14,25,73} \brk{-4 - 4 u_{1354} + 4 u_{1453}}
+ \doublepenta_{14,52,73} \brk{4 u_{1324} - 2 u_{1423}}
\nn\\ & \mspace{50mu}
+ F_1^{1245} F_1^{4567} \brk{-2 u_{1254} + 2 u_{1254} u_{4675} - 2 u_{1254} u_{4567} u_{4675}}
\nn\\ & \mspace{50mu}
+ F_1^{1247} F_1^{1345} \brkleft{
+ 4 u_{1274}
+ 4 u_{1452}
- 4 u_{1453}
+ 4 u_{1254} u_{1374}}
\nn\\ & \mspace{160mu}
- 4 u_{1374} u_{1432}
+ 4 u_{1374} u_{1452}
- 4 u_{1357} u_{1473}
\nn\\ & \mspace{160mu}
\brkright{\mathord{}
- 2 u_{1374} u_{1437} u_{1452}
+ 2 u_{1274} u_{1427} u_{1453}
+ 2 u_{1257} u_{1423} u_{1472}}
\nn\\ & \mspace{50mu}
- 8 \fivedoublebox_{1,23,45} u_{1354}
+   \fivedoublebox_{1,24,35} \brk{6 u_{1354} - 2 u_{1423} + 4 u_{1453}}
\nn\\ & \mspace{50mu}
+   \fivedoublebox_{1,24,36} \brk{-2 u_{1364} - 2 u_{1423} + 4 u_{1326} u_{1463}}
+   \fivedoublebox_{1,24,56} \brk{-4 u_{1425} + 4 u_{1524}}
\nn\\ & \mspace{50mu}
+   \fivedoublebox_{1,45,67} \brk{-2 u_{1675} + 2 u_{1547} u_{1675}}
+   \fivedoublebox_{4,12,35} \brk{4 u_{2435} - 4 u_{2534}}
\nn\\ & \mspace{50mu}
- 4 \fivedoublebox_{4,12,56} u_{1245}
- 4 \fivedoublebox_{4,12,57} u_{1245}
+   \fivedoublebox_{4,15,27} \brk{4 u_{1245} - 4 u_{2457} + 4 u_{2754} + 4 u_{1742} u_{2754}}
\nn\\ & \mspace{50mu}
+ 4 \fivepentabox_{1,23,45}
+   \fivepentabox_{1,24,35} \brk{-4 + 4 u_{2543}}
+   \fivepentabox_{4,15,26} \brk{-4 + 4 u_{1652}}
+ 4 \fivepentabox_{4,56,12}
\nn\\ & \mspace{50mu}
+ F_2^{1243} \brk{-4 + 2 u_{1432}}
+ F_2^{1245} \brk{-4 + 12 u_{1254} + 4 u_{1452}}
\nn\\ & \mspace{50mu}
- 2 F_2^{1425} u_{1254}
+ F_2^{1456} \brk{-4 + 2 u_{1564}}
- 2 F_2^{4152} u_{1254}
\big\rangle_{34}
\,,
\label{eq:f34}
\end{align}
\begin{align}
f_{7}^{(2)}&=7\big\langle
\pentabox_{123,4567} \brkleft{4 + 4 u_{2476} - 4 u_{2573} - 8 u_{2674} - 4 u_{3654}}
\nn\\ & \mspace{100mu}
\brkright{\mathord{}- 4 u_{2576} u_{3456} + 4 u_{2675} u_{3456} - 4 u_{2475} u_{3654} + 8 u_{2574} u_{3654} + 4 u_{2673} u_{3654}}
\nn\\ & \mspace{30mu}
+ \pentabox_{125,3467} \brkleft{-2 + 2 u_{2376} + 2 u_{2475} - 2 u_{3465} + 4 u_{3564}}
\nn\\ & \mspace{100mu}
\brkright{\mathord{}+ 2 u_{2374} u_{3465} + 2 u_{2675} u_{3465} - 4 u_{2375} u_{3564} - 2 u_{2476} u_{3564} - 2 u_{2674} u_{3564}}
\nn\\ & \mspace{30mu}
+ \pentabox_{127,3456} \brk{4 u_{2365} - 4 u_{2563} - 4 u_{2364} u_{3457} + 4 u_{2463} u_{3457} - 4 u_{2465} u_{3754} + 4 u_{2564} u_{3754}}
\nn\\ & \mspace{30mu}
+ \pentabox_{145,2367} \brk{-2 u_{2365} - 2 u_{2764} + 2 u_{2364} u_{3457} + 2 u_{2765} u_{3457} - 2 u_{2465} u_{3754} + 2 u_{2564} u_{3754}}
\nn\\ & \mspace{30mu}
+ \doublepenta_{1,234,567} \brk{4 - 4 u_{1437} - 4 u_{1764} + 4 u_{1436} u_{1764}}
\nn\\ & \mspace{30mu}
+ \doublepenta_{1,234,657} \brk{-4 + 4 u_{1437} + 4 u_{1754} - 4 u_{1435} u_{1754}}
+ \doublepenta_{1,234,756} \brk{4 - 2 u_{1436} - 2 u_{1654}}
\nn\\ & \mspace{30mu}
+ \doublepenta_{1,256,347} \brk{2 - 2 u_{1657} - 2 u_{1746} + 2 u_{1645} u_{1756}}
\nn\\ & \mspace{30mu}
+ \doublepenta_{1,256,437} \brk{-2 + 2 u_{1657} + 2 u_{1736} - 2 u_{1635} u_{1756}}
+ \doublepenta_{1,256,734} \brk{2 - u_{1436} - u_{1654}}
\nn\\ & \mspace{30mu}
+ \doublepenta_{1,324,567} \brk{-4 + 4 u_{1427} + 4 u_{1764} - 4 u_{1426} u_{1764}}
+ \doublepenta_{1,324,657} \brk{4 - 2 u_{1427} - 2 u_{1754}}
\nn\\ & \mspace{30mu}
+ \doublepenta_{1,347,526} \brk{-2 + 2 u_{1627} + 2 u_{1746} - 2 u_{1624} u_{1746}}
+ \doublepenta_{1,347,625} \brk{2 - u_{1527} - u_{1745}}
\nn\\ & \mspace{30mu}
+ \doublepenta_{1,423,567} \brk{4 - 2 u_{1327} - 2 u_{1763}}
+ \doublepenta_{1,437,526} \brk{2 - u_{1627} - u_{1736}}
\nn\\ & \mspace{30mu}
+ F_1^{1234} F_1^{1567} \brkleft{2 u_{1243} + 2 u_{1576} + 4 u_{1273} u_{1546} - 4 u_{1372} u_{1546} + 4 u_{1243} u_{1576}}
\nn\\ & \mspace{100mu}
- 4 u_{1342} u_{1576} - 4 u_{1273} u_{1645} + 2 u_{1372} u_{1645} - 4 u_{1243} u_{1675}
\nn\\ & \mspace{100mu}
+ 2 u_{1342} u_{1675} + 2 u_{1245} u_{2536} + 2 u_{1572} u_{2536} - 2 u_{1246} u_{2635} - 2 u_{1573} u_{2635}
\nn\\ & \mspace{100mu}
\brkright{\mathord{}+ 4 u_{1274} u_{1645} u_{2435} - 4 u_{1274} u_{1546} u_{2436} - 2 u_{1374} u_{1645} u_{2534} + 4 u_{1374} u_{1546} u_{2634}}
\nn\\ & \mspace{30mu}
+ F_1^{1256} F_1^{1347} \brkleft{-3 u_{1265} + u_{1374} + 2 u_{1275} u_{1364} - 6 u_{1265} u_{1374} - 2 u_{1275} u_{1463}}
\nn\\ & \mspace{100mu}
+ 6 u_{1265} u_{1473} - 2 u_{1374} u_{1562} + u_{1473} u_{1562} - 2 u_{1364} u_{1572} + u_{1463} u_{1572}
\nn\\ & \mspace{100mu}
+ u_{1263} u_{2354} + u_{1372} u_{2354} - u_{1264} u_{2453} - u_{1375} u_{2453}
\nn\\ & \mspace{100mu}
\brkright{\mathord{}- u_{1463} u_{1576} u_{2356} + 2 u_{1364} u_{1576} u_{2456} + 2 u_{1276} u_{1463} u_{2653} - 2 u_{1276} u_{1364} u_{2654}}
\nn\\ & \mspace{30mu}
+ B_{123,456} \brkleft{-2 - 2 u_{1436} + 2 u_{1634} + 4 u_{2654} + 2 u_{1234} u_{2456}}
\nn\\ & \mspace{100mu}
\brkright{\mathord{}- 2 u_{1635} u_{2456} + 4 u_{1435} u_{2654} - 2 u_{1534} u_{2654} - 2 u_{1632} u_{2654}}
\nn\\ & \mspace{30mu}
+ B_{124,356} \brkleft{-1 - 2 u_{1245} - u_{1346} + u_{1643} + 4 u_{2653} - 4 u_{1243} u_{2356}}
\nn\\ & \mspace{100mu}
\brkright{\mathord{}+ 2 u_{1645} u_{2356} + 4 u_{1246} u_{2653} - 2 u_{1345} u_{2653} - 4 u_{1642} u_{2653}}
\nn\\ & \mspace{30mu}
+ B_{124,567} \brkleft{-8 + 4 u_{1246} + 4 u_{1547} + 4 u_{1642} + 8 u_{2765} + 4 u_{1245} u_{2567}}
\nn\\ & \mspace{100mu}
\brkright{\mathord{}- 4 u_{1247} u_{2765} - 4 u_{1546} u_{2765} - 4 u_{1742} u_{2765}}
\nn\\ & \mspace{30mu}
+ B_{125,346} \brkleft{2 - 2 u_{1452} - 4 u_{2346} + 2 u_{2643} + 4 u_{1352} u_{2346}}
\nn\\ & \mspace{100mu}
\brkright{\mathord{}+ 2 u_{1654} u_{2346} - 6 u_{1256} u_{2643} - 2 u_{1354} u_{2643} - 2 u_{1652} u_{2643}}
\nn\\ & \mspace{30mu}
+ B_{125,347} \brkleft{-u_{1357} - u_{2347} - u_{1754} u_{2347} - u_{1257} u_{2743}}
\nn\\ & \mspace{100mu}
\brkright{\mathord{}+ 2 u_{1253} u_{2347} + 2 u_{1457} u_{2347} + 2 u_{1354} u_{2743} - 2 u_{1453} u_{2743}}
\nn\\ & \mspace{30mu}
+ \pentabox_{1,23,457} \brk{8 - 4 u_{1437} - 4 u_{1457} - 8 u_{1754} + 4 u_{1435} u_{1754}}
\nn\\ & \mspace{30mu}
+ \pentabox_{1,23,567} \brk{4 - 4 u_{1567} - 4 u_{1765}}
+ \pentabox_{1,24,567} \brk{4 - 4 u_{1547} - 4 u_{1765} + 4 u_{1546} u_{1765}}
\nn\\ & \mspace{30mu}
+ \pentabox_{1,25,347} \brk{2 u_{1347} - 2 u_{1357} - 2 u_{1743} + 2 u_{1753}}
\nn\\ & \mspace{30mu}
+ \pentabox_{1,26,347} \brk{-2 + 2 u_{1347} + 2 u_{1763} - 2 u_{1346} u_{1763}}
\nn\\ & \mspace{30mu}
+ \pentabox_{1,26,457} \brk{-2 + 4 u_{1457} - 2 u_{1467} - 2 u_{1754} + 4 u_{1764} - 2 u_{1456} u_{1764}}
\nn\\ & \mspace{30mu}
+ \pentabox_{1,32,457} \brk{-8 + 4 u_{1427} + 4 u_{1457} + 8 u_{1754} - 4 u_{1425} u_{1754}}
\nn\\ & \mspace{30mu}
+ \pentabox_{1,32,567} \brk{-4 + 4 u_{1567} + 4 u_{1765}}
+ \pentabox_{1,34,256} \brk{-2 - 6 u_{1256} + 2 u_{1652}}
\nn\\ & \mspace{30mu}
+ \pentabox_{1,34,267} \brk{-4 - 4 u_{1267} + 2 u_{1742} + 2 u_{1762} - 2 u_{1246} u_{1762}}
\nn\\ & \mspace{30mu}
+ \pentabox_{1,34,567} \brk{4 - 4 u_{1567} - 4 u_{1765}}
\nn\\ & \mspace{30mu}
+ \pentabox_{1,37,245} \brk{2 + 2 u_{1245} - 4 u_{1275} - 4 u_{1542} + 2 u_{1572} + 2 u_{1247} u_{1572}}
\nn\\ & \mspace{30mu}
+ \pentabox_{1,37,256} \brk{2 - 2 u_{1276} - 2 u_{1652} + 2 u_{1257} u_{1672}}
\nn\\ & \mspace{30mu}
+ \pentabox_{1,42,567} \brk{-4 + 4 u_{1527} + 4 u_{1765} - 4 u_{1526} u_{1765}}
\nn\\ & \mspace{30mu}
+ \pentabox_{1,43,256} \brk{2 + 6 u_{1256} - 2 u_{1652}}
\nn\\ & \mspace{30mu}
+ \pentabox_{1,43,267} \brk{4 + 4 u_{1267} - 2 u_{1732} - 2 u_{1762} + 2 u_{1236} u_{1762}}
\nn\\ & \mspace{30mu}
+ \pentabox_{1,43,567} \brk{-4 + 4 u_{1567} + 4 u_{1765}}
\nn\\ & \mspace{30mu}
+ \pentabox_{1,45,237} \brk{4 - 2 u_{1237} - 2 u_{1257} + 4 u_{1732} + 2 u_{1235} u_{1752}}
\nn\\ & \mspace{30mu}
+ \pentabox_{1,45,267} \brk{-4 - 4 u_{1267} + 2 u_{1752} + 2 u_{1762} - 2 u_{1256} u_{1762}}
\nn\\ & \mspace{30mu}
+ \pentabox_{1,47,256} \brk{2 u_{1256} - 2 u_{1276} - 2 u_{1652} + 2 u_{1672}}
\nn\\ & \mspace{30mu}
+ \doublepenta_{12,37,45} \brk{-2 + 2 u_{1275} - 2 u_{1572}}
+ \doublepenta_{12,37,54} \brk{2 - 2 u_{1274} + 2 u_{1472}}
\nn\\ & \mspace{30mu}
+ \doublepenta_{12,37,56} \brk{-2 + 2 u_{1276} - 2 u_{1672}}
+ \doublepenta_{12,37,65} \brk{2 - 2 u_{1275} + 2 u_{1572}}
\nn\\ & \mspace{30mu}
+ \doublepenta_{14,23,57} \brk{-4 + 4 u_{1437} - 4 u_{1734}}
+ \doublepenta_{14,23,75} \brk{4 - 4 u_{1435} + 4 u_{1534}}
\nn\\ & \mspace{30mu}
+ \doublepenta_{14,25,37} \brk{-1 + u_{1457} - u_{1754}}
+ \doublepenta_{14,25,73} \brk{2 + 2 u_{1354} - 2 u_{1453}}
\nn\\ & \mspace{30mu}
+ \doublepenta_{14,52,73} \brk{-1 - u_{1324} + u_{1423}}
\nn\\ & \mspace{30mu}
+ F_1^{1234} F_1^{1457} \brkleft{2 + 4 u_{1273} - 4 u_{1372} + 2 u_{1475} - 2 u_{1574} - 4 u_{1274} u_{2435}}
\nn\\ & \mspace{100mu}
\brkright{\mathord{}- 2 u_{1375} u_{2435} + 2 u_{1472} u_{2435} + 2 u_{1275} u_{2534} + 4 u_{1374} u_{2534} - 2 u_{1473} u_{2534}}
\nn\\ & \mspace{30mu}
+ F_1^{1237} F_1^{1245} \brkleft{-4 - 2 u_{1253} - 6 u_{1254} + 2 u_{1273} - 2 u_{1274} - 4 u_{1372}}
\nn\\ & \mspace{100mu}
+ 4 u_{1452} - 2 u_{1274} u_{1352} + 6 u_{1254} u_{1372} + 2 u_{1257} u_{1372} + 2 u_{1243} u_{1452}
\nn\\ & \mspace{100mu}
- 2 u_{1273} u_{1452} + 2 u_{1243} u_{1472} - 2 u_{1253} u_{1472} + 2 u_{1257} u_{1472}
\nn\\ & \mspace{100mu}
\brkright{\mathord{}+ 4 u_{1352} u_{1472} - 2 u_{1234} u_{1273} u_{1357} - 6 u_{1237} u_{1254} u_{1372} + 2 u_{1247} u_{1253} u_{1472}}
\nn\\ & \mspace{30mu}
+ F_1^{1245} F_1^{1347} \brkleft{-4 - 4 u_{1254} + 6 u_{1452} - 2 u_{1453} - 2 u_{1472} - 2 u_{1473}}
\nn\\ & \mspace{100mu}
+ u_{1354} u_{1432} + u_{1374} u_{1432} + u_{1257} u_{1472} + u_{1357} u_{1473}
\nn\\ & \mspace{100mu}
\brkright{\mathord{}+ u_{1374} u_{1437} u_{1452} + u_{1274} u_{1427} u_{1453} - u_{1257} u_{1423} u_{1472}}
\nn\\ & \mspace{30mu}
+ \fivedoublebox_{1,23,45} \brk{10 - 8 u_{1324} - 6 u_{1354} + 4 u_{1423} - 6 u_{1453} + 4 u_{1325} u_{1453}}
\nn\\ & \mspace{30mu}
+ \fivedoublebox_{1,23,47} \brk{8 + 8 u_{1374} + 4 u_{1423} - 4 u_{1473} - 4 u_{1327} u_{1473}}
\nn\\ & \mspace{30mu}
+ \fivedoublebox_{1,23,56} \brk{6 - 4 u_{1325} - 6 u_{1365} - 6 u_{1563} + 4 u_{1326} u_{1563}}
\nn\\ & \mspace{30mu}
+ \fivedoublebox_{1,23,57} \brk{-4 + 4 u_{1325} + 4 u_{1573} - 4 u_{1327} u_{1573}}
+ \fivedoublebox_{1,23,67} \brk{2 - 4 u_{1326} - 4 u_{1623}}
\nn\\ & \mspace{30mu}
+ \fivedoublebox_{1,24,37} \brk{-6 + 4 u_{1324} - 2 u_{1374} - 2 u_{1423} + 4 u_{1473} - 2 u_{1327} u_{1473}}
\nn\\ & \mspace{30mu}
+ \fivedoublebox_{1,24,56} \brk{4 - 4 u_{1425} + 4 u_{1524}}
+ \fivedoublebox_{1,24,57} \brk{2 + 4 u_{1524} - 2 u_{1427} u_{1574}}
\nn\\ & \mspace{30mu}
+ \fivedoublebox_{1,25,34} \brk{2 u_{1325} + 6 u_{1345} - 2 u_{1324} u_{1543}}
\nn\\ & \mspace{30mu}
+ \fivedoublebox_{1,25,37} \brk{-2 + 2 u_{1375} + 2 u_{1523} - 2 u_{1327} u_{1573}}
+ \fivedoublebox_{1,25,47} \brk{-3 + 2 u_{1574} - u_{1427} u_{1574}}
\nn\\ & \mspace{30mu}
+ \fivedoublebox_{1,26,34} \brk{-2 + 2 u_{1346} + 2 u_{1623} - 2 u_{1324} u_{1643}}
+ \fivedoublebox_{1,26,37} \brk{-1 + 2 u_{1673} - u_{1327} u_{1673}}
\nn\\ & \mspace{30mu}
+ \fivedoublebox_{1,26,45} \brk{-2 + 2 u_{1426} + 8 u_{1456} + 2 u_{1624} - 4 u_{1425} u_{1654}}
\nn\\ & \mspace{30mu}
+ \fivedoublebox_{1,27,34} \brk{4 u_{1327} - 4 u_{1347} + 2 u_{1723} - 10 u_{1743} + 2 u_{1324} u_{1743}}
\nn\\ & \mspace{30mu}
+ \fivedoublebox_{1,27,45} \brk{2 - 2 u_{1457} - 6 u_{1754}}
+ \fivedoublebox_{1,34,56} \brk{3 - 6 u_{1435} + 3 u_{1436} u_{1564}}
\nn\\ & \mspace{30mu}
+ \fivepentabox_{1,23,45} \brk{4 + 4 u_{2435}}
+ \fivepentabox_{1,23,56} \brk{6 + 2 u_{2536} - 2 u_{2635}}
\nn\\ & \mspace{30mu}
+ \fivepentabox_{1,23,67} \brk{6 + 2 u_{2637} - 2 u_{2736}}
+ \fivepentabox_{1,24,35} \brk{-4 - 4 u_{2345} + 4 u_{2543}}
\nn\\ & \mspace{30mu}
+ \fivepentabox_{1,25,34} \brk{-4 - 4 u_{2354} + 4 u_{2453}}
+ \fivepentabox_{1,25,36} \brk{-2 - 6 u_{2356} + 2 u_{2653}}
\nn\\ & \mspace{30mu}
+ \fivepentabox_{1,26,35} \brk{-4 - 4 u_{2365} + 4 u_{2563}}
+ \fivepentabox_{1,26,37} \brk{-2 - 6 u_{2367} + 2 u_{2763}}
\nn\\ & \mspace{30mu}
+ \fivepentabox_{1,27,36} \brk{-4 - 4 u_{2376} + 4 u_{2673}}
+ \fivepentabox_{1,34,25} \brk{-2 + 2 u_{2354} + 2 u_{2453}}
-2 \fivedoublepenta_{14,567}
\nn\\ & \mspace{30mu}
+ F_1^{1234} F_1^{1237} \brk{-4 - 4 u_{1243} + 4 u_{1342} - 4 u_{1372} + 2 u_{1247} u_{1372} + 2 u_{1342} u_{1372}}
\nn\\ & \mspace{30mu}
+ F_2^{1234} \brk{-8 + 8 u_{1342}}
+ F_2^{1243} \brk{-1 + 3 u_{1234} + u_{1432}}
\nn\\ & \mspace{30mu}
+ F_2^{1245} \brk{-4 + 12 u_{1254} + 4 u_{1452}}
+ F_2^{1247} \brk{-6 - 6 u_{1274} + 6 u_{1472}}
\nn\\ & \mspace{30mu}
+ F_2^{1254} \brk{-1 + 3 u_{1245} + u_{1542}}
+ F_2^{1265} \brk{-2 + 2 u_{1256} + 2 u_{1652}}
\nn\\ & \mspace{30mu}
+ F_2^{1273} \brk{2 - 2 u_{1237} - 10 u_{1732}}
+ F_2^{1275} \brk{-2 - 2 u_{1257} + 2 u_{1752}}
\nn\\ & \mspace{30mu}
- 4 F_2^{1276} u_{1267}
- 16 F_2^{1374} u_{1743}
+ 2 \threedoublebox_{123}
\big\rangle_7
\,,
\label{eq:f7}
\end{align}
Again, $\avg{\cdot}_{25}$, $\avg{\cdot}_{34}$, and $\avg{\cdot}_7$
means averaging over the respective permutation symmetry group
$K_{\mathbold{a}}$, see~\eqref{eq:Kgroups}. The results are expressed
in terms of the conformal integrals~\eqref{eq:integrals} and general
cross ratios~\eqref{eq:uijkl}. To convert to a multiplicatively
independent set of $14$ cross ratios, one can expand all $u_{ijkl}$ in
terms of $x_{ij}^2$, and then pick a conformal frame, for example by
setting $x_7=\infty$ and $x_{12}^2=1$, and solve for the remaining
$x_{ij}^2$ in terms of a basis of $14$ cross ratios, for example
\eqref{eq:ui7} or \eqref{eq:ui72}.

%%%%%%%%%%%%%%%%%%%%%%%%%%%%%%%%%%%%%%%%%%%%%%%%%%%%%%%%%%%%
\subsection{Preliminary \texorpdfstring{$n$}{n}-Point Guesses}
\label{sec:npointguesses}

Based on our explicit two-loop results for $n=5,6,7$, we have tried to
guess general $n$-point expressions for two classes of component
functions, namely $f_{2,n-2}^{(2)}$ and $f_n^{(2)}$. These guesses
are very preliminary and yet incomplete. To fully determine these
$n$-point functions will require more input. Yet, at least a subset of
terms is matched nicely by our guesses. The (preliminary and
incomplete) expression for the component functions $f_{2,n-2}^{(2)}$ is:
\begin{align}
f_{2,n-2}^{\mathrm{guess}\,(2)}&=
4(n-2)\times
\\ & \mspace{-50mu}
\brkleft[a]2{
- \threedoublebox_{123}
+ F_2^{1324}
+ F_2^{1325}
+ 6 F_2^{1234} u_{1243}
+ \half F_2^{1324} u_{1243}
- F_2^{3142} u_{1243}
- \half F_2^{1325} u_{1253}
} \nn\\ & \mspace{-35mu}
- 2 \fivedoublebox_{3,12,45} u_{1234}
- 2 \fivedoublebox_{3,12,4n} u_{1234}
+ 4 \fivedoublebox_{3,14,2n} u_{1234}
+ \fivedoublebox_{1,23,45} (-2 u_{1354}+ 2\delta_{n,5} + 2\stepfcn_{n\geq6} u_{1453})
\nn\\ & \mspace{-35mu}
+   \fivepentabox_{3,12,45}
+ 2 \fivepentabox_{3,45,12}
- 2 \fivepentabox_{3,14,25} u_{1245}
- 2 \fivepentabox_{3,12,45} u_{1425}
+ \fivedoublepenta_{35,124}
\nn\\ & \mspace{-35mu}
+ F_1^{1234} F_1^{123n}(u_{1243} u_{12n3} + 2 u_{12n3} u_{1342} - u_{124n} u_{13n2})
\nn\\ & \mspace{-35mu}
+ \frac{\stepfcn_{n\geq6}}{4}\brkleft[s]2{
- 8 B_{123,456} u_{1234} u_{2456}
+ 8 \pentabox_{3,45,12n} u_{12n3}
- 8 \pentabox_{3,54,12n} u_{12n3}
} \nn\\ & \mspace{-20mu}
- 8 \pentabox_{1,34,256} (u_{1246}- u_{1245} u_{1652})
- 2 \doublepenta_{12,34,65} (u_{1245}-2u_{1542})
\nn\\ & \mspace{-20mu}
- 4 F_1^{12,n-3,n-2} F_1^{3456} u_{12,n-2,n-3}
- 4 F_1^{123n} F_1^{1245} u_{1243} u_{14n2}
\nn\\ & \mspace{-20mu}
+ c_n \brkleft2{
F_1^{123n} F_1^{1245} (u_{1234} u_{12n3} u_{135n}
+3 u_{123n} u_{1254} u_{13n2}
+4 u_{1253} u_{14n2}- u_{124n} u_{1253} u_{14n2})
}
\nn\\ & \mspace{-0mu}
+ \brkright[a]2{\brkright[s]2{\brkright2{
2 B_{134,256} (u_{1246}+3 u_{1243} u_{2365}- u_{1245} u_{2563})
+ 2 \doublepenta_{12,34,56} (u_{1246}-2u_{1642})
}}}_{2,n-2}
\,, \nn
\end{align}
where $\avg{\cdot}_{2,n-2}$ means averaging over the permutation group
$K_{2,n-2}=\grp{S}_2\times\grp{\grp{D}_{n-2}}$ of the external labels $(1,\dots,n)$, and
\begin{equation}
c_n=\begin{cases}
1 & n=6\,,\\
2 & n=7\,.
\end{cases}
\end{equation}
With this expression, we find
\begin{align}
f_{23}^{(2)}&=f_{23}^{\mathrm{guess}\,(2)}
\,,\nn\\
f_{24}^{(2)}&=f_{24}^{\mathrm{guess}\,(2)}
+ 16\,\brk[a]2{
F_1^{1234}F_1^{3456}u_{1243}u_{3564}\brk{1-u_{3456}}
}_{24}
\,,\nn\\
f_{25}^{(2)}&=f_{25}^{\mathrm{guess}\,(2)}
+ 20\,\brkleft[a]2{
- 2 B_{123,457} u_{1237} u_{2754}
- 2 \pentabox_{3,56,124} u_{1243}
+ 2 \pentabox_{3,56,127} u_{1273}
}
\nn\\ & \mspace{50mu}
+   \pentabox_{1,34,267} (-2 u_{1247} + 2 u_{1246} u_{1762})
+ \doublepenta_{12,34,76} (-\half u_{1246} + u_{1642})
\nn\\ & \mspace{50mu}
- F_1^{1237} F_1^{1245} u_{1257} u_{1472}
+ F_1^{1245} F_1^{3456} (u_{1253} u_{2346} - u_{1256} u_{2643})
\nn\\ & \mspace{50mu}
- \brkright[a]2{
2 F_1^{1237} F_1^{3456} u_{1273} (1 + u_{3465} - u_{3564})
}_{25}
\,.
\end{align}
Similarly, our preliminary guess for the component $f_n^{(2)}$ is
(this still involves a lot of guesswork):
\makeatletter
% increment and decrement notation for compactness:
\newcommand{\pone}[1]{\accentset{{\cc@style.}}{#1}}
\newcommand{\mone}[1]{\underaccent{{\cc@style.}}{#1}}
\newcommand{\mtwo}[1]{\underaccent{{\cc@style.\mkern-1.7mu.}}{#1}}
\newcommand{\mthree}[1]{\underaccent{{\cc@style.\mkern-1.7mu.\mkern-1.7mu.}}{#1}}
\makeatother
\begin{align}
f_{n}^{\mathrm{guess}\,(2)}&=
2n\times
\nn \\ & \mspace{-50mu}
\brkleft[a]2{
\threedoublebox_{123}
-\fivedoublepenta_{1\mthree{n}\mtwo{n}\mone{n}n}
+\fivepentabox_{12345} (2+2 u_{2435})
+\fivepentabox_{13425} (-1+u_{2354}+u_{2453})
}
\nn \\ & \mspace{-40mu}
+\sum\nolimits_{k=5}^{n} \brk!{
	 \fivepentabox_{12k3\mone{k}} (-2-2 u_{23k\mone{k}}+2 u_{2\mone{k}k3})
    +\fivepentabox_{12\mone{k}3k} (-2-2 u_{23\mone{k}k}+2 u_{2k\mone{k}3})
}
\nn \\ & \mspace{-40mu}
+\sum\nolimits_{k=6}^{n} \brk!{
     \fivepentabox_{12\mone{k}3k} (1- u_{23\mone{k}k}- u_{2k\mone{k}3})
    +\fivepentabox_{123\mone{k}k} (3+ u_{2\mone{k}3k}- u_{2k3\mone{k}})
}
\nn \\ & \mspace{-40mu}
+\fivedoublebox_{12345} (-4 u_{1324}+2 u_{1423} + \delta_{n5} (-4 u_{1354}+2 u_{1453}+u_{1325} u_{1453}))
\nn \\ & \mspace{-40mu}
+\fivedoublebox_{1243n} (-2-2 u_{13n4}+3 u_{14n3}-u_{132n} u_{14n3})
\nn \\ & \mspace{-40mu}
+\fivedoublebox_{1234n} (4+4 u_{13n4}+2 u_{1423}-2 u_{14n3}-2 u_{132n} u_{14n3})
+\fivedoublebox_{12n34} (2 u_{132n}-2 u_{1n43})
\nn \\ & \mspace{-40mu}
+\sum\nolimits_{k=5}^{n-1} \fivedoublebox_{123\mone{k}k} (-3 u_{1\mone{k}k3}-3 u_{13k\mone{k}}+2 u_{132k} u_{1\mone{k}k3})
+\fivedoublebox_{123\mone{n}n} (-2 u_{1\mone{n}23})
+\fivedoublebox_{12\mone{n}3n} u_{1\mone{n}n3}
\nn \\ & \mspace{-40mu}
+F_2^{1234} (-4+4 u_{1342})
+F_2^{12n\mone{n}} (-2 u_{12\mone{n}n})
+F_2^{12n3} (1-u_{123n}-5 u_{1n32})
\nn \\ & \mspace{-40mu}
+F_2^{12\mone{n}\mtwo{n}} (-1+u_{12\mtwo{n}\mone{n}}+u_{1\mone{n}\mtwo{n}2})
+F_2^{12n\mtwo{n}} (-1-u_{12\mtwo{n}n}+u_{1n\mtwo{n}2})
\nn \\ & \mspace{-40mu}
+F_1^{1234} F_1^{123n} (-2-2 u_{1243}+2 u_{1342}+u_{13n2}(-2+u_{124n}+u_{1342}))
\nn \\ & \mspace{-40mu}
+d\suprm{c}_1 \brk!{F_2^{13n4} (-4 u_{1n43}) +F_2^{1245} (-1+3 u_{1254}+u_{1452})}
\nn \\ & \mspace{-40mu}
+\stepfcn_{n\geq6} \brkleft[s]2{
    F_1^{123n} F_1^{1245} (-u_{12n4}(1+u_{1352}) + u_{125n}(u_{13n2} + u_{14n2}))
    }
    \nn \\ & \mspace{-20mu}
    +F_1^{123n} F_1^{1245} d\suprm{c}_2 \half (-2-2 u_{13n2}-3 u_{1254} (1+(-1+u_{123n}) u_{13n2})
    \nn \\ & \mspace{+20mu}
    +2 u_{1452}-u_{12n3} (-1+u_{1234} u_{135n}+u_{1452})+u_{124n} u_{1253} u_{14n2}+2 u_{1352} u_{14n2})
    \nn \\ & \mspace{-20mu}
    +F_1^{1245} F_1^{134n} (-u_{1453}+\half u_{125n} u_{14n2}+\half u_{135n} u_{14n3})
    \nn \\ & \mspace{-20mu}
    +F_1^{1245} F_1^{134n} d\suprm{c}_2 (-1-u_{1254}+\sfrac{3}{2} u_{1452}+\quarter u_{13n4} u_{143n} u_{1452}
    \nn \\ & \mspace{+20mu}
    +\quarter u_{12n4} u_{142n} u_{1453}-\quarter u_{125n} u_{1423} u_{14n2}-\half u_{14n3})
    \nn \\ & \mspace{-20mu}
    +F_1^{1234} F_1^{145n} (1- u_{15n4}-2 u_{12n4} u_{2435})
    \nn \\ & \mspace{-20mu}
    +F_1^{1234} F_1^{145n} d\suprm{c}_2 (u_{12n3}+\half u_{14n5}+\half u_{14n2} u_{2435}+u_{13n4} u_{2534}-\half u_{14n3} u_{2534})
    \nn \\ & \mspace{-20mu}
    +5 \fivedoublebox_{12345}
    +\fivedoublebox_{1243n} (-1+2 u_{1324}+u_{13n4}-u_{1423}-u_{14n3})
    \nn \\ & \mspace{-20mu}
    +\fivedoublebox_{1253n} (-1+u_{13n5}-u_{132n} u_{15n3})
    +\fivedoublebox_{12n34} (-2 u_{134n}+u_{1n23}-3 u_{1n43}+u_{1324} u_{1n43})
    \nn \\ & \mspace{-20mu}
    +\fivedoublebox_{12\mone{n}\mthree{n}\mtwo{n}} (-1+u_{1\mthree{n}2\mone{n}}+4 u_{1\mthree{n}\mtwo{n}\mone{n}}+u_{1\mone{n}2\mthree{n}}-2 u_{1\mthree{n}2\mtwo{n}} u_{1\mone{n}\mtwo{n}\mthree{n}})
    +d\suprm{c}_2 \fivedoublebox_{1235n} (-u_{132n} u_{15n3})
    \nn \\ & \mspace{-20mu}
    +\sum\nolimits_{k=4}^{n-2} F_2^{12k\mone{k}} (-\half+\sfrac{3}{2} u_{12\mone{k}k}+\half u_{1k\mone{k}2})
    +F_2^{124n} (-3 + \sfrac{3}{2} d\suprm{c}_2 u_{14n2} -3 d\suprm{c}_3 u_{12n4})
    \nn \\ & \mspace{-20mu}
    +B_{124356} (-\half-u_{1245}-\half u_{1346}+\half u_{1643}-2 u_{1243} u_{2356}
    +u_{1645} u_{2356}+2 u_{2653}
    \nn \\ & \mspace{+20mu}
    +2 u_{1246} u_{2653}-u_{1345} u_{2653} -2 u_{1642} u_{2653})
    \nn \\ & \mspace{-20mu}
    +\sum_{k=6}^{n}\sum_{j=3}^{k-3} \brk!{
        \pentabox_{1j\pone{j}2\mone{k}k} (-1-3 u_{12\mone{k}k}+u_{1k\mone{k}2})
        +\pentabox_{1\pone{j}j2\mone{k}k} (+1+3 u_{12\mone{k}k}-u_{1k\mone{k}2})
    }
    \nn \\ & \mspace{-20mu}
    +\sum_{k=4}^{n-2} \brkleft[s]2{
        \sum\nolimits_{j=2}^{k-2} (\pentabox_{1j\pone{j}k\pone{k}n}-\pentabox_{1\pone{j}jk\pone{k}n}) (+2-2 u_{1k\pone{k}+n}-2 u_{1n\pone{k}+k})
        }
        \nn \\ & \mspace{+50mu}
        +\pentabox_{12\mone{k}k\pone{k}n} (+2-2 u_{1k\mone{k}n}-2 u_{1n\pone{k}k}+2 u_{1k\mone{k}\pone{k}} u_{1n\pone{k}k})
        \nn \\ & \mspace{+50mu}
        +\pentabox_{1\mone{k}2k\pone{k}n} (-2+2 u_{1k2n}+2 u_{1n\pone{k}k}-2 u_{1k2\pone{k}} u_{1n\pone{k}k})
        \nn \\ & \mspace{+50mu}
        +\pentabox_{12\pone{k}\mone{k}kn} (u_{1\mone{k}kn}-u_{1\mone{k}\pone{k}n}-u_{1nk\mone{k}}+u_{1n\pone{k}\mone{k}})
        \nn \\ & \mspace{+50mu}
        +\pentabox_{1\mone{k}n2k\pone{k}} (u_{12k\pone{k}}-u_{12n\pone{k}}-u_{1\pone{k}k2}+u_{1\pone{k}n2})
        \nn \\ & \mspace{+50mu}
        +\pentabox_{13n2k\pone{k}} (1- u_{12n\pone{k}}- u_{1\pone{k}k2}+u_{12kn} u_{1\pone{k}n2})
        \nn \\ & \mspace{+50mu}
        +\pentabox_{12\mone{n}\mone{k}kn} (-1+u_{1\mone{k}kn}+u_{1n\mone{n}\mone{k}}-u_{1\mone{k}k\mone{n}} u_{1n\mone{n}k})
        \nn \\ & \mspace{+50mu}
        +\pentabox_{1\mone{k}k2\mone{n}n} (-1+1 u_{12\mone{n}n}+u_{1nk2}-u_{12k\mone{n}} u_{1n\mone{n}2})
        \nn \\ & \mspace{+50mu}
        +\brkright[s]2{
        \pentabox_{1k\mone{k}2\mone{n}n} (+1-1 u_{12\mone{n}n}-u_{1n\mone{k}2}+u_{12\mone{k}  \mone{n}} u_{1n\mone{n}2})
    }
    \nn \\ & \mspace{-20mu}
    +\sum\nolimits_{k= 5}^{ n-1} \brk!{
         \doublepenta_{123n\mone{k}k} (-1+u_{12nk}-u_{1kn2})
        +\doublepenta_{123nk\mone{k}} (1-u_{12n\mone{k}}+u_{1\mone{k}n2})
    }
    \nn \\ & \mspace{-20mu}
    +\doublepenta_{14253n} \half (-1+ u_{145n}- u_{1n54})
    +\doublepenta_{14235n} d\suprm{c}_2 (-1+u_{143n}-u_{1n34})
    \nn \\ & \mspace{-20mu}
    +\brkright[a]2{\brkright[s]2{
     \doublepenta_{1423n5}     (2 d\suprm{c}_3 - d\suprm{c}_2 u_{1435} + 2 u_{1534})
    +\doublepenta_{1425n3} \half (2 d\suprm{c}_3 - d\suprm{c}_2 u_{1453} + 2 u_{1354})
}}_{\mathrm{dihedral}}
\,,
\end{align}
where $\avg{\cdot}_n$ means averaging over the dihedral
group $K_n\equiv{\grp{D}_n}$ of external points, and we use the shorthand notation
\begin{equation}
\mthree{n}=n-3
\,,\quad
\mtwo{n}=n-2
\,,\quad
\mone{n}=n-1
\,,\quad
\pone{n}=n+1
\quad
\text{etc.}
\end{equation}
as well as
\begin{equation}
d\suprm{c}_1=\begin{cases}
0 & n=5 \\ 1 & n=6 \\ 2 & n=7
\end{cases}
\qquad
d\suprm{c}_2=\begin{cases}
1 & n=6 \\ 2 & n=7
\end{cases}
\qquad
d\suprm{c}_3=\begin{cases}
0 & n\leq6 \\ 1 & n=7
\end{cases}
\end{equation}
With this expression, we find
\begin{align}
f_{5}^{(2)}&=f_{5}^{\mathrm{guess}\,(2)}
\,,\nn\\
f_{6}^{(2)}&=f_{6}^{\mathrm{guess}\,(2)}
+ 12\,\brk[a]1{
B_{123,456}\brk{-2+u_{1436}+ u_{2456}\brk{4-u_{1432}-u_{1536}}}
}_6
\,,\nn\\
f_{7}^{(2)}&=f_{7}^{\mathrm{guess}\,(2)}
+14\,\brkleft[a]2{
-\half\doublepenta_{14,52,73}\brk{1+u_{1324}-u_{1423}}
+(12\text{ five-point terms }\fivedoublebox_{\dots})
}
\nn\\ & \mspace{50mu}
+(4\text{ six-point terms }B_{\dots})
+(3\text{ six-point terms }F_1^{\dots}F_1^{\dots})
\nn\\ & \mspace{50mu}
+(12\text{ seven-point terms }\doublepenta_{\dots})
+(4\text{ seven-point terms }\pentabox_{\dots})
\nn\\ & \mspace{50mu}
+\brkright[a]2{
(2\text{ seven-point terms }F_1^{\dots}F_1^{\dots})
}_7
\,.
\end{align}
%

%%%%%%%%%%%%%%%%%%%%%%%%%%%%%%%%%%%%%%%%%%%%%%%%%%%%%%%%%%%%
%%%%%%%%%%%%%%%%%%%%%%%%%%%%%%%%%%%%%%%%%%%%%%%%%%%%%%%%%%%%
\section{Integrals}
\label{app:Integrals}

%%%%%%%%%%%%%%%%%%%%%%%%%%%%%%%%%%%%%%%%%%%%%%%%%%%%%%%%%%%%
\subsection{Asymptotic Expansions}

The goal of this appendix is to review and explain the main idea behind
the method of asymptotic expansions that allows us to obtain the
integrals in the correlators as a series expansion in one cross ratio.
This will be a simple extension of the analysis that has been done for
four-point conformal
integrals~\cite{Eden:2012rr,Goncalves:2016vir,Georgoudis:2017meq}.

A generic two-loop finite conformal integral with at most six external
points has the following form%
\footnote{Instead of the factor of $x_{78}^2$ in the denominator,
there could be a factor $x_{78}^{2n}$, $n\geq1$ in the numerator (see
\eg \figref{fig:extra-integrals}), but such integrals do not occur in
the correlators discussed in this work, and hence we do not consider
them here. Higher powers of $x_{78}^2$ in the denominator do not
appear in our integrals since they are divergent.}
\begin{align}
I = \int \frac{[\dd[d]{x_\ell}]}{x_{78}^2\prod_{i=1}^6(x_{i7}^2)^{a_i}(x_{i8}^2)^{b_i}}
\label{eq:GenericDefinitionIntegral}
\end{align}
with $1+\sum_{i}a_i=d$, $1+\sum_{i}b_i=d$ and
$[\dd[d]{x_\ell}] \equiv \dd[d]{x_7}\dd[d]{x_8}$. Without loss of
generality, it is possible to use conformal symmetry to send one point
to infinity, say $x_6$ (if the integral only has five points, then we
send the point $x_5$ to infinity), and one point to zero, say $x_1$.
The method of asymptotic expansions can be used to obtain a series
expansion of an integral when one variable is small, say
$x_2^2\rightarrow 0$. Within this method, we are instructed to divide
each integration region into two parts, one where the integration
variable is of the size of $x_2^2$ and another where it is much bigger.%
\footnote{In the asymptotic expansion method, we are instructed to
integrate over all space and shift the dimension from $4$ to
$d=4-2\epsilon$ to regulate possible divergences in each region that
should cancel when all contributions are combined, as we shall see.}
In each region, it is possible to simplify the integrand using the
general expansion
\begin{align}
\frac{1}{(x_{ij}^2)^c} = \sum_{n=0}^{\infty} {\binom{-c}{n} }\frac{(x_j^2-2x_i\cdot x_j )^n}{(x_i^2)^{c+n}},  \ \ \ \ x_i\gg x_j\,,
\qquad
{\binom{a}{b}} \equiv \frac{\Gamma(a+1)}{\Gamma(1+a-b)\Gamma(1+b)}
\,.
\end{align}
At two loops, there are four different regions:
\begin{align}
\text{region }1 \; &: \quad x_7,x_8\approx{x_2}
\,,\nn\\
\text{region }2 \; &: \quad x_7\approx{x_2}\,,\quad x_2\ll{x_8}
\,,\nn\\
\text{region }3 \; &: \quad x_8\approx{x_2}\,,\quad x_2\ll{x_7}
\,,\nn\\
\text{region }4 \; &: \quad x_2\ll x_7,x_8
\,.
\end{align}
We will denote the integral~\eqref{eq:GenericDefinitionIntegral} with the
domain of integration restricted to the respective region by
$I_k$, $k=1,\dots,4$, and can expand:
\begin{align}
I_1 =& \sum_{n_1,\dots n_6=0}^{\infty} \prod_{i=1}^3 {\binom{-a_{i+2}}{n_i} }  {\binom{-b_{i+2}}{n_{i+3}} } \int \frac{[\dd[d]{x_\ell}]  }{x_{78}^2\prod_{i=1}^2(x_{i7}^2)^{a_i}(x_{i8}^2)^{b_i}}\prod_{j=3}^5\frac{\prod_{i=0}^1(x_{7+i}^2-2x_j\cdot x_{7+i})^{n_{j-2+3i}}}{(x_{j}^2)^{a_j+b_j+n_{j-2}+n_{j+1}}}\nonumber\\
I_2=&\sum_{n_i=0}^{\infty} {\binom{-b_{2}}{n_4} }\prod_{i=1}^3{\binom{-a_{i+2}}{n_{i}} }\int \frac{[\dd[d]{x_\ell}]  (2x_7\cdot x_8-x_7^2)^{n_5}(2x_2\cdot x_8-x_2^2)^{n_4}}{(x_{8}^2)^{1+b_{1}+b_2+n_4+n_5}\prod_{i=1}^2(x_{i7}^2)^{a_i}}\prod_{j=3}^5\frac{(x_7^2-2x_j\cdot x_7)^{n_{j-2}}}{(x_{8j}^2)^{b_j}(x_j^2)^{a_j+n_{j-2}}}\nonumber\\
I_4=&\sum_{n_1,\dots n_2=0}^{\infty} {\binom{-a_{2}}{n_1} } {\binom{-b_{2}}{n_2} } \int \frac{[\dd[d]{x_\ell}] \,\, (x_2^2-2x_2\cdot x_7)^{n_1}(x_2^2-2x_2\cdot x_8)^{n_2}  }{x_{78}^2(x_{7}^2)^{a_1+a_2+n_1}(x_{8}^2)^{b_1+b_2+n_2}\prod_{i=3}^5(x_{i7}^2)^{a_i}(x_{i8}^2)^{b_i}}
\end{align}
where $I_3$ is obtained from $I_2$ by $a_i\leftrightarrow b_i$. The
regions $1$ and $4$ are expressed in terms of two-loop integrals, while the
regions $2$ and $3$ are given by products of two one-loop integrals. But in all
of the regions the integrals have at most $4$ external points, and thus
are simpler. It is simple to extract the leading (when they go to
zero) dependence on $x_2^2$:
\begin{align}
I_1\approx (x_2^2)^{d-(1+a_1+a_2+b_1+b_2)},I_2 \approx (x_2^2)^{\frac{d}{2}-a_1-a_2},  I_4\approx (x_2^2)^0 .
\end{align}
The factors in the numerators of the integrals $I_k$ can be written as
a combination of integrals with open indices contracted with some
external vectors. As an example, take a two-loop integral depending on
two vectors $x_3$, $x_4$ with $2$ open indices:%
\footnote{The generalization to more external vectors and more indices
is straightforward. This formula follows just by the $\grp{SO}(d)$ symmetry
of the system.}
\begin{align}
\int \frac{[\dd[d]{x_\ell}] \,N^{\mu_1\mu_2}}{D(x_3,x_4)} = x_3^{\mu_1}x_3^{\mu_2}T_0+x_3^{\mu_1}x_4^{\mu_2}T_1+x_3^{\mu_2}x_4^{\mu_1}T_2+x_4^{\mu_1}x_4^{\mu_2}T_3+\delta^{\mu_1\mu_2}T_4
\,,
\end{align}
where $D$ is some denominator, and $T_i$ are scalar integrals that can
be obtained by contracting this equation with
$x_3^{\mu_1}x_3^{\mu_2},\dots,x_4^{\mu_1}x_4^{\mu_2},\delta^{\mu_1\mu_2}
$ and inverting this system of equations.

%%%%%%%%%%%%%%%%%%%%%%%%%%%%%%%%%%%%%%%%%%%%%%%%%%%%%%%%%%%%
\subsubsection*{Example: Five-Point Double Pentaladder}

One of the integrals that appear in the two-loop five-point function
is the $5$-pt double pentaladder integral
\begin{align}
I=\frac{\fivedoublepenta_{23,145}}{x_{14}^2x_{15}^2cx_{45}^2}
=\int \frac{[\dd[4]{x_\ell}] \, x_{27}^2x_{38}^2}{x_{17}^2x_{18}^2x_{28}^2x_{37}^2x_{47}^2x_{48}^2x_{57}^2x_{58}^2x_{78}^2}
\end{align}
that is obtained from~\eqref{eq:GenericDefinitionIntegral} by setting
$a_1=b_1=b_2=a_3=a_4=b_4=a_5=b_5=1, a_2=b_3=-1$. Recall that we can
send one point to infinity, say $x_5$. Let us analyze all four regions
for this integral
\begin{align}
I_1 &= \sum_{n_1,n_2,n_3=0}^{\infty}\int \frac{[\dd[d]{x_\ell}] \, x_{27}^2 x_{38}^2(2x_3\cdot x_7-x_7^2)^{n_1}(2x_4\cdot x_7-x_7^2)^{n_2}(2x_4\cdot x_8-x_8^2)^{n_3}}{x_{7}^2x_{8}^2x_{28}^2(x_{3}^2)^{1+n_1}(x_{4}^2)^{2+n_2+n_3}x_{78}^2}\nonumber\\
I_2& =\sum_{n_i=0}^{\infty}\int \frac{[\dd[d]{x_\ell}] \, x_{27}^2x_{38}^2 (2x_2\cdot x_8-x_2^2)^{n_1}(2x_7\cdot x_3-x_7^2)^{n_2}(2x_7\cdot x_4-x_7^2)^{n_3}(2x_7\cdot x_8-x_7^2)^{n_4}}{x_{7}^2(x_{8}^2)^{2+n_1+n_4}(x_{3}^2)^{1+n_2}(x_{4}^2)^{1+n_3}x_{48}^2} \nonumber\\
I_3& =\sum_{n_i=0}^{\infty}\frac{1}{(x_4^2)^{1+n_1}}\int \frac{[\dd[d]{x_\ell}] \, x_{27}^2x_{38}^2\,(2x_8\cdot x_4-x_8^2)^{n_1}(2x_8\cdot x_7-x_8^2)^{n_2}}{x_{8}^2x_{28}^2\,(x_{7}^2)^{2+n_2}x_{37}^2x_{47}^2} \nonumber\\
I_4&= \sum_{n_1=0}^{\infty}\int \frac{[\dd[d]{x_\ell}] \, x_{27}^2x_{38}^2\, (2x_8\cdot x_2-x_2^2)^{n_1}}{x_{7}^2(x_{8}^2)^{2+n_1}x_{37}^2x_{47}^2x_{48}^2x_{78}^2}.
\end{align}
Some comments are in order here: $I_2$ does not contribute since the
integral in $x_7$ integrates to zero (this is a scaleless
integral~\cite{Smirnov:1994tg}); the leading power of $I_1$ is
$(x_2^2)^{1-2\epsilon}$ and thus this region is subleading compared to
$I_3$ and $I_4$; the terms $x_8^2$ in the numerator of $I_3$ and
$x_2^2$ in the numerator of $I_4$ can be dropped to leading order in
$x_2^2\rightarrow 0$. In the following we will focus on the last two
regions%
\footnote{We have used the one loop integral formula with numerators
\begin{align}
\int \frac{\dd[d]{x_0}\, x_{0}^{\mu_1}\,\dots x_{0}^{\mu_J}}{(x_0^2)^{\alpha}(x_{02}^2)^{\beta}} =\frac{\Gamma(\alpha+\beta-\frac{d}{2})\Gamma(\frac{d}{2}-\alpha+J)\Gamma(\frac{d}{2}-\beta)}{\Gamma(\alpha)\Gamma(\beta)\Gamma(d-\alpha-\beta+J)} \frac{x_2^{\mu_1}\dots x_2^{\mu_J}}{(x_{2}^{2})^{\alpha+\beta-\frac{d}{2}}}+\dots
\end{align}
where the $\dots$ represent subleading terms in $x_2^2$.}
\begin{align}
I_3&= \frac{1}{(x_3^2)^{5-d}} \sum_{n_i=0}^{\infty}\frac{2^{n_1+n_2}\Gamma(\epsilon)\Gamma(1-\epsilon)}{(x_4^2)^{1+n_1}(x_2^2)^{\epsilon}}\bigg[\frac{x_3^2\Gamma(1-\epsilon+n_1+n_2)}{\Gamma(2-2\epsilon+n_1+n_2)}-\frac{2x_3\cdot x_2\Gamma(2-\epsilon+n_1+n_2)}{\Gamma(3-2\epsilon+n_1+n_2)}\bigg]\nonumber\\
&\times (x_2\cdot x_4)^{n_1}\int \frac{\dd[d]{x_7}\, (x_7^2-2x_2\cdot x_7)\,(x_2\cdot x_7)^{n_2}}{(x_{7}^2)^{2+n_2}x_{37}^2x_{47}^2}\nonumber\\
I_4& =\frac{1}{(x_3^2)^{5-d}}  \sum_{n_1=0}^{\infty}2^{n_1}\int \frac{\dd[4]{x_7}\dd[4]{x_8}\, (x_7^2-2x_2\cdot x_7)x_{38}^2\, (x_8\cdot x_2)^{n_1}}{x_{7}^2(x_{8}^2)^{2+n_1}x_{37}^2x_{47}^2x_{48}^2x_{78}^2}
\end{align}
where we have scaled all points by
$x_i^{\mu}\rightarrow |x_3| x_i^{\mu}$. In practice we have to
truncate upper limit of the sum, which translates into evaluating the
integral as expansion around $x_2\rightarrow 0$. The integrals in each
region might have $\epsilon$ divergences which should cancel, for each
order in $x_2$, when all regions are combined. The leading term in
$x_2\rightarrow 0$ comes from truncating both sums to the $n_i=0$ and
furthermore neglecting the $x_2\cdot x_3$ and $x_2\cdot x_7$ in the
numerators of both regions
\begin{align}
&I_3 \rightarrow \frac{1}{(x_{3}^2)^{1+2\epsilon}}\frac{\Gamma(\epsilon)\Gamma^2(1-\epsilon)}{(x_4^2)(x_2^2)^{\epsilon}}\frac{1}{\Gamma(2-2\epsilon)}\int \frac{\dd[d]{x_7}\, }{x_{7}^2x_{37}^2x_{47}^2}\\
&I_4 \rightarrow \frac{1}{(x_{3}^2)^{1+2\epsilon}}\int \frac{[\dd[d]{x_\ell}] \, x_{38}^2\, }{(x_{8}^2)^{2}x_{37}^2x_{47}^2x_{48}^2x_{78}^2}.
\end{align}
Notice that each integral is divergent but when we plug the values of
the integrals and combine them, the divergences disappear,%
\footnote{We have used integration by parts identities (IBPS) to
reduce each three point integral to a sum of master integrals that we
have computed with the \maple package
\hyperint~\cite{Panzer:2014caa}.}
giving rise to
\begin{align}
&I=\frac{1}{8} \left(\tilde{\Delta} \left(4 u_3  \partial_{u_3}\tilde{I}_2-4  \partial_{u_3}\tilde{I}_3-u_4 \partial_{u_4}\tilde{I}_1+4 u_4 \partial_{u_4}\tilde{I}_2\right)+2 u_3 u_4  \partial_{u_3}\tilde{I}_1+4 u_3^2  \partial_{u_3}\tilde{I}_2\right.\\
&- 4(u_4+1)u_3\partial_{u_3}\tilde{I}_2- +4(1-u_3-u_4)\partial_{u_3}\tilde{I}_3+u_3 u_4  \partial_{u_4}\tilde{I}_1+(u_4-1)u_4^2 \partial_{u_4}\tilde{I}_1+4 u_3 u_4  \partial_{u_4}\tilde{I}_2\nonumber\\
&\left.+4(1-u_4) u_4 \partial_{u_4}\tilde{I}_2-8 u_4 \partial_{u_4}\tilde{I}_3+8u_4 \Phi^{(1)}(2-\log \left(u_1 u_4\right))+48 \zeta_3+(u_3 -u_4+1+\tilde{\Delta} )u_4 \mathcal{I}_1\right)\nonumber
\end{align}
where $\tilde{\Delta} =\sqrt{\left(u_4-u_3+1\right)^2-4 u_4} $, we
have used the conformal frame with
$x_{5}\rightarrow \infty$, $x_{14}^2=1$, and
$\tilde{I}_j=\tilde{\Delta}\,\mathcal{I}_j$ with
\begin{align}
\mathcal{I}_1 = \int \frac{\dd[4]{x_7}\dd[4]{x_8}}{x_{17}^2 x_{18}^2 x_{37}^2 x_{38}^2 x_{47}^2 x_{78}^2}
\,,\quad
\mathcal{I}_2 = \int \frac{\dd[4]{x_7}\dd[4]{x_8}}{x_{18}^2 x_{37}^2 x_{38}^2 x_{47}^2 x_{78}^2}
\,,\quad
\mathcal{I}_3 = \int \frac{\dd[4]{x_7}\dd[4]{x_8}}{x_{17}^2 x_{18}^2 x_{37}^2 x_{48}^2 x_{78}^2}
\,.
\label{eq:I123def}
\end{align}

%%%%%%%%%%%%%%%%%%%%%%%%%%%%%%%%%%%%%%%%%%%%%%%%%%%%%%%%%%%%
\subsection{Coefficients of the Konishi Four-Point Function}
\label{app:CoefficientsKonishi}

In the main text, we have written down the expression for the two-loop
four-point correlator involving three $\tp$ operators and one
Konishi operator in terms of a linear combination of conformal integrals. The
coefficients in this linear combination are given by
\begin{align}
&a_{1,0}=\frac{c(z-1) (\bar{z}-1) }{15 z^2 \bar{z}^3}  (24 z^7 (\bar{z}-1)^2 \bar{z}^2 (2 \bar{z}-1)-z^4 (\bar{z}-1) (\bar{z} (\bar{z} (\bar{z} (2 \bar{z} (\bar{z} (48 \bar{z}^2-369 \bar{z}+169)
\nonumber\\
&+1087)-7)-1926)+123)+180)+z^6 \bar{z} (\bar{z} (\bar{z} (161-2 \bar{z} (3 \bar{z} (248 \bar{z}-523)+1028))+74)\nonumber\\
&+51)-z^5 (\bar{z} (\bar{z} (\bar{z} (2 \bar{z} (\bar{z} (6 \bar{z} (16 \bar{z}-277)+3067)-2126)+647)+552)-417)+108)\nonumber\\
&+z^3 (\bar{z} (\bar{z} (\bar{z} (\bar{z} (\bar{z} (5019-2 \bar{z} (\bar{z} (24 \bar{z}+665)-458))-10252)+5208)-1268)+483)+72)\nonumber\\
&+z^2 (\bar{z} (\bar{z} (\bar{z} (\bar{z} (\bar{z} (\bar{z} (\bar{z} (276 \bar{z}+335)-1223)+1919)+2370)-4940)+4983)-1776)-144)\nonumber\\
&-z \bar{z} (\bar{z} (\bar{z} (\bar{z} (\bar{z} (4 \bar{z} (\bar{z} (33 \bar{z}-53)+138)+1473)-2068)+75)+2280)-1152)\nonumber\\
&+3 \bar{z}^2 (\bar{z} (\bar{z} (\bar{z} (\bar{z} (17 \bar{z}+93)-77)-81)+176)-48)) \\
&a_{1,1} =\frac{2c}{15 z^2 \bar{z}^3 }  (z-1) (\bar{z}-1) (48 z^7 (\bar{z}-1)^2 \bar{z}^3-2 z^6 (\bar{z}-1) \bar{z} (\bar{z} (\bar{z} (744 \bar{z}^2-657 \bar{z}+29)+19)\nonumber\\
&-3)-z^5 (\bar{z} (\bar{z} (2 \bar{z} (\bar{z} (\bar{z} (96 \bar{z}^2-966 \bar{z}+1459)-462)+15)+313)-333)+36)\nonumber\\
&+z^4 (\bar{z} (\bar{z} (2 \bar{z} (\bar{z} (\bar{z} (\bar{z} (-48 \bar{z}^2+321 \bar{z}+98)-998)+1145)+1041)-1627)-159)+108)\nonumber\\
&+2 z^2 \bar{z} (\bar{z} (\bar{z} (\bar{z} (\bar{z} (\bar{z} (6 \bar{z}^2+82 \bar{z}-3)+1116)+470)-1047)+876)-180)\nonumber\\
&+2 z^3 (\bar{z} (\bar{z} (\bar{z} (\bar{z} (\bar{z} (\bar{z} (5 \bar{z} (12 \bar{z}-79)-114)+1067)-3251)+1631)-618)+276)-36)\nonumber\\
&+z \bar{z}^2 (\bar{z} (\bar{z} (900-\bar{z} (\bar{z} (4 \bar{z} (9 \bar{z}-8)+469)+1351))+24)-360)\nonumber\\
&+3 \bar{z}^3 (\bar{z} (\bar{z} (\bar{z} (8 \bar{z}+81)-37)+52)-24))\nonumber \\
&a_{1,2} = \frac{c}{15 z^2 \bar{z}^2} (z-1) (\bar{z}-1) (-2 z^5 (\bar{z} (\bar{z} (\bar{z} (\bar{z} (96 \bar{z}^2-630 \bar{z}+517)+73)+200)+137)-183)\nonumber\\
&+24 z^7 (\bar{z}-1)^2 \bar{z} (2 \bar{z}+1)-2 z^6 (\bar{z}-1) (\bar{z} (\bar{z} (3 \bar{z} (248 \bar{z}-91)-55)+23)-15)\nonumber\\
&+z^4 (\bar{z} (\bar{z} (\bar{z} (2 \bar{z} (\bar{z} (3 \bar{z} (51-16 \bar{z})+284)-883)+1557)+2635)-1062)-552)\nonumber\\
&+z^3 (\bar{z} (\bar{z} (\bar{z} (\bar{z} (2 \bar{z} (\bar{z} (72 \bar{z}-107)-109)+1047)-5514)+2385)-456)+576)\nonumber\\
&-z^2 \bar{z} (\bar{z} (\bar{z} (\bar{z} (\bar{z} (6 \bar{z} (6 \bar{z}+5)+145)-2872)+243)+720)+288)+z \bar{z}^2 (\bar{z} (264-\bar{z} (\bar{z} (\bar{z} (12 \bar{z}\nonumber\\
&-25)+496)+687))+576)+3 (\bar{z}-1) \bar{z}^4 (5 \bar{z}+88))\nonumber\\
&b_{1,0} =-\frac{4c}{15 z \bar{z} } (z-1) (\bar{z}-1) (z^7 \bar{z} (6 \bar{z} (\bar{z} (\bar{z} (3 \bar{z}-10)+12)-6)+1)+z^6 (3 \bar{z} (\bar{z} (9-\bar{z} (\bar{z} (4 \bar{z} (5 \bar{z}\nonumber\\
&-9)+15)+3))+1)+1)+3 z^5 (\bar{z} (\bar{z} (\bar{z} (\bar{z} (2 \bar{z} (3 \bar{z} (\bar{z}+6)-17)-57)+42)-5)+9)-12)\nonumber\\
&-z^4 (\bar{z} (\bar{z} (\bar{z} (\bar{z} (3 \bar{z} (5 \bar{z} (4 \bar{z}+3)+57)-916)+804)-126)+9)-72)\nonumber\\
&+z^3 (\bar{z} (\bar{z} (\bar{z} (3 \bar{z} (3 \bar{z} (\bar{z} (8 \bar{z}-1)+14)-268)+916)-171)-45)-60)\nonumber\\
&-3 z^2 (\bar{z} (\bar{z} (\bar{z} (\bar{z} (\bar{z} (3 \bar{z} (4 \bar{z}-3)+5)-42)+57)+34)-36)-6)\nonumber\\
&+z \bar{z} (\bar{z} (\bar{z} (\bar{z} (\bar{z} (\bar{z} (\bar{z}+3)+27)-9)-45)+108)-60)+\bar{z}^2 (\bar{z} (\bar{z} ((\bar{z}-36) \bar{z}+72)-60)+18)) \nonumber\\
&b_{1,1} =-\frac{4c}{15 z \bar{z}}(z-1) (\bar{z}-1) (36 z^7 (\bar{z}-1)^3 \bar{z}^2+z^6 (2-3 (1-2 \bar{z})^2 \bar{z} (5 \bar{z}^2 (2 \bar{z}-1)-1))\nonumber\\
&+z^3 (\bar{z}^2 (\bar{z} (3 \bar{z} (9 \bar{z} (4 \bar{z}^2+\bar{z}+8)-268)+608)+57)-12)+3 z^5 (\bar{z} (\bar{z} (\bar{z} (\bar{z} (4 \bar{z} (3 \bar{z} (\bar{z}+5)\nonumber\\
&+1)-133)+72)-5)+22)-12)-3 z^4 (\bar{z} (\bar{z} (\bar{z} (\bar{z} (\bar{z} (6 \bar{z} (6 \bar{z}+5)+133)-408)+268)-12)\nonumber\\
&+15)-12)-3 z^2 \bar{z} (\bar{z} (\bar{z} (\bar{z} (\bar{z} (4 \bar{z} (3 \bar{z}+1)+5)-12)-19)+72)-12)+3 z \bar{z}^2 ((\bar{z} (\bar{z}+22)\nonumber\\
&-15) \bar{z}^2+12)+2 \bar{z}^3 (\bar{z} ((\bar{z}-18) \bar{z}+18)-6)) \nonumber\\
&b_{1,2} = \frac{-2c}{15 z \bar{z} }  (z-1) (\bar{z}-1) (z^7 (\bar{z}-1) (12 \bar{z}^3 (3 \bar{z}-5)-1)-3 z^2 \bar{z}^2 (\bar{z} (\bar{z} (3 (\bar{z}-2) \bar{z}+41)-36)\nonumber\\
&+48)+z^6 (1-\bar{z} (3 \bar{z} (2 \bar{z} (\bar{z} (4 \bar{z} (5 \bar{z}-6)+5)-2)+3)+8))+3 z^5 \bar{z} (\bar{z} (\bar{z} (4 \bar{z} (\bar{z} (3 \bar{z} (\bar{z}+4)+7)\nonumber\\
&-29)+19)+6)+23)-z^4 (\bar{z} (\bar{z} (\bar{z} (2 \bar{z} (3 \bar{z} (\bar{z} (16 \bar{z}+5)+58)-433)+347)+123)+60)\nonumber\\
&+12)+z^3 \bar{z} (\bar{z} (\bar{z} (\bar{z} (3 \bar{z} (4 \bar{z} (5 \bar{z}+1)+19)-347)+250)+108)+60)-z \bar{z}^3 ((\bar{z}-1) \bar{z} (\bar{z} (\bar{z}+9)\nonumber\\
&-60)-60)+\bar{z}^7+\bar{z}^6-12 \bar{z}^4)\nonumber
\end{align}
where $u_4=z\bar{z}$, $u_3u_5=(1-z)(1-\bar{z})$, and
$c= {\Nc^2 \left(\Nc^2-1\right) }/{(z-\bar{z})^4} $. The
other coefficients can be extracted from the ancillary \mathematica
file \filename{CorrelatorInLimit.m}.

The procedure to obtain these coefficients is the following:
Start by choosing the polarizations of points $x_1$ and $x_2$ according
to~\eqref{eq:SL2polarizations}, then insert the expressions of the
conformal integrals, expanded as explained in the previous subsection,%
\footnote{We provide these expanded
conformal integrals in the ancillary file
\filename{ListFivePtIntegrals.wl}.}
take the
limit $x_2\rightarrow x_1$ and keep terms only to sub-subleading order
(since the integrals
were computed up to sub-subleading order in $x_2\rightarrow x_1$,
one is able to extract information about two primaries),
use the light-cone conformal blocks (or alternatively the conformal
Casimir equations), subtract the contribution of the twenty prime
operator, and finally read off the coefficients by translating powers
of $u_5-1$ and $u_2-1$ to the corresponding structures of a spinning
four-point function. The resulting coefficients are listed in the file.

%%%%%%%%%%%%%%%%%%%%%%%%%%%%%%%%%%%%%%%%%%%%%%%%%%%%%%%%%%%%
\subsection{Simple One-Loop Integrals}
\label{app:OneLoopIntegrals}

In the main text, we have expressed the one-loop correlator involving
half-BPS operators and the Lagrangian in terms of the following
one-loop conformal integrals:
\begin{align}
&\int \frac{\dd[4]{x_0} x_{06}^2}{x_{10}^2x_{20}^2x_{30}^2x_{40}^2x_{50}^2} = t_{12345,6},\\
&\int \frac{\dd[4]{x_0} (x_{07}^2)^2}{x_{10}^2x_{20}^2x_{30}^2x_{40}^2x_{50}^2x_{60}^2} =t_{123456,7}.
\end{align}
The first can be expressed in terms of linear combinations of one-loop
box integrals while the second can be expressed in terms of the first,
once we notice that $x_{07}^2 = \sum_{i=1}^6a_i x_{i0}^2$
\begin{align}
&t_{12345,6} =\frac{1}{\det x_{ij}^2} \sum_{i=1}^{5}a_{i} I_{i}
\end{align}
where $I_i$ are one-loop box integrals where the position $i$ is
absent and $a_{i}$ are polynomials in $x_{kl}^2$ with degree
\begin{align}
&\textrm{degree $1$ in $6$ and $i$ and $2$ for the other points}
\end{align}
for $a_i$. These coefficients are highly constrained by symmetry
(permutation of the points) and the fact that the integral should
reduce to the usual ladder in the limit $x_6\rightarrow x_i$ with
$x_i$ being any of the external points in the denominator of the
integral. For example, we have
\begin{align}
a_{5} = (x_{56}^2 \det x_{ij}^2) -2 \sum_{i=1}^4x_{i5}^2x_{i6}^2  \det\nolimits'_i x_{lk}^2+\sum_{i\neq j =1}^5 x_{i5}^2x_{j6}^2 V_{ij}
\end{align}
where
$V_{12} =
x_{34}^2(x_{14}^2x_{23}^2+x_{13}^2x_{24}^2-x_{12}^2x_{34}^2)$ and the
prime in the determinant means that point $x_i$ has been removed.

%%%%%%%%%%%%%%%%%%%%%%%%%%%%%%
% Bibliography
%%%%%%%%%%%%%%%%%%%%%%%%%%%%%%
\pdfbookmark[1]{\refname}{references}
\addcontentsline{toc}{section}{References}
\bibliographystyle{nb}
% \bibliography{myrefs}
\bibliography{references}

\end{document}